\documentclass[a4paper,11pt]{article}

\usepackage{xcolor}
\definecolor{nosaka}{rgb}{0.7, 0.3, 0.0}
\definecolor{nosaka2}{rgb}{0.3, 0.7, 0.0}
\definecolor{fran}{rgb}{0.040, 0.475, 0.435}
\definecolor{nao}{rgb}{0.55, 0.0, 0.45}

\usepackage{graphicx}
\usepackage{amsmath}
\usepackage{mathrsfs}
\usepackage{arydshln}
\usepackage{color}
\usepackage{amssymb}
\usepackage{ulem}
\usepackage{bm}
\usepackage{xfrac}
\usepackage{nicefrac}
\usepackage{braketmod}

\usepackage{geometry}
\numberwithin{equation}{section}
\allowdisplaybreaks

\begin{document}

\newcommand{\hiduke}[1]{\hspace{\fill}{\small [{#1}]}}
\newcommand{\aff}[1]{${}^{#1}$}
\renewcommand{\thefootnote}{\fnsymbol{footnote}}

\global\long\def\cs{\text{CS}}%

\global\long\def\Iside{\mathrm{vec}}%

\global\long\def\Icross{\mathrm{mat}}%

\global\long\def\bra#1{\Bra{#1}}%

\global\long\def\bbra#1{\Bbra{#1}}%

\global\long\def\ket#1{\Ket{#1}}%

\global\long\def\kket#1{\Kket{#1}}%

\global\long\def\braket#1{\Braket{#1}}%

\global\long\def\bbraket#1{\Bbraket{#1}}%

\global\long\def\brakket#1{\Brakket{#1}}%

\global\long\def\bbrakket#1{\Bbrakket{#1}}%

\newcommand{\fmmn}{
\left[ \text{FMMN} \right]}

\begin{titlepage}
\begin{flushright}
{\footnotesize YITP-22-11}
\end{flushright}
\begin{center}
{\Large\bf
M2-branes and
$\mathfrak{q}$-Painlev\'e equations}\\
\bigskip\bigskip
\bigskip\bigskip
{\large Giulio Bonelli,\footnote{\tt bonelli(at)sissa.it}}\aff{1,2,3}
{\large Fran Globlek,\footnote{\tt fgloblek(at)sissa.it}}\aff{1,2,3}
{\large Naotaka Kubo,\footnote{\tt naotaka.kubo(at)yukawa.kyoto-u.ac.jp}}\aff{4}
{\large Tomoki Nosaka,\footnote{\tt nosaka(at)yukawa.kyoto-u.ac.jp}}\aff{1,2,3,5,6}\\
{\large and Alessandro Tanzini\footnote{\tt tanzini(at)sissa.it}}\aff{1,2,3}\\
\bigskip\bigskip
\aff{1} {\small
\it International School for Advanced Studies (SISSA), Via Bonomea 265, 34136 Trieste, Italy
}\\
\aff{2} {\small
\it Institute for Geometry and Physics, IGAP, via Beirut 2, 34151 Trieste, Italy
}\\
 \aff{3} {\small
 \it INFN Sezione di Trieste, 
 Italy
}\\
\aff{4} {\small
\it Center for Gravitational Physics, Yukawa Institute for Theoretical Physics, Kyoto University, Sakyo-ku, Kyoto 606-8502, Japan
}\\
\aff{5} {\small
\it RIKEN Interdisciplinary Theoretical and Mathematical Sciences (iTHEMS),
Wako, Saitama 351-0198, Japan
}\\
\aff{6} {\small
\it Kavli Institute for Theoretical Sciences and CAS Center for Excellence in Topological Quantum Computation, University of Chinese Academy of Sciences, Beijing, 100190, China
}\\

\bigskip
\end{center}
\bigskip
\bigskip
\begin{abstract}
In this paper we investigate a novel connection between the effective theory of M2-branes on 
$(\mathbb{C}^2/\mathbb{Z}_2\times \mathbb{C}^2/\mathbb{Z}_2)/\mathbb{Z}_k$
and the $\mathfrak{q}$-deformed Painlev\'e equations, by proposing that the grand canonical partition function of the corresponding four-nodes circular quiver 
$\mathcal{N}=4$ Chern-Simons matter theory solves the $\mathfrak{q}$-Painlev\'e VI equation.
 We analyse how this describes the moduli space of the topological string on local  $\text{dP}_5$ and, via geometric engineering, five dimensional $N_f=4$  $\text{SU}(2)$ $\mathcal{N}=1$ gauge theory on a circle.
The results we find extend
the known relation between ABJM theory, $\mathfrak{q}$-Painlev\'e $\text{III}_3$, and topological strings on local
${\mathbb P}^1\times{\mathbb P}^1$.
From the mathematical viewpoint the quiver Chern-Simons theory provides a conjectural Fredholm determinant realisation of the $\mathfrak{q}$-Painlev\'e VI $\tau$-function.
We provide evidence for this proposal by analytic and numerical checks 
and discuss in detail the successive decoupling limits down to $N_f=0$, corresponding to $\mathfrak{q}$-Painlev\'e$\,\,$III${}_3$.

\end{abstract}

\bigskip\bigskip\bigskip

\end{titlepage}

\renewcommand{\thefootnote}{\arabic{footnote}}
\setcounter{footnote}{0}

\tableofcontents

\section{Introduction and Summary}

The correspondence among supersymmetric gauge theories, topological string and isomonodromic deformation problems has led to new insights both on the non-perturbative structure of gauge and string theories on a side and on the other provided new explicit solutions to old mathematical physics problems.
Starting from \cite{Gamayun:2012ma} where a general solution of Painlev\'e equations was proposed in terms of instanton counting in four dimensional $\mathcal{N}=2$ gauge theories \cite{Nekrasov:2002qd}, several further studies followed concerning
Argyres-Douglas sectors \cite{Bonelli:2016qwg}, the relation with blow-up equations \cite{Bershtein:2014yia,Nekrasov:2020qcq,Jeong:2020uxz},
the uplift to $\mathfrak{q}$-difference equations and their relation with  topological strings and five dimensional ${\mathcal N}=1$ gauge theories on ${\mathbb R}^4\times S^1$  \cite{Bonelli:2016idi,Bershtein:2016aef,Bonelli:2017ptp,Bonelli:2017gdk,2001CMaPh.220..165S,Gu:2021ize,Bershtein:2021gkq} and
the relation with BPS spectra of the latter and cluster integrable systems \cite{Bershtein:2017swf,Bershtein:2018srt,Bonelli:2020dcp,DelMonte:2021ytz}.
The five dimensional gauge theory is directly linked to M-theory by geometric 
engineering via compactification on Calabi-Yau threefold \cite{Hollowood:2003cv}. Due to the fact that 
the set of dimensionful couplings of the gauge
theory arise on equal footing as K\"ahler moduli coupled to the dual topological string, the identification of gauge couplings, Coulomb parameters and masses arises only after
a suitable choice of frame is done 
on the K\"ahler cone.

The automorphisms of the BPS quiver governing the 
states counting of the topological string generate 
different sets of ${\mathfrak q}$-difference equations 
according to the specific choice of the K\"ahler parameter
on which the topological string amplitudes is perturbatively expanded \cite{Bonelli:2020dcp}. 
Typically in the literature these are considered 
in the frame where the evolution is defined 
on the gauge coupling of the five dimensional gauge theory, but 
other perturbative expansions are possible.
In this paper we consider the perturbative expansion naturally arising from the correspondence with the
M2-branes' effective theory.
The prototypical example of this is the ABJM theory \cite{Aharony:2008ug,Aharony:2008gk} describing
the topological string on local 
${\mathbb P}^1\times{\mathbb P}^1$ and its relation with $\mathfrak{q}$-Painlev\'e III${}_3$ 
\cite{Bonelli:2017gdk}.
The ABJM theory describes the effective theory 
of M2-branes transverse to a ${\mathbb C}^4/{\mathbb Z}_k$ orbifold space.
The crucial issue for the identification of the two geometrical pictures above is the 
proposed correspondence between
the ABJM grand canonical partition function and a non-perturbative completion of topological strings in terms of the Fredholm determinant of the quantum mirror curve. This proposal 
is dubbed Topological String/Spectral theory (TS/ST) correspondence
\cite{Hatsuda:2013oxa,Grassi:2014zfa}. 
Indeed, the results we present in this paper provide new tests and examples of this correspondence.

The aim of this paper is to enlarge 
this picture to other M2-brane effective theories encompassing more general geometric backgrounds for the topological string. The interest of this is two-fold.
On the physical side, it gives new calculational tools for the topological string and five dimensional gauge theories BPS state counting beyond the ordinary perturbative frames. On the mathematical side, it provides new Fredholm determinant expressions for the $\tau$-functions of $\mathfrak{q}$-Painlev\'e equations. Moreover, it is worth to observe that this viewpoint 
opens a new algebraic characterization of duality symmetries of supersymmetric gauge theories in three dimensions in terms of affine Weyl group associated to the $\mathfrak{q}$-difference equations
\cite{Kubo:2019ejc,Furukawa:2020cjp,Kubo:2021enh,Furukawa:2021pll}.
Finally, we expect this line of thought to shed light on the classification problem of five dimensional superconformal field theories with eight supercharges.

Let us be more specific on few of the issues touched so far.

Painlev\'e equations are second order non-linear ordinary differential equations whose branch points are not movable by tuning the initial conditions (Painlev\'e property). One of the original motivations for their study was to define new special functions as their solutions.
Painlev\'e equations also play an important r\^ole in various problems in mathematical physics.
In \cite{Gamayun:2012ma,Iorgov:2014vla} an explicit expression for the solutions of the Painlev\'e VI equation was obtained in the short time expansion. It goes as follows.
First of all, extending the common technique used for integrable systems to express a non-linear differential equation as a consistency condition of a linear problem, the Painlev\'e VI equation has been reformulated as the isomonodromic deformation of the $sl(2,\mathbb{C})$ linear system on the Riemann sphere with four regular singular points.
In \cite{Gamayun:2012ma,Iorgov:2014vla} it has been proposed that the isomonodromic $\tau$-function of Painlev\'e VI is given by the Fourier transform of four point conformal blocks of the $c=1$ Liouville theory on the sphere, or, more specifically, its multiplicative Zak transform.
Due to AGT correspondence \cite{Alday:2009aq}, the four point conformal block is further related to the Nekrasov partition function on the self-dual $\Omega$-background. This pertains to four dimensional gauge theory. 

One can also consider a $\mathfrak{q}$-difference version of Painlev\'e equations by uplifting the linear system to its discrete version \cite{Grammaticos:1991zz,Ramani:1991zz,Jimbo:9507010,Kajiwara:2017}. A classification of $\mathfrak{q}$-Painlev\'e equations was proposed in
\cite{2001CMaPh.220..165S}
based on algebraic geometric methods. The gauge theoretical counterpart of this generalisation is given
by five dimensional ${\cal N}=1$ gauge theories compactified on $S^1$. The class of the latter admitting a five-dimensional UV-completion was previously studied in \cite{Seiberg:1996bd} by using string theoretical arguments. Strikingly, the two classifications match.

In this context a Fredholm determinant realisation\footnote{
Note that this Fredholm determinant is different from the spectral determinant of the auxiliary linear problem associated with the Painlev\'e equation.
See for example \cite{Bershtein:2021uts}.}
of the $\mathfrak{q}$-Painlev\'e $\text{III}_3$ $\tau$-function was found in terms of the grand partition function of ABJM matrix model \cite{Bonelli:2016idi,Bonelli:2017gdk} building on TS/ST correspondence.
Note that the fugacity $\kappa$ dual to the rank $N$ of the matrix model corresponds to the initial condition of the ($\mathfrak{q}$-)Painlev\'e equation.
In particular, the matrix model can be evaluated exactly for finite values of the rank $N$ ($N=1,2,\cdots$), giving the small $\kappa$ expansion of the grand partition function.
By using these data one can check $\mathfrak{q}$-Painlev\'e equation perturbatively in $\kappa$.
We will adopt this technique in section \ref{sec4}. 

In this paper we study in detail the case of $\mathfrak{q}$-Painlev\'e VI, corresponding to five-dimensional ${\cal N}=1$ $\text{SU}\left(2\right)$ gauge theory with four fundamental hypermultiplets $N_f=4$. Via geometric engineering this corresponds to topological strings on the local $D_5$ del Pezzo Calabi-Yau threefold. 
The key idea of our analysis is to generalise the correspondence between $\mathfrak{q}$-Painlev\'e III${}_3$, topological strings on local
$\mathbb{P}^1\times \mathbb{P}^1$
and ABJM theory.
The latter is three-dimensional ${\cal N}=6$ $\text{U}\left(N\right)_k\times \text{U}\left(N+M\right)_{-k}$ superconformal Chern-Simons matter theory,
where the subscripts denote the Chern-Simons levels.
The evaluation of its partition function on $S^3$ reduces by
supersymmetric localization \cite{Kapustin:2009kz} to a matrix model of rank $2N+M$.
By applying the Fermi gas formalism \cite{Marino:2011eh} one can rewrite this matrix model in terms of $N$ dimensional spectral traces \cite{Awata:2012jb,Honda:2013pea}, whose kernel coincides with the inverse quantum mirror curve of local $\mathbb{P}^1\times \mathbb{P}^1$ \cite{Kashaev:2015wia}. This provides a Fredholm determinant presentation for the $\tau$-function of $\mathfrak{q}$-Painlev\'e $\text{III}_3$ \cite{Bonelli:2017gdk}.
The $\mathfrak{q}$-difference parameter corresponds to the Chern-Simons level $k$, while the time variable $t$ is realized by the rank difference $M$ \cite{Honda:2014npa}.
Hence it is natural to expect that the matrix models corresponding to the higher $\mathfrak{q}$-Painlev\'e equations are realized by the partition functions of more general quiver superconformal Chern-Simons matter theories, where additional Painlev\'e parameters get realised by the relative ranks of the gauge nodes.
Indeed, through a series of intensive studies of the four-nodes theory $\text{U}\left(N\right)_k\times \text{U}\left(N\right)_0\times \text{U}\left(N\right)_{-k}\times \text{U}\left(N\right)_0$ \cite{Marino:2011eh,Moriyama:2014gxa,Moriyama:2014nca} 
which describes $N$ M2-branes placed on $\left(\mathbb{C}^2/\mathbb{Z}_2\times \mathbb{C}^2/\mathbb{Z}_2\right)/\mathbb{Z}_k$ orbifold \cite{Imamura:2008nn} it was found that the large $\mu=\log \kappa$ expansion of the modified grand potential $J\left(\mu\right)$ (related to the grand partition function as $\Xi\left(\kappa\right)=\sum_n e^{J\left(\mu+2\pi in\right)}$) \cite{Moriyama:2017gye,Moriyama:2017nbw,Kubo:2018cqw,Kubo:2019ejc} of this theory is consistent with the refined topological string free energy on local $D_5$ del Pezzo geometry at large radius. 
Later this result was generalized to the case of different ranks $\text{U}\left(N_1\right)_k\times \text{U}\left(N_2\right)_0\times \text{U}\left(N_3\right)_{-k}\times \text{U}\left(N_4\right)_0$, 
finding that the three extra degrees of freedom of the rank differences realize 
a three dimensional sublattice of quantised values in the full K\"ahler moduli space of local $D_5$.
In the present work we show that the five dimensional moduli space can be realized by turning on the Fayet-Iliopoulos parameters of the Chern-Simons matter theory.
    As it is the case for the correspondence between the ABJM theory and $\mathfrak{q}$-Painlev\'e $\text{III}_3$, by using the exact values of the partition function of $\text{U}\left(N_1\right)_k\times \text{U}\left(N_2\right)_0\times \text{U}\left(N_3\right)_{-k}\times \text{U}\left(N_4\right)_0$ theory 
at fixed moduli we can check the $\tau$-form of the $\mathfrak{q}$-Painlev\'e VI equation in the small $\kappa$ expansion, see Tab.\ref{220119_tableexactvalues} for a summary of our results.

This paper is organized as follows.
In section \ref{sec_background}, we recall some background material and fix our notations. In section \ref{tre},
we perform a detailed analysis of the matrix model of the quiver superconformal Chern-Simons theory and the related quantum curve. In section \ref{sec4}, we extensively check that the grand partition function of the above theory satisfies $\mathfrak{q}$-Painlev\'e equations, thus providing a conjectural Fredholm determinant representation for the corresponding $\tau$-functions. In section \ref{five} we describe the coalescence limits from the viewpoint of the analysis of matrix models and quantum curves, providing matrix model realizations of the $\mathfrak{q}$-Painlev\'e $\tau$-functions. In section \ref{sei} we discuss the coalescence limit from the viewpoint of $\mathfrak{q}$-difference equations
by considering both the perturbative gauge theory 
realisation of the $\tau$-function and the 
magnetic matrix model one. Finally,
in section \ref{sec_discussion} we discuss some open questions for further investigation.
We collect in the appendices some relevant definitions and details of some computations.

 \section{
Five dimensional gauge theory, $\mathfrak{q}$-Painlev\'e and TS/ST 
correspondence
}
\label{sec_background}
\subsection{Five dimensional gauge theory and $\mathfrak{q}$-deformed PVI equations in bilinear form}
\label{sec_5dPainlevegauge}
In this section we briefly review some aspects of Painlev\'e/gauge theory correspondence that are needed to fix the notation.
The relation between Painlev\'e VI differential equation and two dimensional Liouville CFT with $c=1$ was first noticed in \cite{Gamayun:2012ma}.
This connection arises from the formulation of Painlev\'e VI equation as the isomonodromic deformation problem of an auxiliary $sl(2,\mathbb{C})$ linear system on the Riemann sphere with four regular punctures.
This linear system is solved in terms of the degenerate five point conformal block of $c=1$ Liouville CFT \cite{Iorgov:2014vla}. 
By AGT correspondence \cite{Alday:2009aq} the isomonodromic $\tau$ function is indeed given as the Fourier transform of the full Nekrasov partition function, introduced in \cite{Nekrasov:2003rj}.
In this sense the Painlev\'e equation can be viewed as a non-trivial identity among Nekrasov partition functions. Indeed, Painlev\'e/gauge theory
correspondence can be proven in this case in terms of blow-up equations \cite{Bershtein:2014yia,Nekrasov:2020qcq}.

The $\mathfrak{q}$-deformed Painlev\'e VI equation is defined through the $\mathfrak{q}$-difference version of the analogue isomonodromic deformation problem \cite{Jimbo:9507010}.
This suggests a connection between $\mathfrak{q}$-PVI system and $\mathfrak{q}$-Virasoro algebra.
Indeed, 
it was shown in \cite{Jimbo:2017ael} that the associated linear system can be solved in terms of $\mathfrak{q}$-Virasoro five point conformal blocks.

By using the five dimensional uplift of AGT correspondence \cite{Awata:2009ur,Awata:2011ce} $\mathfrak{q}$-Virasoro four point conformal block 
can be computed in terms of the
five dimensional Nekrasov-Okounkov \cite{Nekrasov:2003rj} (NO) partition function (with Chern-Simons level zero), as:
\begin{align}
\tau\left(\theta_0,\theta_1,\theta_t,\theta_\infty;s,\sigma,t\right)&=\sum_{n\in\mathbb{Z}}s^nt^{\left(\sigma+n\right)^2-\theta_t^2-\theta_0^2}
C\left(\theta_0,\theta_1,\theta_t,\theta_\infty;\sigma+n\right)
Z\left(\theta_0,\theta_1,\theta_t,\theta_\infty;\sigma+n,t\right),\label{tau}\\
C\left(\theta_0,\theta_1,\theta_t,\theta_\infty;\sigma\right)&=\frac{\prod_{\epsilon,\epsilon'=\pm}
G_{\mathfrak{q}}\left(1+\epsilon\theta_\infty-\theta_1+\epsilon'\sigma\right)
G_{\mathfrak{q}}\left(1+\epsilon\sigma-\theta_t+\epsilon'\theta_0\right)
}
{G_{\mathfrak{q}}\left(1+2\sigma\right)G_{\mathfrak{q}}\left(1-2\sigma\right)}, \\
Z\left(\theta_0,\theta_1,\theta_t,\theta_\infty;\sigma,t\right)&=\sum_{\lambda_+,\lambda_-}t^{|\lambda_+|+|\lambda_-|}\frac{\prod_{\epsilon,\epsilon'=\pm}N_{\phi,\lambda_{\epsilon'}}\left(\mathfrak{q}^{\epsilon\theta_\infty-\theta_1-\epsilon'\sigma}\right)
N_{\lambda_{\epsilon},\phi}\left(\mathfrak{q}^{\epsilon\sigma-\theta_t-\epsilon'\theta_0}\right)}{\prod_{\epsilon,\epsilon'}N_{\lambda_\epsilon,\lambda_{\epsilon'}}\left(\mathfrak{q}^{\left(\epsilon-\epsilon'\right)\sigma}\right)}.
\label{5dNO}
\end{align}
Here $C$,$Z$ are respectively the one-loop determinant and the instanton partition function of the 5d ${\cal N}=1$ $\text{SU}\left(2\right)$ Yang-Mills with $N_f=4$, whose Seiberg-Witten curve reads
\begin{align}
&\left(w- m_1'\right)\left(w- m_2'\right)v^2\nonumber \\
&\quad
+\left[
-\left(\left( m_1' m_2'\right)^{\frac{1}{2}}
+q'\left(\frac{1}{ m_3' m_4'}\right)^{\frac{1}{2}}
\right)w^2
+Ew
- m_1'
 m_2'
\left(\left(\frac{1}{ m_1' m_2'}\right)^{\frac{1}{2}}
+q'\left(m_3' m_4'\right)^{\frac{1}{2}}
\right)
\right]v
\nonumber \\
&\quad +q'\left(\frac{
 m_1'
 m_2'
}{
 m_3'
 m_4'
}\right)^{\frac{1}{2}}
\left(w- m_3'\right)\left(w- m_4'\right)=0,
\label{quantumD5delPezzocurve}
\end{align}
with
\begin{align}
\mathfrak{q}^{\theta_0}=\left(\frac{ m_1'}{ m_3'}\right)^{\frac{1}{2}},\quad
\mathfrak{q}^{\theta_1}=\left( m_2' m_4'\right)^{\frac{1}{2}},\quad
\mathfrak{q}^{\theta_t}=\left( m_1' m_3'\right)^{-\frac{1}{2}},\quad
\mathfrak{q}^{\theta_\infty}=\left(\frac{ m_4'}{ m_2'}\right)^{\frac{1}{2}},\quad
t=q'\left(\frac{
 m_2'
 m_4'
}
{
 m_1'
 m_3'
}
\right)^{\frac{1}{2}}.
\label{qPVparametersandrescaled5dparameters}
\end{align}
The Omega-background is chosen to be self-dual $\epsilon_2=-\epsilon_1$, and $\mathfrak{q}=e^{-\beta\epsilon_1}$, where $\beta$ is the radius of $S^1$ on which the five dimensional theory is compactified.
This parameter identification can be obtained by comparing the results in \cite{Bao:2011rc} and \cite{Jimbo:2017ael}.
Note that solving $\mathfrak{q}$-PVI equations does not fix uniquely the choice $C$ in \eqref{5dNO}, as it was also the case for $\mathfrak{q}$-P$\text{III}_3$ \cite{Bershtein:2016aef,Bonelli:2017gdk}.
We will address this point further in section \ref{sec_discussion}.
This $\tau\left(\theta_0,\theta_1,\theta_t,\theta_\infty;s,\sigma,t\right)$ is a $\mathfrak{q}$-uplift of the isomonodromic $\tau$ function of the Painlev\'e VI differential equation.
The relation between the $\tau$-function and the $\mathfrak{q}$-Painlev\'e transcendents is not obvious.
Nevertheless, building on results in the differential case one can define them via some identities satisfied by the $\mathfrak{q}$-uplifted $\tau$-function \eqref{tau}.
Specifically, for $\mathfrak{q}$-PVI we define $y,z$ as
\begin{align}
y=\mathfrak{q}^{-2\theta_1-1}\cdot t\frac{\tau_3\tau_4}{\tau_1\tau_2},\quad
z=-\mathfrak{q}^{\theta_t-\theta_1-1}\cdot t\frac{\underline{\tau}_7\tau_8}{\underline{\tau}_5\tau_6},
\end{align}
where $\tau_i,\underline{\tau}_i,\bar{\tau}_i$ are defined from a single $\tau$-function $\tau\left(\theta_0,\theta_1,\theta_t,\theta_\infty;s,\sigma,t\right)$ as
\begin{align}
&\tau_1\left(t\right)=\tau\left(\theta_0,\theta_1,\theta_t,\theta_\infty+\frac{1}{2};s,\sigma,t\right),\quad
&&\tau_2\left(t\right)=\tau\left(\theta_0,\theta_1,\theta_t,\theta_\infty-\frac{1}{2};s,\sigma,t\right),\label{tau1and2}\\
&\tau_3\left(t\right)=\tau\left(\theta_0+\frac{1}{2},\theta_1,\theta_t,\theta_\infty;s,\sigma+\frac{1}{2},t\right),\quad
&&\tau_4\left(t\right)=\tau\left(\theta_0-\frac{1}{2},\theta_1,\theta_t,\theta_\infty;s,\sigma-\frac{1}{2},t\right),\\
&\tau_5\left(t\right)=\tau\left(\theta_0,\theta_1-\frac{1}{2},\theta_t,\theta_\infty;s,\sigma,t\right),\quad
&&\tau_6\left(t\right)=\tau\left(\theta_0,\theta_1+\frac{1}{2},\theta_t,\theta_\infty;s,\sigma,t\right),\\
&\tau_7\left(t\right)=\tau\left(\theta_0,\theta_1,\theta_t-\frac{1}{2},\theta_\infty;s,\sigma+\frac{1}{2},t\right),\quad
&&\tau_8\left(t\right)=\tau\left(\theta_0,\theta_1,\theta_t+\frac{1}{2},\theta_\infty;s,\sigma-\frac{1}{2},t\right),\\
&\underline{\tau}_i\left(t\right)=\tau_i\left(\mathfrak{q}^{-1}t\right),\quad
\bar{\tau}_i\left(t\right)=\tau_i\left(\mathfrak{q}t\right).
\label{tau1to8underlinebar}
\end{align}
The $\mathfrak{q}$-PVI equations follow from the (conjectured) bilinear identities satisfied by the five dimensional Nekrasov-Okounkov partition function \cite{Jimbo:2017ael}:
\begin{align}
\tau_1\tau_2-\mathfrak{q}^{-2\theta_1}\cdot t\tau_3\tau_4-\left(1-\mathfrak{q}^{-2\theta_1}\cdot t\right)\tau_5\tau_6&=0,
\label{qPVIbilineareq1}
\\
\tau_1\tau_2-t\tau_3\tau_4-\left(1-\mathfrak{q}^{-2\theta_t}\cdot t\right)\underline{\tau}_5\bar{\tau}_6&=0,
\\
\tau_1\tau_2-\tau_3\tau_4+\left(1-\mathfrak{q}^{-2\theta_1}\cdot t\right)\mathfrak{q}^{2\theta_t}\underline{\tau}_7\bar{\tau}_8&=0,\label{qPVIbilineareq3}\\
\tau_1\tau_2-\mathfrak{q}^{2\theta_t}\tau_3\tau_4+\left(1-\mathfrak{q}^{-2\theta_t}\cdot t\right)\mathfrak{q}^{2\theta_t}\tau_7\tau_8&=0,\label{qPVIbilineareq4}\\
\underline{\tau}_5\tau_6+\mathfrak{q}^{-\theta_1-\theta_\infty+\theta_t-\frac{1}{2}}\cdot t\underline{\tau}_7\tau_8-\underline{\tau}_1\tau_2&=0,\\
\underline{\tau}_5\tau_6+\mathfrak{q}^{-\theta_1+\theta_\infty+\theta_t-\frac{1}{2}}\cdot t\underline{\tau}_7\tau_8-\tau_1\underline{\tau}_2&=0,\\
\underline{\tau}_5\tau_6+\mathfrak{q}^{\theta_0+2\theta_t}\underline{\tau}_7\tau_8-\mathfrak{q}^{\theta_t}\underline{\tau}_3\tau_4&=0,\\
\underline{\tau}_5\tau_6+\mathfrak{q}^{-\theta_0+2\theta_t}\underline{\tau}_7\tau_8-\mathfrak{q}^{\theta_t}\tau_3\underline{\tau}_4&=0.
\label{qPVIbilineareq}
\end{align}
These eight equations provide the $\tau$-form of $\mathfrak{q}$-Painlev\'e VI system.

\subsection{Topological string/spectral theory correspondence}
\label{tsst}

Let's
consider the topological string on a local toric Calabi-Yau threefold $X$, whose mirror curve
can be written as $W_X\left(e^x,e^p\right)=\sum_{m,n}a_{m,n}e^{mx+np}=0$
in $\mathbb{C}^*\times\mathbb{C}^*$.
$W_X$ is the Newton polynomial of the CY3 $X$ and the sum over $m,n$ runs 
over a finite set in $\mathbb{Z}^2$ determined by the toric fun of $X$, known as the Newton polygon of the curve
\cite{Katz:1996fh,Batyrev,Chiang:1999tz,Hori:2000kt}.
Let us consider the quantum version of this curve, ${\widehat {\cal O}}=\sum_{m,n}a_{m,n}e^{m{\widehat x}+n{\widehat p}}$, with $\left[{\widehat x},{\widehat p}\right]=i\hbar$.
It has been conjectured in \cite{Grassi:2014zfa} that the operator ${\widehat {\cal O}}^{-1}$ is trace class, that is $\text{tr}{\widehat {\cal O}}^{-n}$ are finite for $n=1,2,\cdots$. 
This has been checked in some relevant cases in
\cite{Marino:2015ixa,Kashaev:2015wia}
and our analysis in this paper extends the verification to other cases.
In particular, the following Fredholm operator admits a well defined spectral determinant:
\begin{align}
\Xi(\kappa)=\det\left(1+\kappa{\widehat {\cal O}}^{-1}\right).
\label{XiinsecTSST}
\end{align}
In the topological string/spectral theory (TS/ST) correspondence \cite{Grassi:2014zfa} this spectral determinant is conjectured to 
provide a non-perturbative completion of the free energy of topological strings on $X$.
Indeed, the perturbative expansion of the topological string free energy displays infinitely many poles, while the spectral determinant \eqref{XiinsecTSST} is finite for an arbitrary value of the topological string coupling $\hbar$.
Interestingly, from the viewpoint of the topological string free energy the right analytic properties of the spectral determinant are achieved by cancelling these poles by a proper combination of refined and unrefined topological string amplitudes \cite{Hatsuda:2012dt}.

A first instance of TS/ST correspondence was discussed for $X$ being the local 
$\mathbb{P}^1\times \mathbb{P}^1$ threefold in
\cite{Hatsuda:2013oxa}, 
based on the large $N$ analysis of the partition function of ABJM theory and it has been extended in 
\cite{Codesido:2015dia,Codesido:2016ixn}
to higher genus 
mirror curves.
While a general proof of the TS/ST conjecture is still missing, it has been shown to hold
in a double scaling limit of $\hbar$ and the K\"ahler parameters in \cite{Bonelli:2016idi}
by using the theory of Painlev\'e equations.

Via geometric engineering \cite{Katz:1996fh,Aharony:1997bh}, M-theory compactified on a
local Calabi-Yau threefold defines a five dimensional ${\cal N}=1$ gauge theory, whose 
Seiberg-Witten curve is
identified with the mirror curve, while the topological string partition function gets related to the Nekrasov partition function in a self-dual Omega background \cite{Nekrasov:2002qd}. Indeed
TS/ST correspondence suggests a link between the spectral determinant of the quantum Seiberg-Witten curve and the five dimensional NO partition function \cite{Bonelli:2017gdk}.

When $X$ is the local $D_5$ del Pezzo, one obtains by geometric engineering
5d supersymmetric gauge theory with $\text{SU}(2)$ gauge group minimally coupled to $N_f=4$
fundamental hypermultiplets. The quantum mirror curve is given by
\begin{align}
{\widehat {\cal O}}&=
q\left(\frac{
 m_1
 m_2
}{
 m_3
 m_4
}\right)^{\frac{1}{2}}
e^{-{\widehat x}+{\widehat p}}
-\left(\left(m_1 m_2\right)^{\frac{1}{2}}
+q\left(\frac{1}{ m_3 m_4}\right)^{\frac{1}{2}}
\right)e^{{\widehat p}}
+e^{{\widehat x}+{\widehat p}}\nonumber \\
&\quad -q\left(\frac{
 m_1
 m_2
}{
 m_3
 m_4
}\right)^{\frac{1}{2}}\left( m_3+ m_4\right)e^{-{\widehat x}}
+E
-\left( m_1+ m_2\right)e^{{\widehat x}}\nonumber \\
&\quad 
+q\left(
 m_1
 m_2
 m_3
 m_4
\right)^{\frac{1}{2}}
e^{-{\widehat x}-{\widehat p}}
- m_1
 m_2
\left(\left(\frac{1}{ m_1 m_2}\right)^{\frac{1}{2}}
+q\left(m_3 m_4\right)^{\frac{1}{2}}
\right)e^{-{\widehat p}}
+ m_1 m_2e^{{\widehat x}-{\widehat p}}.
\label{quantumD5delPezzocurve2}
\end{align}
Here $ m_i,q$ are related to $ m_i',q'$ in \eqref{quantumD5delPezzocurve} through the following rescaling \cite{Kashaev:2015wia,Bonelli:2016idi}:
\begin{align}
\log  m_i'=\frac{2\pi}{\hbar} \log  m_i,\quad
\log q'=\frac{2\pi}{\hbar} \log q.
\label{rescaledmass}
\end{align}
The Planck constant is related to the Omega deformation parameter $\mathfrak{q}=e^{-\beta\epsilon_1}$ as 
\begin{align}
\mathfrak{q}=e^{\frac{4\pi^2i}{\hbar}}.
\label{qandhbar}
\end{align}
Together with \eqref{qPVparametersandrescaled5dparameters}, we obtain the following relation between Painlev\'e parameters $\left(\theta_0,\theta_1,\theta_t,\theta_\infty,t\right)$ and the coefficients of the quantum mirror curve $\left(m_1,m_2,m_3,m_4,q\right)$
\begin{align}
&\theta_0=\frac{1}{4\pi i}\log\frac{m_1}{m_3},\quad
\theta_1=\frac{1}{4\pi i}\log\left(m_2m_4\right),\quad
\theta_t=\frac{1}{4\pi i}\log\frac{1}{m_1m_3},\quad
\theta_\infty=\frac{1}{4\pi i}\log\frac{m_4}{m_2},\\
&\frac{\log t}{\log\mathfrak{q}}=\frac{1}{4\pi i}\log\left(\frac{q^2m_2m_4}{m_1m_3}\right).
\end{align}
The Coulomb vev $\sigma$ is related to $\kappa$.
Let us notice that in the correspondence the spectral determinant
computes the $s=1$ NO partition function.

The TS/ST correspondence suggests that the spectral determinant $\Xi(\kappa)$ \eqref{XiinsecTSST}, with ${\widehat {\cal O}}$ given as \eqref{quantumD5delPezzocurve2}, is equal to $\tau(\theta_0,\theta_1,\theta_t,\theta_\infty;1,\sigma,t)$ \eqref{tau} up to a $\kappa$-independent overall factor and upon an appropriate relation between $\kappa$ and $\sigma$.
Combining this with the five-dimensional uplift of the Painlev\'e/gauge theory correspondence \cite{Bonelli:2017gdk} it follows that $\Xi(\kappa)$ should solve the $\mathfrak{q}$-Painlev\'e equations \eqref{tau1and2}-\eqref{qPVIbilineareq}.
Notice that while the Nekrasov-Okounkov partition function is written as a small $t$ expansion, the spectral determinant is manifestly given as a small $\kappa$ expansion, and hence 
solves $\mathfrak{q}$-Painlev\'e equations in a different regime.
The difficult part of this program is to invert the operator ${\widehat {\cal O}}$ for a generic set of the coefficients $ m_1, m_2, m_3, m_4,q$.
It is here that the three dimensional Chern-Simons matter theory enters the story.

\section{Chern-Simons matter matrix model and quantum curve}\label{tre}
\begin{figure}
\begin{center}
\includegraphics[width=16cm]{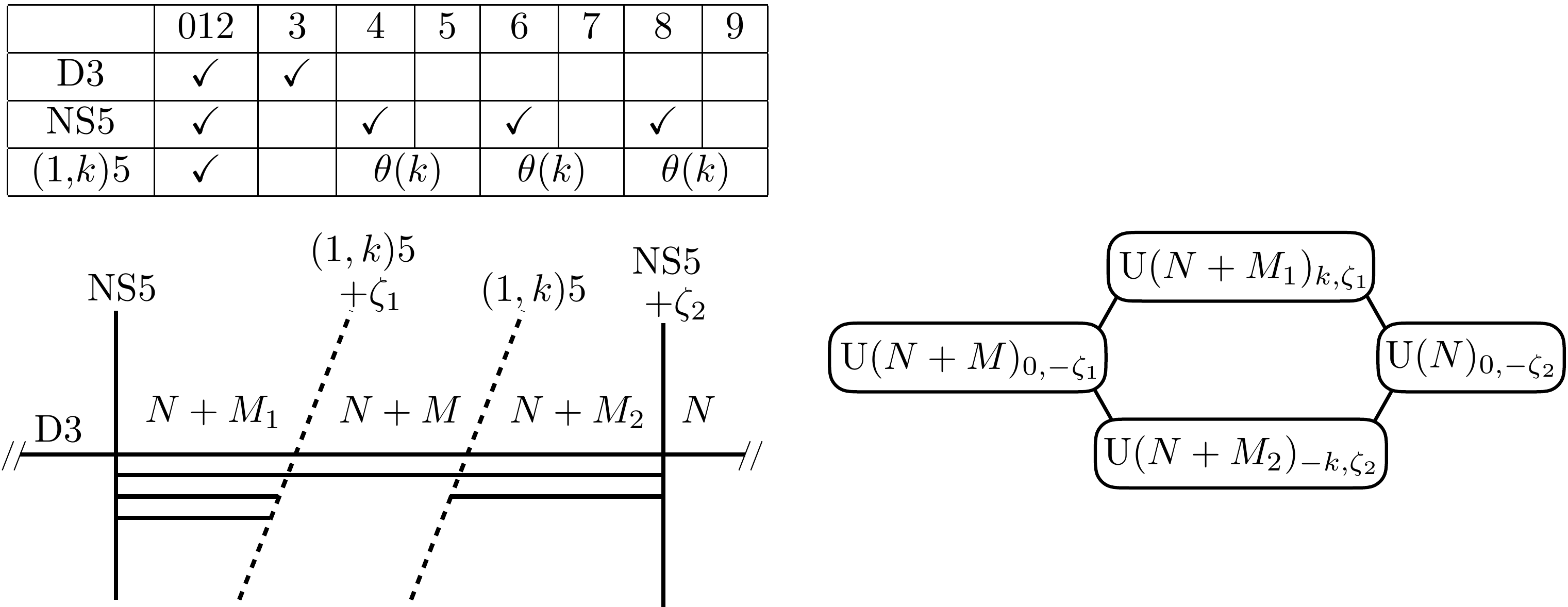}
\end{center}
\caption{
Left: Type IIB brane setup of the three dimensional superconformal Chern-Simons matter theory \eqref{N1N2N3N4zeta1zeta2theory}, where $\theta(k)$ in the row of $(1,k)$5-brane stands for the direction with an angle $\arctan(k)$ from the first axis in each of the pairs; Right: The quiver diagram of the three dimensional Chern-Simons matter theory realized by the brane setup.
}
\label{fig_IIBbranesetup}
\end{figure}

In \cite{Moriyama:2017gye,Moriyama:2017nbw,Kubo:2018cqw,Kubo:2019ejc} 
it was found that  ${\cal N}=4$ $\text{U}\left(N_1\right)_{k}\times \text{U}\left(N_2\right)_0\times \text{U}\left(N_3\right)_{-k}\times \text{U}\left(N_4\right)_0$ superconformal Chern-Simons (CS) matter theory 
is related to the $D_5$ del Pezzo geometry \eqref{quantumD5delPezzocurve}.
More precisely, it was conjectured that the partition function of the CS matter theory
computes the fermionic spectral traces of the inverse quantum mirror curve of $D_5$ del Pezzo geometry \eqref{quantumD5delPezzocurve2}, \begin{align}
Z_k(N_1,N_2,N_3,N_4)=\frac{Z_k(N=0)}{N!}\int_{-\infty}^\infty \frac{d^Nx}{(2\pi)^N}\det_{i,j}^N\left\langle x_i\left|{\widehat {\cal O}}^{-1}\right|x_j\right\rangle,
\end{align}
where 
we parametrize the four ranks as $N_1=N+M_1$, $N_2=N+M$, $N_3=N+M_2$, $N_4=N$, and assume $N,M_1,M_2,M\ge 0$. This is represented in Fig.\ref{fig_IIBbranesetup} where we also describe the IIB D-brane set-up.
Therefore the grand partition function of this theory gives, after resummation, the spectral determinant \eqref{XiinsecTSST}:
\begin{align}
\sum_{N=0}^\infty \kappa^N\frac{Z_k(N_1,N_2,N_3,N_4)}{Z_k(N=0)}=\det\left(1+\kappa{\widehat {\cal O}}^{-1}\right).
\label{211223Fredholm}
\end{align}
Here ${\widehat {\cal O}}$ is the quantum mirror curve \eqref{quantumD5delPezzocurve2} with $\hbar=2\pi k$, where the three rank differences $N_i-N$ correspond to a three dimensional subspace of the five mass parameters of the curve, while the overall rank $N$ is dual to the true\footnote{Here we follow the standard terminology already used in \cite{Bonelli:2017gdk} (see section 2.1)
distinguishing the complex moduli of the mirror curve into a "true" modulus which we call $\kappa$ 
and mass parameters.} modulus.
In order to turn on the remaining mass parameters in this paper we further introduce Fayet-Iliopoulos (FI) terms for each gauge node in the following way\footnote{
We could consider more general choices of FI parameters and also introduce hypermultiplet's masses.
However, such extra deformations  either do not preserve the Newton polygon of the quantum curve \eqref{quantumD5delPezzocurve}  or do not affect the quantum curve at all.}
\begin{align}
\text{U}\left(N_1\right)_k\times \text{U}\left(N_2\right)_0\times \text{U}\left(N_3\right)_{-k}\times \text{U}\left(N_4\right)_0\rightarrow
\text{U}\left(N_1\right)_{k,\zeta_1}\times \text{U}\left(N_2\right)_{0,-\zeta_1}\times \text{U}\left(N_3\right)_{-k,\zeta_2}\times \text{U}\left(N_4\right)_{0,-\zeta_2}.
\label{N1N2N3N4zeta1zeta2theory}
\end{align}
The partition function of the resulting gauge theory on $S^3$ is reduced by supersymmetric localisation to the following integral \cite{Kapustin:2009kz} 
\begin{align}
&Z_k\left(N;M_1,M_2,M,\zeta_1,\zeta_2\right)\nonumber \\
&=\frac{i^{-\frac{N_1^2}{2}+\frac{N_3^2}{2}}}{N_1!N_2!N_3!N_4!}\int_{-\infty}^\infty 
\prod_{i=1}^{N_1}\frac{d\lambda^{\left(1\right)}_i}{2\pi}
\prod_{i=1}^{N_2}\frac{d\lambda^{\left(2\right)}_i}{2\pi}
\prod_{i=1}^{N_3}\frac{d\lambda^{\left(3\right)}_i}{2\pi}
\prod_{i=1}^{N_4}\frac{d\lambda^{\left(4\right)}_i}{2\pi}\nonumber \\
&\quad \times e^{\frac{ik}{4\pi}\sum_{i=1}^{N_1}\left(\lambda_i^{\left(1\right)}\right)^2-\frac{ik}{4\pi}\sum_{i=1}^{N_3}\left(\lambda_i^{\left(3\right)}\right)^2}
e^{-i\zeta_1\left(\sum_{i=1}^{N_1}\lambda_i^{\left(1\right)}-\sum_{i=1}^{N_2}\lambda_i^{\left(2\right)}\right)
-i\zeta_2\left(\sum_{i=1}^{N_3}\lambda_i^{\left(3\right)}-\sum_{i=1}^{N_4}\lambda_i^{\left(4\right)}\right)}\nonumber \\
&\quad \times \prod_{a=1}^4\frac{\prod_{i<j}^{N_a}\left(2\sinh\frac{\lambda_i^{\left(a\right)}-\lambda_j^{\left(a\right)}}{2}\right)^2}{\prod_{i=1}^{N_a}\prod_{j=1}^{N_{a+1}}2\cosh\frac{\lambda_i^{\left(a\right)}-\lambda_j^{\left(a+1\right)}}{2}},
\label{eq_MM_Def}
\end{align}
 where $N_5=N_1$
and $\lambda_j^{\left(5\right)}=\lambda_j^{\left(1\right)}$.

\subsection{The quantum curve}
In appendix B we show that the matrix model \eqref{eq_MM_Def} can be written in the following form (see \eqref{220112ZkVIfourblocks})
\begin{align}
 & Z_k\left(N;M_{1},M_{2},M,\zeta_{1},\zeta_{2}\right)\nonumber \\
 & =
\frac{e^{i\Theta_{k}\left(M_{1},M_{2},M,\zeta_{1},\zeta_{2}\right)}Z_{k}^{\left(\cs\right)}\left(M_{1}\right)Z_{k}^{\left(\cs\right)}\left(M_{2}\right)}{N!}
\int_{-\infty}^\infty \prod_{n=1}^{N}\frac{d\mu_{n}}{2\pi}\nonumber \\
&\quad \times \det\begin{pmatrix}
\left[\braket{\mu_{m}|\widehat{D}_{1}^{\text{VI}}\widehat{D}_{2}^{\text{VI}}|\mu_{n}}\right]_{m,n}^{N\times N}
&\left[\brakket{\mu_{m}|\widehat{D}_{1}^{\text{VI}}\widehat{d}_{2}^{\text{VI}}|-t_{0,M,s}}\right]_{m,s}^{N\times M}\\
\left[\bbraket{t_{0,M,r}|\widehat{d}_{1}^{\text{VI}}\widehat{D}_{2}^{\text{VI}}|\mu_{n}}\right]_{r,n}^{M\times N}
&
\left[\bbrakket{t_{0,M,r}|\widehat{d}_{1}^{\text{VI}}\widehat{d}_{2}^{\text{VI}}|-t_{0,M,s}}\right]_{r,s}^{M\times M}
\end{pmatrix},
\label{pippo}
\end{align}
where we have used the notations for one-dimensional quantum mechanics which are summarized in the beginning of appendix B.
This is manifestly an ideal Fermi gas partition function only for $M=0$.
Indeed, in this case \eqref{pippo} reduces to 
\begin{align}
 & Z_{k}\left(N;M_{1},M_{2},0,\zeta_{1},\zeta_{2}\right)\nonumber \\
 & =Z_{k}\left(0;M_{1},M_{2},0,\zeta_{1},\zeta_{2}\right)\frac{1}{N!}\int_{-\infty}^\infty \prod_{n=1}^{N}d\mu_{n}\det\left(\left[\braket{\mu_{m}|\widehat{\rho}_{k}\left(M_{1},M_{2},0,\zeta_{1},\zeta_{2}\right)|\mu_{n}}\right]_{m,n}^{N\times N}\right),\label{eq:FGF-M0-Form}
\end{align}
with the density matrix ${\widehat\rho}_k(M_1,M_2,0,\zeta_1,\zeta_2)$ given as \eqref{eq:DensM-M0}
\begin{align}
&{\widehat\rho}_k\left(M_1,M_2,0,\zeta_1,\zeta_2\right)\nonumber \\
&=
e^{\pi\zeta_2}
e^{\left(-\frac{i\zeta_1}{k}+\frac{1}{2}-\frac{M_1}{2k}\right){\widehat x}}
\frac{
\Phi_b\left(\frac{{\widehat x}}{2\pi b}-\frac{iM_1}{2b}+\frac{ib}{2}\right)
}{
\Phi_b\left(\frac{{\widehat x}}{2\pi b}+\frac{iM_1}{2b}-\frac{ib}{2}\right)
}
\frac{1}{2\cosh\frac{{\widehat p}}{2}}e^{\left(\frac{i\zeta_1}{k}+\frac{M_1+M_2}{2k}\right){\widehat x}}
\frac{
\Phi_b\left(\frac{{\widehat x}}{2\pi b}+\frac{iM_1}{2b}\right)
}{
\Phi_b\left(\frac{{\widehat x}}{2\pi b}-\frac{iM_1}{2b}\right)
}\nonumber \\
&\quad \times \frac{
\Phi_b\left(\frac{{\widehat x}}{2\pi b}+\frac{iM_2}{2b}+\frac{\zeta_2}{b}\right)
}{
\Phi_b\left(\frac{{\widehat x}}{2\pi b}-\frac{iM_2}{2b}+\frac{\zeta_2}{b}\right)
}
\frac{1}{2\cosh\frac{{\widehat p}}{2}}
e^{\left(\frac{1}{2}-\frac{M_2}{2k}\right){\widehat x}}
\frac{
\Phi_b\left(\frac{{\widehat x}}{2\pi b}-\frac{iM_2}{2b}+\frac{ib}{2}+\frac{\zeta_2}{b}\right)
}{
\Phi_b\left(\frac{{\widehat x}}{2\pi b}+\frac{iM_2}{2b}-\frac{ib}{2}+\frac{\zeta_2}{b}\right)
},
\label{211015rhoM1M2integerM30}
\end{align}
where $b=\sqrt{k}$.

Although an explicit Fermi gas formalism is presently not available in the case $M>0$, we 
conjecture a formula for the corresponding quantum curve.
In \cite{Kubo:2019ejc} 
the following formula for the quantum curve at $M>0$
and vanishing FI parameters 
$\zeta_{1}=0$ and $\zeta_{2}=0$
was proposed\footnote{There are a couple of notational differences. First, the sign of the Chern-Simons level between an NS5-brane and a $\left(1,k\right)$5-brane in this paper is opposite to that in \cite{Kubo:2019ejc}. Second, according to \cite{Kubo:2019ejc}, we should care of the position of the node whose rank is the lowest. In this paper, $N_{4}$ is always the smallest.} 
\begin{align}
 & \widehat{\rho}_{k}^{-1}\left(M_{1},M_{2},M,0,0\right)\nonumber \\
 & =e^{\frac{\pi i(-M_{1}+M_{2})}{2}}e^{-{\widehat{x}}+{\widehat{p}}}+[e^{\frac{\pi i(-M_{1}-M_{2})}{2}+\pi ik}+e^{\frac{\pi i(M_{1}+M_{2})}{2}-\pi ik}]e^{{\widehat{p}}}+e^{\frac{\pi i(M_{1}-M_{2})}{2}}e^{{\widehat{x}}+{\widehat{p}}}\nonumber \\
 & \quad+[e^{\frac{\pi i(-M_{1}-M_{2}+2M)}{2}}+e^{\frac{\pi i(M_{1}+M_{2}-2M)}{2}}]e^{-{\widehat{x}}}+E+[e^{\frac{\pi i(-M_{1}-M_{2}+2M)}{2}}+e^{\frac{\pi i(M_{1}+M_{2}-2M)}{2}}]e^{{\widehat{x}}}\nonumber \\
 & \quad+e^{\frac{\pi i(M_{1}-M_{2})}{2}}e^{-{\widehat{x}}-{\widehat{p}}}+[e^{\frac{\pi i(-M_{1}-M_{2})}{2}+\pi ik}+e^{\frac{\pi i(M_{1}+M_{2})}{2}-\pi ik}]e^{-{\widehat{p}}}+e^{\frac{\pi i(-M_{1}+M_{2})}{2}}e^{{\widehat{x}}-{\widehat{p}}}.\label{eq:QC-KMconj}
\end{align}
Here $E$ is a constant depending on $k$, $M_{1}$, $M_{2}$ and $M$.\footnote{
The procedure in \cite{Kubo:2019ejc} cannot decide the overall phase, the constant term $E$ and the constant shift of $\widehat{x}$ and $\widehat{p}$. We fixed the overall phase and the shift by comparing to the exact result \eqref{eq:QC-M0-1}. We will also fix the form of $E$ in \eqref{eq:ConstTerm-Def}.} Notice that the parameter dependence of the coefficients is multiplicative. In other words, the functions in the exponents are linear combinations of the parameters. This is also the case for the exact result \eqref{eq:QC-M0-1}. Therefore, it is natural to assume that the full parameter dependence is also multiplicative. With this assumption, we can uniquely combine \eqref{eq:QC-KMconj} and \eqref{eq:QC-M0-1} into\footnote{
While this draft was in preparation, the quantum mirror curve for general values of $(M_1,M_2,M,\zeta_1,\zeta_2)$ also appeared in \cite{Furukawa:2021pll}.
Our formula for the quantum mirror curve is their eq.~(A.1) combined with eq.~(3.5),
where $m_i^{\fmmn}$ and $z_i^{\fmmn}$ in \cite{Furukawa:2021pll} 
being related to the rank differences $M_1,M_2,M$ and the FI parameters $\zeta_1,\zeta_2$ as
\begin{align}
&m_1^{\fmmn}=e^{\pi i(-M_1+M_2)},\quad
m_2^{\fmmn}=e^{\pi i(M-k)},\quad
m_3^{\fmmn}=e^{\pi i(M_1+M_2-M-k)},\\
&z_1^{\fmmn}=e^{-2\pi\zeta_1},\quad
z_3^{\fmmn}=e^{-2\pi\zeta_2}.
\end{align}
We thank Prof.~Moriyama for pointing it out.
}
\begin{align}
 & \widehat{\rho}_{k}^{-1}\left(M_{1},M_{2},M,\zeta_{1},\zeta_{2}\right)\nonumber \\
 & =e^{\frac{\pi i(-M_{1}+M_{2})}{2}+\pi(\zeta_{1}-\zeta_{2})}e^{-{\widehat{x}}+{\widehat{p}}}+[e^{\frac{\pi i(-M_{1}-M_{2})}{2}+\pi(\zeta_{1}+\zeta_{2})+\pi ik}+e^{\frac{\pi i(M_{1}+M_{2})}{2}+\pi(\zeta_{1}-\zeta_{2})-\pi ik}]e^{{\widehat{p}}}\nonumber \\
 & \quad+e^{\frac{\pi i(M_{1}-M_{2})}{2}+\pi(\zeta_{1}+\zeta_{2})}e^{{\widehat{x}}+{\widehat{p}}}\nonumber \\
 & \quad+[e^{\frac{\pi i(-M_{1}-M_{2}+2M)}{2}+\pi(\zeta_{1}-\zeta_{2})}+e^{\frac{\pi i(M_{1}+M_{2}-2M)}{2}+\pi(-\zeta_{1}-\zeta_{2})}]e^{-{\widehat{x}}}+E\nonumber \\
 & \quad+[e^{\frac{\pi i(-M_{1}-M_{2}+2M)}{2}+\pi(-\zeta_{1}+\zeta_{2})}+e^{\frac{\pi i(M_{1}+M_{2}-2M)}{2}+\pi(\zeta_{1}+\zeta_{2})}]e^{{\widehat{x}}}\nonumber \\
 & \quad+e^{\frac{\pi i(M_{1}-M_{2})}{2}+\pi(-\zeta_{1}-\zeta_{2})}e^{-{\widehat{x}}-{\widehat{p}}}+[e^{\frac{\pi i(-M_{1}-M_{2})}{2}+\pi(-\zeta_{1}-\zeta_{2})+\pi ik}+e^{\frac{\pi i(M_{1}+M_{2})}{2}+\pi(-\zeta_{1}+\zeta_{2})-\pi ik}]e^{-{\widehat{p}}}\nonumber \\
 & \quad+e^{\frac{\pi i(-M_{1}+M_{2})}{2}+\pi(-\zeta_{1}+\zeta_{2})}e^{{\widehat{x}}-{\widehat{p}}},\label{22modelrhoinverse}
\end{align}
where
\begin{align}
E & =e^{\frac{\pi i(-M_{1}+M_{2})}{2}+\pi(-\zeta_{1}-\zeta_{2})+\pi ia_{1}M}+e^{\frac{\pi i(-M_{1}+M_{2})}{2}+\pi(\zeta_{1}+\zeta_{2})+\pi ia_{2}M}\nonumber \\
 & \quad+e^{\frac{\pi i(M_{1}-M_{2})}{2}+\pi(-\zeta_{1}+\zeta_{2})+\pi ia_{3}M}+e^{\frac{\pi i(M_{1}-M_{2})}{2}+\pi(\zeta_{1}-\zeta_{2})+\pi ia_{4}M},\label{eq:ConstTerm-Def}
\end{align}
with unknown constant parameters $a_{i}$.
Notice that, when $M=0$, the quantum curve \eqref{22modelrhoinverse} -- or equivalently, \eqref{eq:QC-M0-1}
-- is factorized into the product of the two quantum curves appearing in \eqref{eq:Hyper-QC}.
These quantum curves are the ones associated to the ABJM theory.
Therefore for $M=0$ the quantum curve of the (2,2) model is factorized into a product of two ABJM quantum curves.
On the other hand, when $M\neq 0$, this factorization does not occur, and this makes the inversion of 
\eqref{22modelrhoinverse}
more difficult. 

\begin{figure}
\begin{center}
\includegraphics[width=7cm]{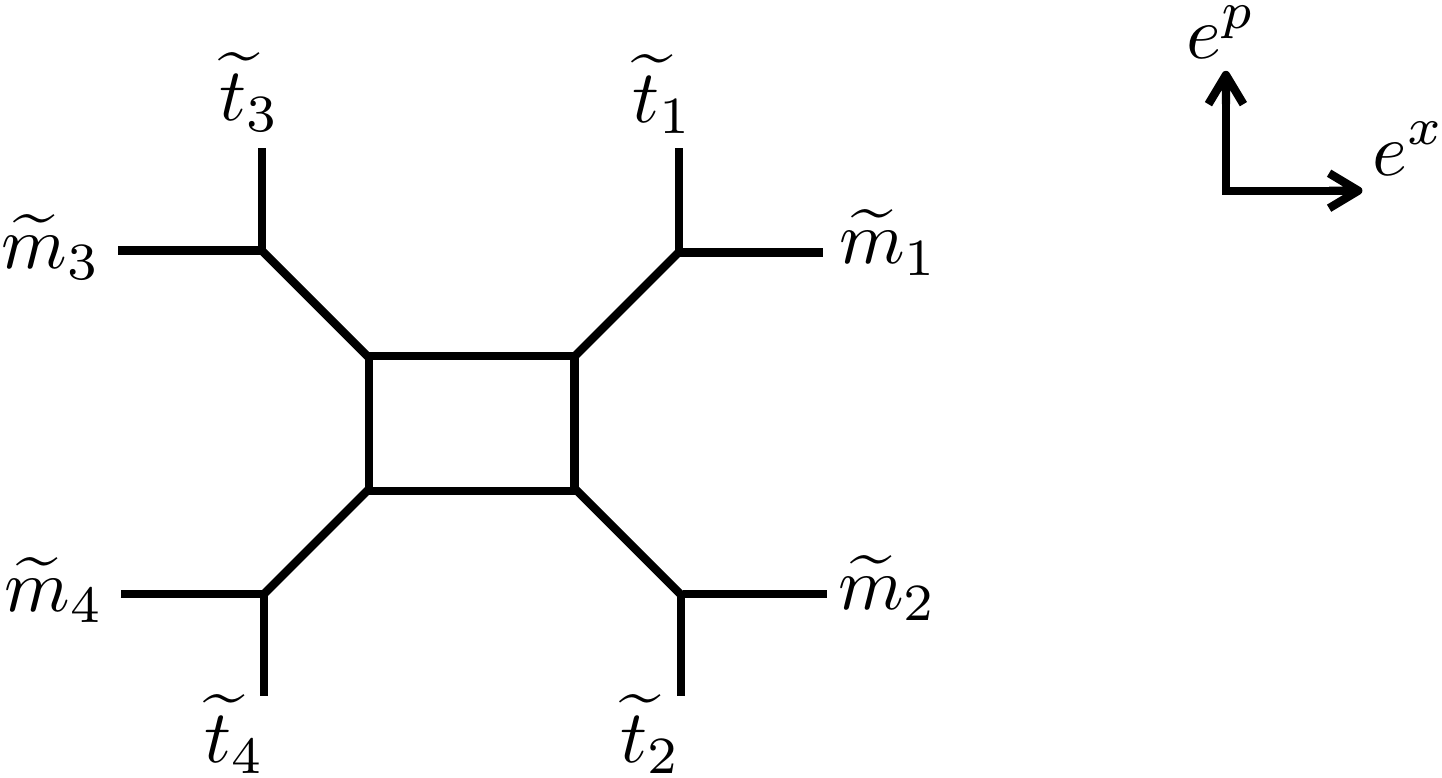}
\end{center}
\caption{
Five-brane web diagram corresponding to the classical limit of the quantum spectral curve \eqref{22modelrhoinverse}.
$ \tilde{m}_i$ and $\tilde{t}_i$ are the asymptotic positions of the 5-branes in the limiting classical curve.
}
\label{fig_asymptoticpoints}
\end{figure}

By comparing the coefficients of \eqref{22modelrhoinverse} with the coefficients of the quantum Seiberg-Witten curve  \eqref{quantumD5delPezzocurve2} we can read off the parameters of the five dimensional gauge theory in the following way.
First we compare the equation ${\widehat\rho}^{-1}|_{{\widehat x},{\widehat p}\rightarrow x,p}=0$ with the Seiberg-Witten curve at the four asymptotic regions $x=\pm\infty,p=\pm\infty$, see Fig.\ref{fig_asymptoticpoints}:
\begin{align}
x \rightarrow\infty:&\quad
e^p= {\widetilde m}_1, {\widetilde m}_2=-e^{\pi i\left(M_2-M\right)},-e^{\pi i\left(-M_1+M\right)-2\pi\zeta_1},\label{220115mtildeandttilde1}\\
x \rightarrow -\infty:&\quad
e^p= {\widetilde m}_3, {\widetilde m}_4=-e^{\pi i\left(M_1-M\right)-2\pi\zeta_1},\quad -e^{\pi i\left(-M_2+M\right)},\label{220115mtildeandttilde2}\\
p \rightarrow \infty:&\quad
e^x={\widetilde t}_1,{\widetilde t}_3=-e^{\pi iM_2-2\pi\zeta_2-\pi ik},\quad -e^{-\pi iM_1+\pi ik},\label{220115mtildeandttilde3}\\
p\rightarrow -\infty:&\quad
e^x={\widetilde t}_2,{\widetilde t}_4=-e^{\pi iM_1-\pi ik},\quad -e^{-\pi iM_2-2\pi\zeta_2+\pi ik}.
\label{220115mtildeandttilde}
\end{align}

The quantities with tilde can be rescaled as 
\begin{align}
\left(
 {\widetilde m}_1,
 {\widetilde m}_2,
 {\widetilde m}_3,
 {\widetilde m}_4,
{\widetilde t}_1,
{\widetilde t}_2,
{\widetilde t}_3,
{\widetilde t}_4\right)
\rightarrow
\left(
\alpha {\widetilde m}_1,
\alpha {\widetilde m}_2,
\alpha {\widetilde m}_3,
\alpha {\widetilde m}_4,
\beta {\widetilde t}_1,
\beta {\widetilde t}_2,
\beta {\widetilde t}_3,
\beta {\widetilde t}_4\right),
\end{align}
with arbitrary non zero complex numbers $\alpha,\beta$ associated to the coordinates translation $\left(p,x\right)\rightarrow \left(p-\log \alpha,x-\log \beta\right)$.
The gauge coupling $q$ is given by
\begin{align}
q=\left(\frac{{\widetilde t}_3{\widetilde t}_4}{{\widetilde t}_1{\widetilde t}_2}\right)^{\frac{1}{2}},
\end{align}
which is by itself rescaling invariant.
In order to match the parameters here with those in \eqref{quantumD5delPezzocurve2} correctly, we have to set $\alpha=\left(\frac{{\widetilde t}_1}{ {\widetilde m}_1 {\widetilde m}_2{\widetilde t}_2}\right)^{\frac{1}{2}}$ so that
\begin{align}
& m_1=\alpha {\widetilde m}_1=\left(\frac{ {\widetilde m}_1{\widetilde t}_1}{ {\widetilde m}_2{\widetilde t}_2}\right)^{\frac{1}{2}},\quad
 m_2=\alpha {\widetilde m}_2=\left(\frac{ {\widetilde m}_2{\widetilde t}_1}{ {\widetilde m}_1{\widetilde t}_2}\right)^{\frac{1}{2}},\\
& m_3=\alpha {\widetilde m}_3=\left(\frac{ {\widetilde m}_3{\widetilde t}_4}{ {\widetilde m}_4{\widetilde t}_3}\right)^{\frac{1}{2}},\quad
 m_4=\alpha {\widetilde m}_4=\left(\frac{ {\widetilde m}_4{\widetilde t}_4}{ {\widetilde m}_3{\widetilde t}_3}\right)^{\frac{1}{2}}.
\end{align}
By substituting the explicit expressions of ${\widetilde m}_i,{\widetilde t}_i$ we obtain
\begin{align}
& m_1=e^{\pi i\left(M_2-M\right)+\pi\left(\zeta_1-\zeta_2\right)},\quad
 m_2=e^{\pi i\left(-M_1+M\right)+\pi\left(-\zeta_1-\zeta_2\right)},\quad
 m_3=e^{\pi i\left(M_1-M\right)+\pi\left(-\zeta_1-\zeta_2\right)},\label{mtildeandqBPTYinMandzeta}
\\
& m_4=e^{\pi i\left(-M_2+M\right)+\pi\left(\zeta_1-\zeta_2\right)},\quad
q=e^{\pi i\left(-M_1-M_2\right)+2\pi ik}.
\label{mtildeandqBPTYinMandzeta2}
\end{align}


\subsection{Parameter identification with $\mathfrak{q}$-PVI system}
According to the discussion at the end of subsection \ref{tsst},
the relation between the $\mathfrak{q}$-PVI tau function and the grand partition function of the superconformal Chern-Simons quiver theory \eqref{N1N2N3N4zeta1zeta2theory} follows by
combining the above results \eqref{qPVparametersandrescaled5dparameters}, 
\eqref{XiinsecTSST}, \eqref{rescaledmass},
\eqref{mtildeandqBPTYinMandzeta}, \eqref{mtildeandqBPTYinMandzeta2}:
\begin{align}
&\tau\left(\theta_0,\theta_1,\theta_t,\theta_\infty;1,\kappa,t\right)\nonumber \\
&=
\frac{
F\left(M_1,M_2,M,\zeta_1,\zeta_2\right)
}{Z_k\left(0;M_1,M_2,M,\zeta_1,\zeta_2\right)}
\sum_{N=0}^\infty \left(\Omega\left(M_1,M_2,M,\zeta_1,\zeta_2\right)\kappa\right)^NZ_k\left(N;M_1,M_2,M,\zeta_1,\zeta_2\right)\nonumber \\
&=F\left(M_1,M_2,M,\zeta_1,\zeta_2\right)\det\left[1+\Omega\left(M_1,M_2,M,\zeta_1,\zeta_2\right)\kappa {\widehat\rho}_k\left(M_1,M_2,M,\zeta_1,\zeta_2\right)\right],\label{eq:tau-GPF}
\end{align}
where $F\left(M_1,M_2,M,\zeta_1,\zeta_2\right)$ and $\Omega\left(M_1,M_2,M,\zeta_1,\zeta_2\right)$ are some functions independent of $\kappa$, with the following parameter identification:
\begin{align}
\begin{pmatrix}
\theta_0\\
\theta_1\\
\theta_t\\
\theta_\infty\\
\frac{\log t}{\log \mathfrak{q}}
\end{pmatrix}
=\frac{1}{4}\begin{pmatrix}
-1& 1& 0&2& 0\\
-1&-1& 2&0&-2\\
-1&-1& 2&0& 2\\
 1&-1& 0&2& 0\\
-4&-4& 4&0& 0
\end{pmatrix}
\begin{pmatrix}
M_1-k\\
M_2-k\\
M-k\\
-i\zeta_1\\
-i\zeta_2
\end{pmatrix}.
\label{identificationwithoutWeyl}
\end{align}
As already discussed in the previous section, an explicit formula for the spectral density matrix
${\widehat\rho}_k$ is known to us only for $M=0$. We therefore can calculate only the $\tau$-functions generated by the action of the affine Weyl group transformations which fix the $M=0$ condition. It is therefore useful to rewrite \eqref{identificationwithoutWeyl} after a suitable change of basis obtained by a  
Weyl group transformation $W\left(D_5\right)$ 
so that we can realize as many of the shifts in \eqref{tau1and2}-\eqref{tau1to8underlinebar} as possible without varying $M$.
Let us compute the relevant change of basis.

The full Weyl group is generated by the fundamental elements $s_a, \, a=1,\ldots, 5$ associated to each node of $D_5$ Dynkin diagram which linearly realise the group action on the parameters
$\left(M_1-k,M_2-k,M-k,-i\zeta_1,-i\zeta_2\right)$
as follows\footnote{
The explicit expressions for $s_1,s_2,s_3,s_4,s_5$ are written in \cite{Kubo:2019ejc} but in a different basis $\left(\log \bar{h}_1,\log\bar{h}_2,\log e_1,\log e_3,\log e_5\right)$, which is related to the current basis $\left(M_1-k,M_2-k,M-k,-i\zeta_1,-i\zeta_2\right)$ as
\begin{align}
\begin{pmatrix}
\log \bar{h}_1\\
\log \bar{h}_2\\
\log e_1\\
\log e_3\\
\log e_5\\
\end{pmatrix}=\pi i\begin{pmatrix}
2&0& 0& 0&0\\
0&2&-2& 0&0\\
1&1&-2& 2&0\\
1&1& 0& 0&-2\\
1&1&-2&-2&0
\end{pmatrix}
\begin{pmatrix}
M_1-k\\
M_2-k\\
M-k\\
-i\zeta_1\\
-i\zeta_2
\end{pmatrix}.
\end{align}
}
\begin{align}
&s_1=
\begin{pmatrix}
 \frac{1}{2}&-\frac{1}{2}&0&0&-1\\
-\frac{1}{2}&\frac{1}{2} &0&0&-1\\
-\frac{1}{2}&-\frac{1}{2}&1&0&-1\\
           0&           0&0&1&0\\
-\frac{1}{2}&-\frac{1}{2}&0&0&0
\end{pmatrix},\quad
s_2=
\begin{pmatrix}
 \frac{1}{2}&-\frac{1}{2}&0&0&1\\
-\frac{1}{2}& \frac{1}{2}&0&0&1\\
-\frac{1}{2}&-\frac{1}{2}&1&0&1\\
           0&           0&0&1&0\\
\frac{1}{2}&\frac{1}{2}&0&0&0
\end{pmatrix},\quad
s_3=
\begin{pmatrix}
1& 0&0&0&0\\
0&-1&0&0&0\\
0&-1&1&0&0\\
0& 0&0&1&0\\
0& 0&0&0&1
\end{pmatrix},\\
&s_4=
\begin{pmatrix}
0&0& 1&0&0\\
1&1&-1&0&0\\
1&0& 0&0&0\\
0&0& 0&1&0\\
0&0& 0&0&1
\end{pmatrix},\quad
s_5=
\begin{pmatrix}
1           &0           &0&0&0\\
0           &1           &0&0&0\\
 \frac{1}{2}&\frac{1}{2} &0&1&0\\
-\frac{1}{2}&-\frac{1}{2}&1&0&0\\
0           &0           &0&0&1
\end{pmatrix}.
\end{align}
Since each of these fundamental Weyl group transformation
induces a similarity transformation $s:{\widehat\rho}^{-1}\rightarrow {\widehat U}{\widehat\rho}^{-1}{\widehat U}^{-1}$ on the quantum spectral curve,
the spectral determinant \eqref{XiinsecTSST} is invariant. We may therefore choose an arbitrary element $w\in W(D_5)$ to identify the $\mathfrak{q}$-PVI parameters as
\begin{align}\label{treventi}
\begin{pmatrix}
\theta_0\\
\theta_1\\
\theta_t\\
\theta_\infty\\
\frac{\log t}{\log \mathfrak{q}}
\end{pmatrix}
=\frac{1}{4}\begin{pmatrix}
-1& 1& 0&2& 0\\
-1&-1& 2&0&-2\\
-1&-1& 2&0& 2\\
 1&-1& 0&2& 0\\
-4&-4& 4&0& 0
\end{pmatrix}
w
\begin{pmatrix}
M_1-k\\
M_2-k\\
M-k\\
-i\zeta_1\\
-i\zeta_2
\end{pmatrix},
\end{align}
instead of \eqref{identificationwithoutWeyl}.
In particular, if we choose
\begin{align}
w=s_3s_2s_1s_3 s_4s_5s_1s_3 s_4s_2s_3s_2=\begin{pmatrix}
-\frac{1}{2}&\frac{1}{2} &0&-1&0\\
-\frac{1}{2}&\frac{1}{2} &0&1&0\\
           0&           1&0&0&0\\
           0&           0&0&0&1\\
-\frac{1}{2}&-\frac{1}{2}&1&0&0
\end{pmatrix},
\end{align}
we obtain
\begin{align}
&\theta_0=\frac{-i\zeta_1-i\zeta_2}{2},\quad
\theta_1=\frac{M_1+M_2-M-k}{2},\quad
\theta_t=\frac{M-k}{2},\quad
\theta_\infty=\frac{i\zeta_1-i\zeta_2}{2},\quad
t=\mathfrak{q}^{M_1-k}=e^{\frac{2\pi iM_1}{k}},\label{theta01tinftytMzetaindentification_convenient} \\ &\mathfrak{q}=e^{\frac{2\pi i}{k}}.
\label{theta01tinftytMzetaindentification_convenient2}
\end{align}
With this parameter identification, the allowed $\tau$-functions are manifestly the ones not 
induced by a shift in $\theta_t$, namely all of them but  $\tau_{7,8},\underline{\tau}_{7,8},\bar{\tau}_{7,8}$
according to \eqref{tau1and2}-\eqref{tau1to8underlinebar}.
Note that for $M=0$, $\mathfrak{q}^{\theta_t}=-1$.

By using \eqref{theta01tinftytMzetaindentification_convenient} and \eqref{theta01tinftytMzetaindentification_convenient2} we find that $\tau_i,\underline{\tau}_i,\bar{\tau}_i$ for $M=0$ are given by
\begin{align}
\tau_{1}&=\tau\left(M_1,M_2,0,\zeta_1-\frac{i}{2},\zeta_2+\frac{i}{2}\right),\quad
\tau_{2}=\tau\left(M_1,M_2,0,\zeta_1+\frac{i}{2},\zeta_2-\frac{i}{2}\right),\label{tauin22modelFredholmdet1}
\\
\tau_{3}&=\tau\left(M_1,M_2,0,\zeta_1+\frac{i}{2},\zeta_2+\frac{i}{2}\right),\quad
\tau_{4}=\tau\left(M_1,M_2,0,\zeta_1-\frac{i}{2},\zeta_2-\frac{i}{2}\right),\\
\tau_{5}&=\tau\left(M_1,M_2-1,0,\zeta_1,\zeta_2\right),\quad
\tau_{6}=\tau\left(M_1,M_2+1,0,\zeta_1,\zeta_2\right),\\
\underline{\tau}_{1}&=\tau\left(M_1-1,M_2+1,0,\zeta_1-\frac{i}{2},\zeta_2+\frac{i}{2}\right),\quad
\underline{\tau}_{2}=\tau\left(M_1-1,M_2+1,0,\zeta_1+\frac{i}{2},\zeta_2-\frac{i}{2}\right),\\
\underline{\tau}_{3}&=\tau\left(M_1-1,M_2+1,0,\zeta_1+\frac{i}{2},\zeta_2+\frac{i}{2}\right),\quad
\underline{\tau}_{4}=\tau\left(M_1-1,M_2+1,0,\zeta_1-\frac{i}{2},\zeta_2-\frac{i}{2}\right),\\
\underline{\tau}_5&=\tau\left(M_1-1,M_2,0,\zeta_1,\zeta_2\right),\\
\bar{\tau}_{1}&=\tau\left(M_1+1,M_2-1,0,\zeta_1-\frac{i}{2},\zeta_2+\frac{i}{2}\right),\quad
\bar{\tau}_{2}=\tau\left(M_1+1,M_2-1,0,\zeta_1+\frac{i}{2},\zeta_2-\frac{i}{2}\right), \\
\bar{\tau}_{3}&=\tau\left(M_1+1,M_2-1,0,\zeta_1+\frac{i}{2},\zeta_2+\frac{i}{2}\right),\quad
\bar{\tau}_{4}=\tau\left(M_1+1,M_2-1,0,\zeta_1-\frac{i}{2},\zeta_2-\frac{i}{2}\right),\\
\bar{\tau}_6&=\tau\left(M_1+1,M_2,0,\zeta_1,\zeta_2\right).
\label{tauin22modelFredholmdet}
\end{align}
Here we have abbreviated $\tau(\theta_0,\theta_1,\theta_t,\theta_\infty;1,\kappa,t)$ in \eqref{eq:tau-GPF} with the substitution of \eqref{theta01tinftytMzetaindentification_convenient} simply as $\tau(M_1,M_2,0,\zeta_1,\zeta_2)$.
Since the variables $\tau_{7,8},\underline{\tau}_{7,8},\bar{\tau}_{7,8}$ are obstructed to us, we can check the system \eqref{qPVIbilineareq1}-\eqref{qPVIbilineareq} only after their elimination.
This provides a subsystem of six equations in six variables out of \eqref{qPVIbilineareq1}-\eqref{qPVIbilineareq} given by the first two equations, which explicitly do not involve $\tau_{7,8},\underline{\tau}_{7,8},\bar{\tau}_{7,8}$ and other four.
Actually, by eliminating $\tau_{7,8},\underline{\tau}_{7,8},\bar{\tau}_{7,8}$ we can obtain from the remaining six equations \eqref{qPVIbilineareq3}-\eqref{qPVIbilineareq} three bilinear equations and a quartic equation. All in all, we then get the system
\begin{align}
&\tau_1\tau_2-\mathfrak{q}^{-2\theta_1}\cdot t\tau_3\tau_4-\left(1-\mathfrak{q}^{-2\theta_1}\cdot t\right)\tau_5\tau_6=0,
\label{211019_bilineareqofFredholmdet1} 
\\
&\tau_1\tau_2-t\tau_3\tau_4-\left(1-\mathfrak{q}^{-2\theta_t}\cdot t\right)\underline{\tau}_5\bar{\tau}_6=0,
\label{211019_bilineareqofFredholmdet2}\\
&\underline{\tau}_1\tau_2-\mathfrak{q}^{-2\theta_\infty}\tau_1\underline{\tau}_2-\left(1-\mathfrak{q}^{-2\theta_\infty}\right)\underline{\tau}_5\tau_6=0,\label{211019_bilineareqofFredholmdet3} \\
&\underline{\tau}_3\tau_4-\mathfrak{q}^{2\theta_0}\tau_3\underline{\tau}_4-\mathfrak{q}^{-\theta_t}\left(1-\mathfrak{q}^{2\theta_0}\right)\underline{\tau}_5\tau_6=0,\label{211019_bilineareqofFredholmdet4} \\
&\underline{\tau}_1\tau_2-\mathfrak{q}^{-\theta_0-\theta_1-\theta_\infty-\frac{1}{2}}\cdot t\underline{\tau}_3\tau_4-\left(1-\mathfrak{q}^{-\theta_0-\theta_1-\theta_t-\theta_\infty-\frac{1}{2}}\cdot t\right)\underline{\tau}_5\tau_6=0,\label{211019_bilineareqofFredholmdet5} \\
&\left(\tau_1\tau_2-\tau_3\tau_4\right)\left(\tau_1\tau_2-\mathfrak{q}^{2\theta_t}\tau_3\tau_4\right)-\left(1-\mathfrak{q}^{2\theta_1}\cdot t^{-1}\right)\left(1-\mathfrak{q}^{2\theta_t}\cdot t^{-1}\right)\mathfrak{q}^{2\theta_\infty}\left(\underline{\tau}_1\tau_2-\underline{\tau}_5\tau_6\right)\left(\tau_1\bar{\tau}_2-\tau_5\bar{\tau}_6\right)=0.
\label{211019_quarticeqofFredholmdet}
\end{align}
We found that these hold provided the factors $F\left(M_1,M_2,0,\zeta_1,\zeta_2\right)$ and $\Omega\left(M_1,M_2,0,\zeta_1,\zeta_2\right)$ satisfy the following relations:
\begin{align}
&F_1F_2
=F_3F_4
=F_5F_6
=\underline{F}_5\bar{F}_6,\quad
\underline{F}_1F_2
=F_1\underline{F}_2
=-\underline{F}_3F_4
=-F_3\underline{F}_4
=\underline{F}_5F_6,\label{Fcondition} \\
&-i\Omega_1
=-i\Omega_2
=i\Omega_3
=i\Omega_4
=\Omega_5
=-\Omega_6
=i\underline{\Omega}_1
=i\underline{\Omega}_2
=-i\underline{\Omega}_3
=-i\underline{\Omega}_4
=-\underline{\Omega}_5\nonumber \\
&\quad =i\bar{\Omega}_1
=i\bar{\Omega}_2
=-i\bar{\Omega}_3
=-i\bar{\Omega}_4
=\bar{\Omega}_6.
\label{211223FandOmegacondition}
\end{align}
Needless to say, here we adopt the very same notation as in \eqref{tauin22modelFredholmdet1}-\eqref{tauin22modelFredholmdet} for the tau functions, both for the labels $1,2,\cdots,6$, and for the under/over-line as shifts of the parameters $\left(M_1,M_2,\zeta_1,\zeta_2\right)$ .
The following choice is consistent with all of the conditions \eqref{Fcondition}, \eqref{211223FandOmegacondition}:
\begin{align}
F\left(M_1,M_2,0,\zeta_1,\zeta_2\right)=e^{\frac{\pi\left(M_2-M_1\right)\left(\zeta_1+\zeta_2\right)}{2}},\quad
\Omega\left(M_1,M_2,0,\zeta_1,\zeta_2\right)=e^{-\frac{\pi i\left(M_1-M_2\right)}{2}+\pi\left(\zeta_1+\zeta_2\right)},
\end{align}
and we adopt it from now on.

In the next section we explain how to check 
equations 
\eqref{211019_bilineareqofFredholmdet1}-\eqref{211019_quarticeqofFredholmdet}.

\section{Checks of the $\mathfrak{q}$-Painlev\'e equations}
\label{sec4}
In this subsection we provide non-trivial evidence that the $\tau$-functions \eqref{tauin22modelFredholmdet1}-\eqref{tauin22modelFredholmdet} defined as Fredholm determinants satisfy the bilinear and quartic equations \eqref{211019_bilineareqofFredholmdet1}-\eqref{211019_quarticeqofFredholmdet}.
In order to do this, 
first we prove some symmetry properties of ${\widehat\rho}_k$ under a given set of linear transformations of the parameters $\left(M_1,M_2,\zeta_1,\zeta_2\right)$ 
which are a subset of the full $W\left(D_5\right)$ symmetry acting on the quantum mirror curve \eqref{22modelrhoinverse}.
Next, by using this symmetry property we provide two types of non-trivial checks of the bilinear/quartic equations: check around the symmetric points under transformations and the direct proof of the equations at the sub-leading order in $\kappa$ with $\left(M_1,M_2,\zeta_1,\zeta_2\right)$ kept unfixed.
Lastly, by reducing the problem using the above symmetries, we provide non-trivial checks of the equations at higher order in $\kappa$ around a discrete set of points $\left(M_1,M_2,\zeta_1,\zeta_2\right)$.

\subsection{Relation among bilinear equations under Weyl transformations}
In this section we consider the discrete symmetry of the spectral density matrix 
${\widehat\rho}_k\left(M_1,M_2,0,\zeta_1,\zeta_2\right)$
given in \eqref{211015rhoM1M2integerM30} under similarity transformations which do not change the spectral determinant \eqref{XiinsecTSST}.
In particular, under a suitable 
similarity transformation (cyclic permutation) we can rewrite the equivalent density matrix 
\begin{align}
{\widehat\rho}_k\left(M_1,M_2,0,\zeta_1,\zeta_2\right)&=
e^{\pi\zeta_2}
e^{\left(-\frac{i\zeta_1}{k}+1-\frac{M_1+M_2}{2k}\right){\widehat x}}
\frac{
\Phi_b\left(\frac{{\widehat x}}{2\pi b}-\frac{iM_1}{2b}+\frac{ib}{2}\right)
}{
\Phi_b\left(\frac{{\widehat x}}{2\pi b}+\frac{iM_1}{2b}-\frac{ib}{2}\right)
}
\frac{
\Phi_b\left(\frac{{\widehat x}}{2\pi b}-\frac{iM_2}{2b}+\frac{ib}{2}+\frac{\zeta_2}{b}\right)
}{
\Phi_b\left(\frac{{\widehat x}}{2\pi b}+\frac{iM_2}{2b}-\frac{ib}{2}+\frac{\zeta_2}{b}\right)
}
\frac{1}{2\cosh\frac{{\widehat p}}{2}}\nonumber \\
&\quad \times
e^{\left(\frac{i\zeta_1}{k}+\frac{M_1+M_2}{2k}\right){\widehat x}}
\frac{
\Phi_b\left(\frac{{\widehat x}}{2\pi b}+\frac{iM_1}{2b}\right)
}{
\Phi_b\left(\frac{{\widehat x}}{2\pi b}-\frac{iM_1}{2b}\right)
}
\frac{
\Phi_b\left(\frac{{\widehat x}}{2\pi b}+\frac{iM_2}{2b}+\frac{\zeta_2}{b}\right)
}{
\Phi_b\left(\frac{{\widehat x}}{2\pi b}-\frac{iM_2}{2b}+\frac{\zeta_2}{b}\right)
}
\frac{1}{2\cosh\frac{{\widehat p}}{2}},
\label{rhowritteninPhib2}
\end{align}
which for simplicity we still call ${\widehat\rho}_k$ (though they are not precisely equal to each other as operators).\footnote{
Explicitly, ${\widehat\rho}_k$ in \eqref{211015rhoM1M2integerM30} and ${\widehat\rho}_k$ in \eqref{rhowritteninPhib2} are related by the following similarity transformation:
\begin{align}
{\widehat\rho}_k^{\eqref{rhowritteninPhib2}}
={\widehat U}{\widehat\rho}_k^{\eqref{211015rhoM1M2integerM30}}{\widehat U}^{-1},
\end{align}
with
\begin{align}
{\widehat U}=e^{\left(\frac{1}{2}-\frac{M_2}{2k}\right){\widehat x}}
\frac{
\Phi_b\left(\frac{{\widehat x}}{2\pi b}+\frac{\zeta_2}{b}-\frac{iM_2}{2b}+\frac{ib}{2}\right)}{
\Phi_b\left(\frac{{\widehat x}}{2\pi b}+\frac{\zeta_2}{b}+\frac{iM_2}{2b}-\frac{ib}{2}\right)}.
\end{align}
}
We now consider the symmetry properties of the above density matrix among different values $(M_1,M_2,\zeta_1,\zeta_2)$ while $M=0$ being preserved.

First, by reordering the quantum dilogarithms $\Phi_b\left(\frac{{\widehat x}}{2\pi b}+\cdots\right)$ within each group of four $\Phi_b$ separated by $\frac{1}{2\cosh\frac{{\widehat p}}{2}}$ we obtain
\begin{align}
{\widehat\rho}_k\left(M_1,M_2,0,\zeta_1,\zeta_2\right)\sim {\widehat\rho}_k\left(\frac{M_1+M_2}{2}-i\zeta_2,\frac{M_1+M_2}{2}+i\zeta_2,0,\zeta_1,\frac{i\left(M_1-M_2\right)}{2}\right),\label{Weyl_directlyshownonrhohat1}\\
{\widehat\rho}_k\left(M_1,M_2,0,\zeta_1,\zeta_2\right)\sim {\widehat\rho}_k\left(\frac{M_1+M_2}{2}+i\zeta_2,\frac{M_1+M_2}{2}-i\zeta_2,0,\zeta_1,-\frac{i\left(M_1-M_2\right)}{2}\right).\label{Weyl_directlyshownonrhohat2}
\end{align}
Here ``$\sim$'' stands for the coincidence up to a similarity transformation, i.e., ${\widehat\rho}_k^{(\text{lhs})}={\widehat U}{\widehat\rho}_k^{(\text{rhs})}{\widehat U}^{-1}$ with some ${\widehat U}$.
Explicitly, ${\widehat U}=e^{(\frac{M_1-M_2}{4k}+\frac{i\zeta_2}{2k}){\widehat p}}$ for \eqref{Weyl_directlyshownonrhohat1} and ${\widehat U}=e^{(-\frac{M_1-M_2}{4k}+\frac{i\zeta_2}{2k}){\widehat p}}$ for \eqref{Weyl_directlyshownonrhohat2}.
Second, by exchanging the first line and the second line in \eqref{rhowritteninPhib2}, which is a cyclic permutation and can be realized by a similarity transformation, we obtain
\begin{align}
{\widehat\rho}_k\left(M_1,M_2,0,\zeta_1,\zeta_2\right)\sim {\widehat\rho}_k\left(k-M_1,k-M_2,0,-\zeta_1,\zeta_2\right).\label{Weyl_directlyshownonrhohat3}
\end{align}
Lastly, we also find (see appendix \ref{sec_proofofzeta1zeta2symmetry})
\begin{align}
{\widehat\rho}_k\left(M_1,M_2,0,\zeta_1,\zeta_2\right)\sim {\widehat\rho}_k\left(M_1,M_2,0,\zeta_2,\zeta_1\right).\label{Weyl_directlyshownonrhohat4}
\end{align}
Let us denote the Weyl reflections 
 \eqref{Weyl_directlyshownonrhohat1}-\eqref{Weyl_directlyshownonrhohat4} 
as 
\begin{align}
&r_1:\,\left(M_1,M_2,\zeta_1,\zeta_2\right)\rightarrow \left(\frac{M_1+M_2}{2}-i\zeta_2,\frac{M_1+M_2}{2}+i\zeta_2,\zeta_1,\frac{i(M_1-M_2)}{2}\right),\label{r1} \\
&r_2:\,\left(M_1,M_2,\zeta_1,\zeta_2\right)\rightarrow \left(\frac{M_1+M_2}{2}+i\zeta_2,\frac{M_1+M_2}{2}-i\zeta_2,\zeta_1,-\frac{i(M_1-M_2)}{2}\right), \\
&r_3:\,\left(M_1,M_2,\zeta_1,\zeta_2\right)\rightarrow \left(k-M_1,k-M_2,-\zeta_1,\zeta_2\right), \\
&r_4:\,\left(M_1,M_2,\zeta_1,\zeta_2\right)\rightarrow \left(M_1,M_2,\zeta_2,\zeta_1\right).
\label{211216r1r2r3r4}
\end{align}
It turns out that $r_2=r_1r_3r_1r_3r_1$, while $r_1,r_3,r_4$ are independent and generate all of the elements in $W(D_5)$ preserving $M$, which is isomorphic to $W(B_3)$.

Notice that by using these symmetries one can generate the five bilinear equations \eqref{211019_bilineareqofFredholmdet1}-\eqref{211019_bilineareqofFredholmdet5}
just from the two equations \eqref{211019_bilineareqofFredholmdet1}, \eqref{211019_bilineareqofFredholmdet5}.
By using $r_1,r_2,r_3,r_4$ \eqref{r1}-\eqref{211216r1r2r3r4} We find
\begin{align}
&\eqref{211019_bilineareqofFredholmdet2}=-t\cdot \left[\eqref{211019_bilineareqofFredholmdet1}\right]_{\left(r_2r_1\right)\left(M_1,M_2,\zeta_1,\zeta_2\right)},\quad
\eqref{211019_bilineareqofFredholmdet3}=\eqref{211019_bilineareqofFredholmdet5}-\mathfrak{q}^{-2\theta_\infty}\cdot\left[\eqref{211019_bilineareqofFredholmdet5}\right]_{r_4\left(M_1,M_2,\zeta_1,\zeta_2\right)},\label{211127_expressbilinear234with151}\\
&\eqref{211019_bilineareqofFredholmdet4}=\mathfrak{q}^{\theta_0+\theta_1-\theta_\infty+\frac{1}{2}}\left(\left[\left[\eqref{211019_bilineareqofFredholmdet5}\right]_{r_3\left(M_1,M_2,\zeta_1,\zeta_2\right)}\right]_{t\rightarrow \mathfrak{q}^{-1}t}-\left[\eqref{211019_bilineareqofFredholmdet5}\right]_{r_4\left(M_1,M_2,\zeta_1,\zeta_2\right)}\right),
\label{211127_expressbilinear234with15}
\end{align}
where $\left[\left(\cdots\right)\right]_{r_i(M_1,M_2,\zeta_1,\zeta_2)}$ stands for the left-hand side of each equation with $(M_1,M_2,\zeta_1,\zeta_2)$ substituted by their images under $r_i$ \eqref{211216r1r2r3r4}.
We will use \eqref{211127_expressbilinear234with15} in section \ref{sec_orderkappaanalyticproof} to check the bilinear equations at first order in $\kappa$.
\subsection{Special checks around symmetric points}
Taking into account \eqref{r1}-\eqref{211216r1r2r3r4} and \eqref{211127_expressbilinear234with151},\eqref{211127_expressbilinear234with15} we notice that \eqref{211019_bilineareqofFredholmdet3},\eqref{211019_bilineareqofFredholmdet4},\eqref{211019_bilineareqofFredholmdet5} are trivially satisfied for some special choices of parameters $\left(M_1,M_2,\zeta_1,\zeta_2\right)$.

For example, by restricting on the subspace $\zeta_1=\zeta_2$
which implies $\theta_\infty=0$ by \eqref{theta01tinftytMzetaindentification_convenient}
and noticing that this is the fixed locus of the reflection $r_4$ in 
\eqref{211216r1r2r3r4}, we see that eq.\eqref{211019_bilineareqofFredholmdet3}
is satisfied.

\noindent
In the same way we can also show the following by using Weyl transformations.

For $\left(M_1,M_2,\zeta_1,\zeta_2\right)=\left(M_1,M_2,\zeta_1,-\zeta_1\right)$, the coefficient of the third term in \eqref{211019_bilineareqofFredholmdet4} vanishes, the coefficient of the second term is $-1$ and also
$\tau_3\left(\kappa/\Omega_3\right)/F_3=\tau_4\left(\kappa/\Omega_4\right)/F_4,\underline{\tau}_3\left(\kappa/\underline{\Omega}_3\right)/\underline{F}_3=\underline{\tau}_4\left(\kappa/\underline{\Omega}_4\right)/\underline{F}_4$ (recall our definition of $\tau$-functions in terms of the spectral determinant \eqref{eq:tau-GPF}) follow due to the Weyl symmetry $r_3r_4r_3$
\begin{align}
r_3r_4r_3:\,(M_1,M_2,\zeta_1,\zeta_2)\rightarrow (M_1,M_2,-\zeta_1,-\zeta_2).
\end{align}
Together with $\underline{F}_3F_4=F_3\underline{F}_4$ and $\Omega_3=\Omega_4=-\underline{\Omega}_3=-\underline{\Omega}_4$ \eqref{211223FandOmegacondition} these implies $\underline{\tau}_3\tau_4=\tau_3\underline{\tau}_4$, hence \eqref{211019_bilineareqofFredholmdet4} is trivially satisfied.

For $\left(M_1,M_2,\zeta_1,\zeta_2\right)=\left(M_1,M_2,\zeta_1,\frac{i\left(M_1-M_2-1\right)}{2}\right)$, the coefficient of the third term in \eqref{211019_bilineareqofFredholmdet5} vanishes, the coefficient of the second term is $1$ and also
$\underline{\tau}_1\left(\kappa/\underline{\Omega}_1\right)/\underline{F}_1=\tau_4\left(\kappa/\Omega_4\right)/F_4,\tau_2\left(\kappa/\Omega_2\right)/F_2=\underline{\tau}_3\left(\kappa/\underline{\Omega}_3\right)/\underline{F}_3$ follow due to the Weyl symmetry $r_1$.
Hence, together with $\underline{\Omega}_1=\Omega_4,\underline{\Omega}_3=\Omega_2$ and $F_1\underline{F}_2=-\underline{F}_3F_4$ \eqref{211223FandOmegacondition}, we find that \eqref{211019_bilineareqofFredholmdet5} is trivially satisfied.

Moreover, we notice that \eqref{211019_bilineareqofFredholmdet2} at $\left(M_1,M_2,\zeta_1,\zeta_2\right)$ is equivalent to \eqref{211019_bilineareqofFredholmdet1} evaluated at a different point obtained by a Weyl transformation $\left(M_1',M_2',\zeta_1',\zeta_2'\right)=\left(k-M_2,k-M_1,-\zeta_1,-\zeta_2\right)$, hence
the equation (l.h.s.~of \eqref{211019_bilineareqofFredholmdet1})$-$(l.h.s.~of \eqref{211019_bilineareqofFredholmdet2})$=0$ is trivially satisfied at the fixed points of this Weyl transformation, which are $\left(M_1,M_2,\zeta_1,\zeta_2\right)=\left(M_1,k-M_1,0,0\right)$.

\subsection{Analysis to first order in $\kappa$}
\label{sec_orderkappaanalyticproof}
Due to the values of the overall coefficients of each term, the bilinear equations \eqref{211019_bilineareqofFredholmdet1}-\eqref{211019_quarticeqofFredholmdet} are trivially satisfied at order $\kappa^0$, while the quartic equation \eqref{211019_quarticeqofFredholmdet} is trivial also at first order in $\kappa$. At higher order in $\kappa$ these equations are non-trivial.
For the bilinear equations \eqref{211019_bilineareqofFredholmdet1}-\eqref{211019_bilineareqofFredholmdet5}, we can analytically check the equations at order $\kappa$ for an arbitrary choice of the parameters $\left(M_1,M_2,\zeta_1,\zeta_2\right)$.
First let us consider \eqref{211019_bilineareqofFredholmdet1}.
At order $\kappa$ the bilinear equation reduces to the following linear relation on ${\widehat\rho}_k(M_1,M_2,0,\zeta_1,\zeta_2)$:
\begin{align}
&\text{tr}\left[
i{\widehat\rho}_k\left(M_1,M_2,0,\zeta_1-\frac{i}{2},\zeta_2+\frac{i}{2}\right)
+i{\widehat\rho}_k\left(M_1,M_2,0,\zeta_1+\frac{i}{2},\zeta_2-\frac{i}{2}\right)\right.\nonumber \\
&\quad +\mathfrak{q}^{-2\theta_1}\left(i{\widehat\rho}_k\left(M_1,M_2,0,\zeta_1+\frac{i}{2},\zeta_2+\frac{i}{2}\right)
+i{\widehat\rho}_k\left(M_1,M_2,0,\zeta_1-\frac{i}{2},\zeta_2-\frac{i}{2}\right)\right)\nonumber \\
&\left.\quad -\left(1-\mathfrak{q}^{-2\theta_1}t\right)\left(
{\widehat\rho}_k\left(M_1,M_2-1,0,\zeta_1,\zeta_2\right)
-{\widehat\rho}_k\left(M_1,M_2+1,0,\zeta_1,\zeta_2\right)
\right)
\right]=0.
\label{orderk}
\end{align} 
Notice that for all the shifts of the parameters in \eqref{orderk}, the arguments of the quantum dilogarithm in ${\widehat\rho}_k$ are different only by units of $\frac{i}{b}$.
Hence by using the recursive relation \eqref{211107quantumdilogrecursive} we can express all the ${\widehat\rho}_k$s entering \eqref{orderk} by using only one of them up to rational factors:
\begin{align}
{\widehat\rho}_k\left(M_1,M_2,0,\zeta_1-\frac{i}{2},\zeta_2+\frac{i}{2}\right)&= 
ie^{-\frac{1}{2k}{\widehat x}}\left(1-e^{\frac{2\pi\zeta_2}{k}+\frac{\pi iM_2}{k}}e^{\frac{{\widehat x}}{k}}\right){\widehat I}_1e^{\frac{1}{2k}{\widehat x}}\left(1+e^{\frac{2\pi\zeta_2}{k}-\frac{\pi iM_2}{k}}e^{\frac{{\widehat x}}{k}}\right){\widehat I}_2, \\
{\widehat\rho}_k\left(M_1,M_2,0,\zeta_1+\frac{i}{2},\zeta_2-\frac{i}{2}\right)&= 
-ie^{\frac{1}{2k}{\widehat x}}\left(1-e^{\frac{2\pi\zeta_2}{k}-\frac{\pi iM_2}{k}}e^{\frac{{\widehat x}}{k}}\right){\widehat I}_1e^{-\frac{1}{2k}{\widehat x}}\left(1+e^{\frac{2\pi\zeta_2}{k}+\frac{\pi iM_2}{k}}e^{\frac{{\widehat x}}{k}}\right){\widehat I}_2, \\
{\widehat\rho}_k\left(M_1,M_2,0,\zeta_1+\frac{i}{2},\zeta_2+\frac{i}{2}\right)&=
ie^{\frac{1}{2k}{\widehat x}}\left(1-e^{\frac{2\pi\zeta_2}{k}+\frac{\pi iM_2}{k}}e^{\frac{{\widehat x}}{k}}\right){\widehat I}_1e^{-\frac{1}{2k}{\widehat x}}\left(1+e^{\frac{2\pi\zeta_2}{k}-\frac{\pi iM_2}{k}}e^{\frac{{\widehat x}}{k}}\right){\widehat I}_2, \\
{\widehat\rho}_k\left(M_1,M_2,0,\zeta_1-\frac{i}{2},\zeta_2-\frac{i}{2}\right)&=
-ie^{-\frac{1}{2k}{\widehat x}}\left(1-e^{\frac{2\pi\zeta_2}{k}-\frac{\pi iM_2}{k}}e^{\frac{{\widehat x}}{k}}\right){\widehat I}_1e^{\frac{1}{2k}{\widehat x}}\left(1+e^{\frac{2\pi\zeta_2}{k}+\frac{\pi iM_2}{k}}e^{\frac{{\widehat x}}{k}}\right){\widehat I}_2, \\
{\widehat\rho}_k\left(M_1,M_2-1,0,\zeta_1,\zeta_2\right)&=
e^{\frac{1}{2k}{\widehat x}}{\widehat I}_1e^{-\frac{1}{2k}{\widehat x}}\left(1+e^{\frac{2\pi\zeta_2}{k}-\frac{\pi iM_2}{k}}e^{\frac{{\widehat x}}{k}}\right)\left(1+e^{\frac{2\pi\zeta_2}{k}+\frac{\pi iM_2}{k}}e^{\frac{{\widehat x}}{k}}\right){\widehat I}_2, \\
{\widehat\rho}_k\left(M_1,M_2+1,0,\zeta_1,\zeta_2\right)&=
e^{-\frac{1}{2k}{\widehat x}}
\left(1-e^{\frac{2\pi\zeta_2}{k}+\frac{\pi iM_2}{k}}e^{\frac{{\widehat x}}{k}}\right)\left(1-e^{\frac{2\pi\zeta_2}{k}-\frac{\pi iM_2}{k}}e^{\frac{{\widehat x}}{k}}\right)
{\widehat I}_1
e^{-\frac{1}{2k}{\widehat x}}
{\widehat I}_2,
\end{align}
where ${\widehat I}_1,{\widehat I}_2$ are
\begin{align}
{\widehat I}_1&=
e^{\pi\zeta_2}
e^{\left(-\frac{i\zeta_1}{k}+1-\frac{M_1+M_2}{2k}\right){\widehat x}}
\frac{
\Phi_b\left(\frac{{\widehat x}}{2\pi b}-\frac{iM_1}{2b}+\frac{ib}{2}\right)
\Phi_b\left(\frac{{\widehat x}}{2\pi b}-\frac{iM_2}{2b}+\frac{ib}{2}+\frac{\zeta_2}{b}+\frac{i}{2b}\right)
}{
\Phi_b\left(\frac{{\widehat x}}{2\pi b}+\frac{iM_1}{2b}-\frac{ib}{2}\right)
\Phi_b\left(\frac{{\widehat x}}{2\pi b}+\frac{iM_2}{2b}-\frac{ib}{2}+\frac{\zeta_2}{b}-\frac{i}{2b}\right)
}
\frac{1}{2\cosh\frac{{\widehat p}}{2}},\label{orderkeachtermsimplifiedbyrecursiverelation1} \\
{\widehat I}_2&=e^{\left(\frac{i\zeta_1}{k}+\frac{M_1+M_2}{2k}\right){\widehat x}}
\frac{
\Phi_b\left(\frac{{\widehat x}}{2\pi b}+\frac{iM_1}{2b}\right)
\Phi_b\left(\frac{{\widehat x}}{2\pi b}+\frac{iM_2}{2b}+\frac{\zeta_2}{b}+\frac{i}{2b}\right)
}{
\Phi_b\left(\frac{{\widehat x}}{2\pi b}-\frac{iM_1}{2b}-\frac{ib}{2}\right)
\Phi_b\left(\frac{{\widehat x}}{2\pi b}-\frac{iM_2}{2b}+\frac{\zeta_2}{b}-\frac{i}{2b}\right)
}
\frac{1}{2\cosh\frac{{\widehat p}}{2}}.
\label{orderkeachtermsimplifiedbyrecursiverelation}
\end{align}
Summing \eqref{orderkeachtermsimplifiedbyrecursiverelation1},\eqref{orderkeachtermsimplifiedbyrecursiverelation} with the coefficients in \eqref{orderk} we find
\begin{align}
&i{\widehat\rho}_k\left(M_1,M_2,0,\zeta_1-\frac{i}{2},\zeta_2+\frac{i}{2}\right)
+i{\widehat\rho}_k\left(M_1,M_2,0,\zeta_1+\frac{i}{2},\zeta_2-\frac{i}{2}\right)\nonumber \\
&\quad +\mathfrak{q}^{-2\theta_1}\left(i{\widehat\rho}_k\left(M_1,M_2,0,\zeta_1+\frac{i}{2},\zeta_2+\frac{i}{2}\right)
+i{\widehat\rho}_k\left(M_1,M_2,0,\zeta_1-\frac{i}{2},\zeta_2-\frac{i}{2}\right)\right)\nonumber \\
&\quad -\left(1-\mathfrak{q}^{-2\theta_1}t\right)\left(
{\widehat\rho}_k\left(M_1,M_2-1,0,\zeta_1,\zeta_2\right)
-{\widehat\rho}_k\left(M_1,M_2+1,0,\zeta_1,\zeta_2\right)
\right)=0,
\label{211107orderkprooffinal}
\end{align}
hence \eqref{211019_bilineareqofFredholmdet1} is satisfied at order $\kappa$.
We can also show
\eqref{211019_bilineareqofFredholmdet5}
at order $\kappa$ by a completely parallel calculation.
So far we have proved that \eqref{211019_bilineareqofFredholmdet1} and \eqref{211019_bilineareqofFredholmdet5} hold at order $\kappa$ for any values of $(M_1,M_2,\zeta_1,\zeta_2)$.
Since the other three bilinear equations \eqref{211019_bilineareqofFredholmdet2},\eqref{211019_bilineareqofFredholmdet3},\eqref{211019_bilineareqofFredholmdet4} can be written in some combinations of these two equations  by using \eqref{211127_expressbilinear234with151},\eqref{211127_expressbilinear234with15}, our calculation also proves that \eqref{211019_bilineareqofFredholmdet2},\eqref{211019_bilineareqofFredholmdet3},\eqref{211019_bilineareqofFredholmdet4} also hold at order $\kappa$ for arbitrary $(M_1,M_2,\zeta_1,\zeta_2)$.

\subsection{Exact values of $\text{tr}{\widehat\rho}_k\left(M_1,M_2,0,\zeta_1,\zeta_2\right)^n$ 
at integer points}
\label{220114sec_exact}
In order to check the equations at higher order, we need to calculate the exact expressions for the higher traces of the spectral density matrix.
In this subsection we explain that this can be done systematically when $k\in\mathbb{N}$ and $M_1,M_2,2i\zeta_1\in\mathbb{Z}$.
For $M_1,M_2\in\mathbb{Z}$, the ratios of quantum dilogarithm in the density matrix ${\widehat\rho}_k$ for $M=0$ \eqref{211015rhoM1M2integerM30} reduce to products of hyperbolic functions:
\begin{align}
{\widehat\rho}_k\left(M_1,M_2,0,\zeta_1,\zeta_2\right)&=
i^{M_1+M_2}
e^{-\frac{i\zeta_1{\widehat x}}{k}}
\frac{1}{2\cosh\frac{{\widehat x}+\pi iM_1}{2}}
\left(\prod_{r=1}^{M_1}2\sinh\frac{{\widehat x}-2\pi i\sigma_{1,r}}{2k}\right)
\frac{1}{2\cosh\frac{{\widehat p}}{2}}\nonumber \\
&\quad \times \left(\prod_{r=1}^{M_1}\frac{1}{2\cosh\frac{{\widehat x}-2\pi i\sigma_{1,r}}{2k}}\right)
e^{\frac{i\zeta_1{\widehat x}}{k}}
e^{\frac{i\zeta_2{\widehat p}}{k}}
\left(\prod_{r=1}^{M_2}\frac{1}{2\cosh\frac{{\widehat x}-2\pi i\sigma_{2,r}}{2k}}\right)
\frac{1}{2\cosh\frac{{\widehat p}}{2}}\nonumber \\
&\quad \times \left(\prod_{r=1}^{M_2}2\sinh\frac{{\widehat x}-2\pi i\sigma_{2,r}}{2k}\right)
\frac{1}{2\cosh\frac{{\widehat x}+\pi iM_2}{2}}
e^{-\frac{i\zeta_2{\widehat p}}{k}},
\end{align}
with $\sigma_{i,r}=\frac{M_i+1}{2}-r$.
See appendix \ref{sec:FGF} for details.
For $k\in\mathbb{N}$,
by using the following formula
\begin{align}
\frac{\prod_{r=1}^{n}2\sinh\frac{x-2\pi i\left(\frac{n+1}{2}-r\right)}{2k}}{2\cosh\frac{x+\pi in}{2}}=\frac{i^{-n}}{\prod_{r=1}^{k-n}2\cosh\frac{x+2\pi i\left(\frac{k-n+1}{2}-r\right)}{2k}},\quad
\left(0\le n\le k\right)
\end{align}
which can be obtained by the standard formula $x^n-y^n=\prod_{j=1}^n\left(x-e^{\frac{2\pi ij}{n}}y\right)$, we can
further rewrite ${\widehat\rho}_k\left(M_1,M_2,0,\zeta_1,\zeta_2\right)$ as
\begin{align}
{\widehat\rho}_k\left(M_1,M_2,0,\zeta_1,\zeta_2\right)&=
e^{-\frac{i\zeta_1{\widehat x}}{k}}
\frac{1}{\prod_{r=1}^{k-M_1}2\cosh\frac{{\widehat x}+t_{0,k-M_1,r}}{2}}
\frac{1}{2\cosh\frac{{\widehat p}}{2}}
\frac{1}{\prod_{r=1}^{M_1}2\cosh\frac{{\widehat x}+t_{0,M_1,r}}{2}}
e^{\frac{i\zeta_1{\widehat x}}{k}}
e^{\frac{i\zeta_2{\widehat p}}{k}}\nonumber \\
&\quad \times \frac{1}{\prod_{r=1}^{M_2}2\cosh\frac{{\widehat x}+t_{\zeta_2,M_2,r}}{2}}
\frac{1}{2\cosh\frac{{\widehat p}}{2}}
\frac{1}{\prod_{r=1}^{k-M_2}2\cosh\frac{{\widehat x}+t_{\zeta_2,k-M_2,r}}{2}}
e^{-\frac{i\zeta_2{\widehat p}}{k}},
\end{align}
where we have defined $t_{\zeta,n,r}$ as \eqref{eq:t-Def}.
By performing some similarity transformations we obtain
\begin{align}
{\widehat\rho}_k\left(M_1,M_2,0,\zeta_1,\zeta_2\right)\sim {\widehat\rho}'={\widehat\rho}'_1{\widehat\rho}'_2,
\label{210812overallfactor}
\end{align}
with
\begin{align}
{\widehat\rho}'_1&=\sqrt{A\left({\widehat x}\right)}\frac{1}{2\cosh\frac{{\widehat p}}{2}}\sqrt{B\left({\widehat x}\right)}, \\
{\widehat\rho}'_2&=\sqrt{B\left({\widehat x}\right)}\frac{1}{2\cosh\frac{{\widehat p}}{2}}\sqrt{A\left({\widehat x}\right)}, \\
A\left(x\right)&=\frac{e^{-\frac{i\zeta_1}{k}x}}{\left(\prod_{r=1}^{k-M_1}2\cosh\frac{x+t_{0,k-M_1,r}}{2k}\right)\left(\prod_{r=1}^{k-M_2}2\cosh\frac{x+t_{\zeta_2,k-M_2,r}}{2k}\right)},\\
B\left(x\right)&=\frac{e^{\frac{i\zeta_1}{k}x}}{\left(\prod_{r=1}^{M_1}2\cosh\frac{x+t_{0,M_1,r}}{2k}\right)\left(\prod_{r=1}^{M_2}2\cosh\frac{x+t_{\zeta_2,M_2,r}}{2k}\right)}.
\label{210830rho1rho2AB}
\end{align}
The operator ${\widehat\rho}'$ has the following property:
\begin{align}
e^{\frac{{\widehat x}}{k}}{\widehat\rho}'
-{\widehat\rho}'e^{\frac{{\widehat x}}{k}}
=\sum_{a=1}^2{\widehat C}_a|0\rangle\!\rangle\langle\!\langle 0|{\widehat D}_a,
\label{211018TWPYstructureofrho'}
\end{align}
with
\begin{align}
{\widehat C}_1=e^{\frac{{\widehat x}}{2k}}\sqrt{A\left({\widehat x}\right)},\quad
{\widehat C}_2=-{\widehat\rho}'_1e^{\frac{{\widehat x}}{2k}}\sqrt{B\left({\widehat x}\right)},\quad
{\widehat D}_1=\sqrt{B\left({\widehat x}\right)}e^{\frac{{\widehat x}}{2k}}{\widehat\rho}'_2,\quad
{\widehat D}_2=\sqrt{A\left({\widehat x}\right)}e^{\frac{{\widehat x}}{2k}}.
\end{align}
Here we have defined the position eigenstates $|x\rangle$ and the momentum eigenstates as \eqref{eq:QM-Def1},\eqref{eq:QM-Def}
From \eqref{211018TWPYstructureofrho'} we can show
\begin{align}
\text{tr}\left({\widehat\rho}'\right)^n&=\frac{k}{2}\int_{-\infty}^\infty \frac{dx}{2\pi}e^{-\frac{x}{k}}\sum_{\ell=0}^{n-1}\sum_{a=1}^2
\left(
\frac{d}{dx}\left(\langle x|\left({\widehat\rho}'\right)^\ell{\widehat C}_a|0\rangle\!\rangle\right)
\langle\!\langle 0|{\widehat D}_a\left({\widehat\rho}'\right)^{n-1-\ell}|x\rangle\nonumber\right. \\
&\quad \left.-\langle x|\left({\widehat\rho}'\right)^\ell{\widehat C}_a|0\rangle\!\rangle
\frac{d}{dx}\left(\langle\!\langle 0|{\widehat D}_a\left({\widehat\rho}'\right)^{n-1-\ell}|x\rangle\right)
\right).
\label{211015trrho'}
\end{align}
If we define $\phi_{a,\ell}\left(x\right)$ as
\begin{align}
\phi_{1,\ell}\left(x\right)=\frac{1}{\sqrt{A\left(x\right)}e^{\frac{x}{2k}}}\langle x|\left({\widehat\rho}'\right)^\ell\sqrt{A\left({\widehat x}\right)}e^{\frac{{\widehat x}}{2k}}|0\rangle\!\rangle,\quad
\phi_{2,\ell}\left(x\right)=\frac{1}{\sqrt{A\left(x\right)}e^{\frac{x}{2k}}}\langle x|\left({\widehat\rho}'\right)^\ell{\widehat\rho}'_1\sqrt{B\left({\widehat x}\right)}e^{\frac{{\widehat x}}{2k}}|0\rangle\!\rangle,
\end{align}
we can write the matrix elements with insertions of ${\widehat C}_a$ as
\begin{align}
\langle x|\left({\widehat\rho}'\right)^\ell {\widehat C}_1|0\rangle\!\rangle=\sqrt{A\left(x\right)}e^{\frac{x}{2k}}\phi_{1,\ell}\left(x\right),\quad
\langle x|\left({\widehat\rho}'\right)^\ell {\widehat C}_2|0\rangle\!\rangle=-\sqrt{A\left(x\right)}e^{\frac{x}{2k}}\phi_{2,\ell}\left(x\right).
\label{211015Cinphi}
\end{align}
By using the fact that $\langle x|{\widehat\rho}'_1|y\rangle=\langle y|{\widehat \rho}'_2|x\rangle$ and $\langle x|{\widehat\rho}'|y\rangle=\langle y|{\widehat\rho}'|x\rangle$ we can also write the matrix elements with ${\widehat D}_a$ in terms of $\phi_{a,\ell}\left(x\right)$, namely:
\begin{align}
\langle \!\langle 0|{\widehat D}_1\left({\widehat\rho}'\right)^\ell|x\rangle=\sqrt{A\left(x\right)}e^{\frac{x}{2k}}\phi_{2,\ell}\left(x\right),\quad
\langle \!\langle 0|{\widehat D}_2\left({\widehat\rho}'\right)^\ell|x\rangle=\sqrt{A\left(x\right)}e^{\frac{x}{2k}}\phi_{1,\ell}\left(x\right).
\label{211015Dinphi}
\end{align}
By using \eqref{211015Cinphi} and \eqref{211015Dinphi} we can rewrite 
\eqref{211015trrho'} as
\begin{align}
\text{tr}\left({\widehat\rho}'\right)^\ell=k\int_{-\infty}^\infty \frac{dx}{2\pi}A\left(x\right)\sum_{\ell=0}^{n-1}\left(
\frac{d\phi_{1,\ell}\left(x\right)}{dx}\phi_{2,n-1-\ell}\left(x\right)
-\phi_{1,\ell}\left(x\right)
\frac{d\phi_{2,n-1-\ell}\left(x\right)}{dx}
\right).
\end{align}
Note that $\phi_{i,\ell}\left(x\right)$ can be calculated recursively as
\begin{align}
\phi_{1,0}\left(x\right)&=\frac{1}{\sqrt{k}},\quad
\phi_{2,0}\left(x\right)=\frac{1}{k}\int_{-\infty}^\infty \frac{dy}{2\pi}\frac{e^{\frac{y}{k}}}{e^{\frac{x}{k}}+e^{\frac{y}{k}}}B\left(y\right)\frac{1}{\sqrt{k}}, \\
{\widetilde \phi}_{i,\ell}\left(x\right)&=\frac{1}{k}\int_{-\infty}^\infty \frac{dy}{2\pi}\frac{e^{\frac{y}{k}}}{e^{\frac{x}{k}}+e^{\frac{y}{k}}}A\left(y\right)\phi_{i,\ell}\left(y\right), \\
\phi_{i,\ell+1}\left(x\right)&=\frac{1}{k}\int_{-\infty}^\infty \frac{dy}{2\pi}\frac{e^{\frac{y}{k}}}{e^{\frac{x}{k}}+e^{\frac{y}{k}}}B\left(y\right){\widetilde \phi}_{i,\ell}\left(y\right).
\label{210812recursionrelation}
\end{align}

If we further assume $M_1+M_2+2i\zeta_1\in \mathbb{Z}$, we can apply the same technique as in \cite{Putrov:2012zi} to evaluate these integrals.
First we introduce a new variable $u=e^{\frac{x}{2k}}$, to rewrite $A\left(x\right)$ and $B\left(x\right)$ as
\begin{align}
A\left(u\right)&=e^{-\frac{\pi\zeta_2\left(k-M_2\right)}{k}}\frac{u^{2k-q}}{\prod_{r=1}^{k-M_1}\left(u^2+e^{-\frac{1}{k}t_{0,k-M_1,r}}\right)\prod_{r=1}^{k-M_2}\left(u^2+e^{-\frac{1}{k}t_{\zeta_2,k-M_2,r}}\right)}, \\
B\left(u\right)&=e^{-\frac{\pi\zeta_2M_2}{k}}\frac{u^q}{\prod_{r=1}^{M_1}\left(u^2+e^{-\frac{1}{k}t_{0,M_1,r}}\right)\prod_{r=1}^{M_2}\left(u^2+e^{-\frac{1}{k}t_{\zeta_2,M_2,r}}\right)},
\end{align}
where $q=M_1+M_2+2i\zeta_1$.
The integration of $\phi_{2,0}\left(x\right)$ can be evaluated as \cite{Putrov:2012zi}
\begin{align}
\phi_{2,0}\left(u\right)&=
\frac{1}{\pi}e^{-\frac{\pi\zeta_2M_2}{k}}\int_0^\infty dv
\frac{1}{u^2+v^2}\frac{v^{q+1}}{\prod_{r=1}^{M_-}\left(v^2+e^{-\frac{1}{k}t_{0,M_-,r}}\right)\prod_{r=1}^{M_+}\left(v^2+e^{-\frac{1}{k}t_{\zeta_2,M_+,r}}\right)}
\frac{1}{\sqrt{k}}\nonumber \\
&=\frac{1}{\pi}e^{-\frac{\pi\zeta_2M_2}{k}}\left(-2\pi i\right)\sum_{w\in\mathbb{C}\backslash\mathbb{R}_{\ge 0}}\text{Res}\left[\frac{1}{u^2+v^2}\frac{v^{q+1}}{\prod_{r=1}^{M_-}\left(v^2+e^{-\frac{1}{k}t_{0,M_-,r}}\right)\prod_{r=1}^{M_+}\left(v^2+e^{-\frac{1}{k}t_{\zeta_2,M_+,r}}\right)}\right.\nonumber \\
&\left.\quad \times \frac{1}{\sqrt{k}}B_1\left(\frac{\log^{\left(+\right)}v}{2\pi i}\right),v\rightarrow w\right].
\end{align}
Here $\log^{\left(+\right)}u$ is the logarithm with the branch cut chosen as $u\in\mathbb{R}_{\ge 0}$, and $B_j\left(x\right)$ are the Bernoulli polynomials.\footnote{
We can also chose $B_j\left(x\right)$ as any polynomials satisfying
\begin{align}
B_{j+1}\left(x+1\right)-B_{j+1}\left(x\right)=\left(j+1\right)x^j.
\end{align}
}
The poles $w$ contributing to $\phi_{2,0}\left(u\right)$ can be listed explicitly as
\begin{align}
w=
\begin{cases}
\pm iu\\
\pm ie^{-\frac{1}{2k}t_{0,M_1,r}},\quad \left(r=1,2,\cdots,M_1\right)\\
\pm ie^{-\frac{1}{2k}t_{\zeta_2,M_2,r}},\quad \left(r=1,2,\cdots,M_2\right)\\
\end{cases}.
\end{align}
In the same way the recursion relations of $\phi_{i,\ell}$ can be rewritten as
\begin{align}
\phi_{i,\ell}\left(u\right)&=\sum_{j\ge 0}\phi_{i,\ell}^{\left(j\right)}\left(u\right)\left(\log u\right)^j, \\
{\widetilde\phi}_{i,\ell}\left(u\right)&=\frac{1}{\pi}e^{-\frac{\pi\zeta_2\left(k-M_2\right)}{k}}\sum_{j\ge 0}\left(-\frac{\left(2\pi i\right)^{j+1}}{j+1}\right)\sum_{w\in\mathbb{C}\backslash\mathbb{R}_{\ge 0}}\text{Res}\left[\phantom{\left(\frac{\log^{\left(+\right)}v}{2\pi i}\right)}\right.\nonumber \\
&\quad \frac{1}{u^2+v^2}\frac{v^{2k-q+1}\phi_{i,\ell}^{\left(j\right)}\left(v\right)}{\prod_{r=1}^{k-M_1}\left(v^2+e^{-\frac{1}{k}t_{0,k-M_1,r}}\right)\prod_{r=1}^{k-M_2}\left(v^2+e^{-\frac{1}{k}t_{\zeta_2,k-M_2,r}}\right)}\nonumber \\
&\quad \times \left.B_{j+1}\left(\frac{\log^{\left(+\right)}v}{2\pi i}\right),v\rightarrow w\right],\label{210911poletobecollected1} \\
{\widetilde \phi}_{i,\ell}\left(u\right)&=\sum_{j\ge 0}{\widetilde \phi}_{i,\ell}^{\left(j\right)}\left(u\right)\left(\log u\right)^j, \\
\phi_{i,\ell+1}\left(u\right)&=\frac{1}{\pi}e^{-\frac{\pi\zeta_2M_2}{k}}\sum_{j\ge 0}\left(-\frac{\left(2\pi i\right)^{j+1}}{j+1}\right)\sum_{w\in\mathbb{C}\backslash\mathbb{R}_{\ge 0}}\text{Res}\left[\phantom{\left(\frac{\log^{\left(+\right)}v}{2\pi i}\right)}\right.\nonumber \\
&\left.\quad \frac{1}{u^2+v^2}\frac{v^{q+1}{\widetilde \phi}_{i,\ell}^{\left(j\right)}\left(v\right)}{\prod_{r=1}^{M_1}\left(v^2+e^{-\frac{1}{k}t_{0,M_1,r}}\right)\prod_{r=1}^{M_2}\left(v^2+e^{-\frac{1}{k}t_{\zeta_2,M_2,r}}\right)}B_{j+1}\left(\frac{\log^{\left(+\right)}v}{2\pi i}\right),v\rightarrow w\right],\label{210911poletobecollected2}
\end{align}
where the poles contributing in \eqref{210911poletobecollected1} are
\begin{align}
w=
\begin{cases}
\pm iu\\
\pm ie^{-\frac{1}{2k}t_{0,k-M_1,r}},\quad \left(r=1,2,\cdots,k-M_1\right)\\
\pm ie^{-\frac{1}{2k}t_{\zeta_2,k-M_2,r}},\quad \left(r=1,2,\cdots,k-M_2\right)\\
\text{poles of }\phi_{i,\ell}^{\left(j\right)}\left(w\right)
\end{cases},
\end{align}
and the poles contributing in \eqref{210911poletobecollected2} are
\begin{align}
w=
\begin{cases}
\pm iu\\
\pm ie^{-\frac{1}{2k}t_{0,M_1,r}},\quad \left(r=1,2,\cdots,M_1\right)\\
\pm ie^{-\frac{1}{2k}t_{\zeta_2,M_2,r}},\quad \left(r=1,2,\cdots,M_2\right)\\
\text{poles of }{\widetilde \phi}_{i,\ell}^{\left(j\right)}\left(w\right)
\end{cases}.
\end{align}
We can show by induction that the poles of $\phi_{i,\ell}$ and ${\widetilde\phi}_{i,\ell}$ satisfy the following inclusions
\begin{align}
&\left\{\text{poles of }\phi_{i,\ell}\left(u\right)\right\}\nonumber \\
&\quad \subset
\left\{\pm e^{-\frac{1}{2k}t_{0,M_1,r}}\right\}_{r=1}^{M_1}
\cup \left\{\pm e^{-\frac{1}{2k}t_{\zeta_2,M_2,r}}\right\}_{r=1}^{M_2}
\cup \left\{\pm ie^{-\frac{1}{2k}t_{0,k-M_1,r}}\right\}_{r=1}^{k-M_1}
\cup \left\{\pm ie^{-\frac{1}{2k}t_{\zeta_2,k-M_2,r}}\right\}_{r=1}^{k-M_2},\label{210911polesofphiandphitilde1}\\
&\left\{\text{poles of }{\widetilde \phi}_{i,\ell}\left(u\right)\right\}\nonumber \\
&\quad \subset
\left\{\pm ie^{-\frac{1}{2k}t_{0,M_1,r}}\right\}_{r=1}^{M_1}
\cup \left\{\pm ie^{-\frac{1}{2k}t_{\zeta_2,M_2,r}}\right\}_{r=1}^{M_2}
\cup \left\{\pm e^{-\frac{1}{2k}t_{0,k-M_1,r}}\right\}_{r=1}^{k-M_1}
\cup \left\{\pm e^{-\frac{1}{2k}t_{\zeta_2,k-M_2,r}}\right\}_{r=1}^{k-M_2},
\label{210911polesofphiandphitilde}
\end{align}
for both $i=1,2$ and any $\ell\ge 0$.
Once we obtain $\left\{\phi_{i,\ell}\right\}_{\ell=0}^{n-1}$, we can calculate $\text{tr}\left({\widehat\rho}'\right)^n$ as
\begin{align}
&\sum_{\ell=0}^{n-1}\left(
\frac{d\phi_{1,\ell}}{du}\phi_{2,n-1-\ell}
-\phi_{1,\ell}\frac{d\phi_{2,n-1-\ell}}{du}\right)=
\sum_{j\ge 0}\Psi_n^{\left(j\right)}\left(u\right)\left(\log u\right)^j,\nonumber \\
&\text{tr}\left({\widehat\rho}'\right)^n=\frac{k}{2\pi}e^{-\frac{\pi\zeta_2\left(k-M_2\right)}{k}}\sum_{j\ge 0}\left(-\frac{\left(2\pi i\right)^{j+1}}{j+1}\right)\sum_{w\in\mathbb{C}\backslash \mathbb{R}_{\ge 0}}\text{Res}\left[\phantom{\left(\frac{\log^{\left(+\right)}v}{2\pi i}\right)}\right.\nonumber \\
&\left.\quad \frac{u^{2k-q}}{\prod_{r=1}^{k-M_1}\left(u^2+e^{-\frac{1}{k}t_{0,k-M_1,r}}\right)\prod_{r=1}^{k-M_2}\left(u^2+e^{-\frac{1}{k}t_{\zeta_2,k-M_2,r}}\right)}\Psi_n^{\left(j\right)}\left(u\right)B_{j+1}\left(\frac{\log^{\left(+\right)}u}{2\pi i}\right),u\rightarrow w\right].
\end{align}
Here the poles $w$ contributing to $\text{tr}\left({\widehat\rho}'\right)^n$ are the ones in the set $\left\{\text{poles of }\phi_{i,\ell}\left(w\right)\right\}$ listed in \eqref{210911polesofphiandphitilde1}.
By using these results, we are able to perform the checks of the bilinear and quartic equations for the $\tau$-functions at higher order that we list in Table \ref{220119_tableexactvalues}.

\begin{table}
\begin{center}
\begin{tabular}{|c|c|c|c|c|c|c|}
\hline
$\left(k,M_1,M_2,\zeta_1\right)$                  &\eqref{211019_bilineareqofFredholmdet1}&\eqref{211019_bilineareqofFredholmdet2}&\eqref{211019_bilineareqofFredholmdet3}&\eqref{211019_bilineareqofFredholmdet4}&\eqref{211019_bilineareqofFredholmdet5}&\eqref{211019_quarticeqofFredholmdet}\\ \hline
$\left(2,1,0,0\right)$                            &                                    &$7$                                    &                                       &                                       &                                       &\\ \hline
$\left(2,1,1,-\frac{i}{2}\right)$                 &$3$                                    &$3$                                    &$3$                                    &$3$                                    &$3$                                    &$3$\\ \hline
$\left(2,1,1,0\right)$                            &$2$                                    &$2$                                    &$2$                                    &$2$                                    &$2$                                    &$2$\\ \hline
$\left(2,1,1,\frac{i}{2}\right)$                  &$3$                                    &$3$                                    &$3$                                    &$3$                                    &$3$                                    &$3$\\ \hline
$\left(3,1,1,-i\right)$          &                                       &                                       &$2$                                    &$2$                                    &$2$                                    &$2$\\ \hline
$\left(3,1,1,-\frac{i}{2}\right)$                 &$2$                                    &$2$                                    &$2$                                    &$2$                                    &$2$                                    &$2$\\ \hline
$\left(3,1,1,0\right)$           &$2$                                    &$2$                                    &$2$                                    &$2$                                    &$2$                                    &$2$\\ \hline
$\left(3,1,1,\frac{i}{2}\right)$                  &$2$                                    &$2$                                    &$2$                                    &$2$                                    &$2$                                    &$2$\\ \hline
$\left(3,1,1,i\right)$           &                                       &                                       &$2$                                    &$2$                                    &$2$                                    &$2$\\ \hline
$\left(3,1,2,-\frac{i}{2}\right)$&                                       &$2$                                    &                                       &                                       &                                       &$2$\\ \hline
$\left(3,1,2,0\right)$                            &$2$                                    &$2$                                    &$2$                                    &$2$                                    &$2$                                    &$2$\\ \hline
$\left(3,1,2,\frac{i}{2}\right)$ &                                       &$2$                                    &                                       &                                       &                                       &\\ \hline
$\left(3,2,1,0\right)$                            &$2$                                    &$2$                                    &$2$                                    &$2$                                    &$2$                                    &$2$\\ \hline
$\left(3,2,1,\frac{i}{2}\right)$ &                                       &                                       &                                       &                                       &                                       &$2$\\ \hline
$\left(3,2,2,-i\right)$          &                                       &$2$                                    &                                       &                                       &                                       &$2$\\ \hline
$\left(3,2,2,0\right)$           &$2$                                    &$2$                                    &                                       &                                       &                                       &$2$\\ \hline
$\left(3,2,2,\frac{i}{2}\right)$                  &$2$                                    &$2$                                    &$2$                                    &$2$                                    &$2$                                    &$2$\\ \hline
$\left(3,2,2,i\right)$           &                                       &                                       &                                       &                                       &                                       &$2$\\ \hline
$\left(4,1,1,-\frac{i}{2}\right)$                 &$2$                                    &$2$                                    &$2$                                    &$2$                                    &$2$                                    &$2$\\ \hline
$\left(4,1,1,\frac{i}{2}\right)$                  &$2$                                    &$2$                                    &$2$                                    &$2$                                    &$2$                                    &\\ \hline
$\left(4,1,2,0\right)$                            &$2$                                    &$2$                                    &$2$                                    &$2$                                    &$2$                                    &$2$\\ \hline
\end{tabular}
\end{center}
\caption{
The list of $\left(k,M_-,M_+,\zeta_1\right)$ for which we have checked that bilinear equations \eqref{211019_bilineareqofFredholmdet1}, \eqref{211019_bilineareqofFredholmdet2}, \eqref{211019_bilineareqofFredholmdet3}, \eqref{211019_bilineareqofFredholmdet4}, \eqref{211019_bilineareqofFredholmdet5}, and the quartic equation \eqref{211019_quarticeqofFredholmdet} for the Fredholm determinant hold.
Each number in the table means that we have confirmed the bilinear/quartic equation at least up to a possible ${\cal O}\left(\kappa^{\#+1}\right)$ correction.
Blank cells stand for the cases where we could not check the equations beyond the first order in $\kappa$ (we could check some of them with fixed values of $\zeta_2$ to higher order in $\kappa$).
}
\label{220119_tableexactvalues}
\end{table}

\section{Coalescence limits: matrix models and quantum curves}\label{five}
The $\mathfrak{q}$-Painlev\'e equations were classified by their symmetry type in \cite{2001CMaPh.220..165S}, where their coalescence patterns are also discussed. In particular we are interested in the following one concerning $\mathfrak{q}$-Painlev\'e $\text{VI}$ equation
\begin{equation}
\text{VI}\rightarrow\text{V}\rightarrow\text{III}_{1}\rightarrow\text{III}_{2}\rightarrow\text{III}_{3}.\label{eq:Coal-pattern}
\end{equation}
The $\mathfrak{q}$-Painlev\'e $\text{III}_{3}$ equation, which is the end point of the above coalescence diagram, is related to the ABJM theory \cite{Bonelli:2017gdk}.
In this section, we study the above coalescence limits in terms of the matrix model and the quantum curve. The degeneration pattern of the quantum curves matches the one in \eqref{eq:Coal-pattern}. 
Since the coalescence limit is similar at all the steps, we study in detail the first one and omit the details for the others.

The coalescence can be seen as a degeneration of the tau function.
In \eqref{eq:tau-GPF}, we saw the relation between the $\mathfrak{q}$-Painlev\'e $\text{VI}$ tau function and the grand partition function of the quiver superconformal Chern-Simons matter theory displayed in Fig.\ref{fig_IIBbranesetup}.
In \eqref{eq:tau-GPF} we normalized the grand partition function by $Z_{k}\left(0;M_{1},M_{2},M,\zeta_{1},\zeta_{2}\right)$.
In this section, we adopt a slightly different normalization factor, namely we define
\begin{align}
Z_{k}^{\text{VI}}\left(N;M_{1},M_{2},M,\zeta_{1},\zeta_{2}\right)
&=\frac{Z_{k}\left(N;M_{1},M_{2},M,\zeta_{1},\zeta_{2}\right)}{e^{i\Theta_{k}\left(M_{1},M_{2},M,\zeta_{1},\zeta_{2}\right)}Z_{k}^{\left(\cs\right)}\left(M_{1}\right)Z_{k}^{\left(\cs\right)}\left(M_{2}\right)}\nonumber \\
 & =\frac{1}{N!\left(N+M\right)!}\int_{-\infty}^\infty \prod_{n=1}^{N}\frac{d\mu_{n}}{2\pi}\prod_{n=1}^{N+M}\frac{d\nu_{n}}{2\pi}\det\left(\begin{array}{c}
\left[\braket{\mu_{m}|\widehat{D}_{1}^{\text{VI}}|\nu_{n}}\right]_{m,n}^{N\times\left(N+M\right)}\\
\left[\bbraket{t_{0,M,r}|\widehat{d}_{1}^{\text{VI}}|\nu_{n}}\right]_{r,n}^{M\times\left(N+M\right)}
\end{array}\right)\nonumber \\
 & \quad\times\det\left(\begin{array}{cc}
\left[\braket{\nu_{m}|\widehat{D}_{2}^{\text{VI}}|\mu_{n}}\right]_{m,n}^{\left(N+M\right)\times N} & \left[\brakket{\nu_{m}|\widehat{d}_{2}^{\text{VI}}|-t_{0,M,r}}\right]_{m,r}^{\left(N+M\right)\times M}\end{array}\right),
\label{eq:ZVI-Def}
\end{align}
where we used \eqref{eq:FGF-Res-Gen}.
Notice that the integrand of $Z_k^{\text{VI}}$ can also be written  explicitly as
\begin{align}
Z_{k}^{\text{VI}}\left(N;M_1,M_2,M,\zeta_1,\zeta_2\right) & =\frac{1}{N!\left(N+M\right)!}
\int_{-\infty}^\infty \prod_{n=1}^{N}\frac{d\mu_{n}}{2\pi k}
\prod_{n=1}^{N+M}\frac{d\nu_{n}}{2\pi k}\nonumber \\
 & \quad\times\prod_{n=1}^{N}e^{\left(-\frac{i\zeta_1}{k}+\frac{2k-M-M_2}{2k}\right)\mu_n}
 \frac{\Phi_{b}\left(\frac{\mu_{n}}{2\pi b}-\frac{iM_{1}}{2b}+\frac{i}{2}b\right)}
 {\Phi_{b}\left(\frac{\mu_{n}}{2\pi b}+\frac{iM_{1}}{2b}-\frac{i}{2}b\right)}
 \frac{\Phi_{b}\left(\frac{\mu_{n}}{2\pi b}-\frac{iM_2-2\zeta_2}{2b}+\frac{i}{2}b\right)}{\Phi_b\left(\frac{\mu_n}{2\pi b}+\frac{iM_2+2\zeta_2}{2b}-\frac{i}{2}b\right)}\nonumber \\
 & \quad\times\prod_{n=1}^{N+M}e^{\left(\frac{i\zeta_1}{k}+\frac{M_1+M_2}{2k}\right)\nu_n}
 \frac{\Phi_{b}\left(\frac{\nu_{n}}{2\pi b}+\frac{iM_{1}}{2b}\right)}
 {\Phi_{b}\left(\frac{\nu_{n}}{2\pi b}-\frac{iM_{1}}{2b}\right)}
 \frac{\Phi_b\left(\frac{\nu_n}{2\pi b}+\frac{iM_2+2\zeta_2}{2b}\right)}
 {\Phi_{b}\left(\frac{\nu_{n}}{2\pi b}-\frac{iM_2-2\zeta_2}{2b}\right)}\nonumber \\
 & \quad\times\left(\frac{\prod_{m<m'}^{N}2\sinh\frac{\mu_{m}-\mu_{m'}}{2k}\prod_{n<n'}^{N+M}2\sinh\frac{\nu_{n}-\nu_{n'}}{2k}}{\prod_{m=1}^{N}\prod_{n=1}^{N+M}2\cosh\frac{\mu_{m}-\nu_{n}}{2k}}\right)^{2}.
\label{220116_ZkVIproductform}
\end{align}

The normalization factor in the first line of \eqref{eq:ZVI-Def}
is the prefactor appearing in \eqref{eq:FGF-Res-Gen} and is independent of $N$.
This normalization factor provides a result consistent with the known result in \cite{Bonelli:2017gdk} at the end of the coalescence, as
we will comment later.
Note that this definition does not contradict our previous analysis since 
for $M=0$
the two normalization factors coincide 
(see \eqref{eq:FGF-Res-M0}).

For clarity, we will study the coalescence limit by treating at the 
same time the matrix model and the
quantum curve. In such a way we can provide the relation between them while flowing along the coalescence.
For this purpose, we first clarify the relation between the matrix model and the quantum curve for $\mathfrak{q}$-Painlev\'e $\text{VI}$.
The conjecture \eqref{22modelrhoinverse} implies that
\begin{align}
 & Z_{k}^{\text{VI}}\left(N;M_{1},M_{2},M,\zeta_{1},\zeta_{2}\right)\nonumber \\
 & =Z_{k}^{\text{VI}}\left(0;M_{1},M_{2},M,\zeta_{1},\zeta_{2}\right)\frac{1}{N!}\int_{-\infty}^\infty \prod_{n=1}^{N}\frac{d\mu_{n}}{2\pi}\det\left(\left[\braket{\mu_{m}|\widehat{\rho}_{k}^{\text{VI}}\left(M_{1},M_{2},M,\zeta_{1},\zeta_{2}\right)|\mu_{n}}\right]_{m,n}^{N\times N}\right),\label{eq:MM-QC-VI}
\end{align}
where $\widehat{\rho}_{k}^{\text{VI}}$ is the conjectural form of the inverse of ${\widehat\rho}_k^{-1}$ in \eqref{22modelrhoinverse}.
This is our starting point of the section.

\begin{figure}
\begin{centering}
\includegraphics[scale=0.35]{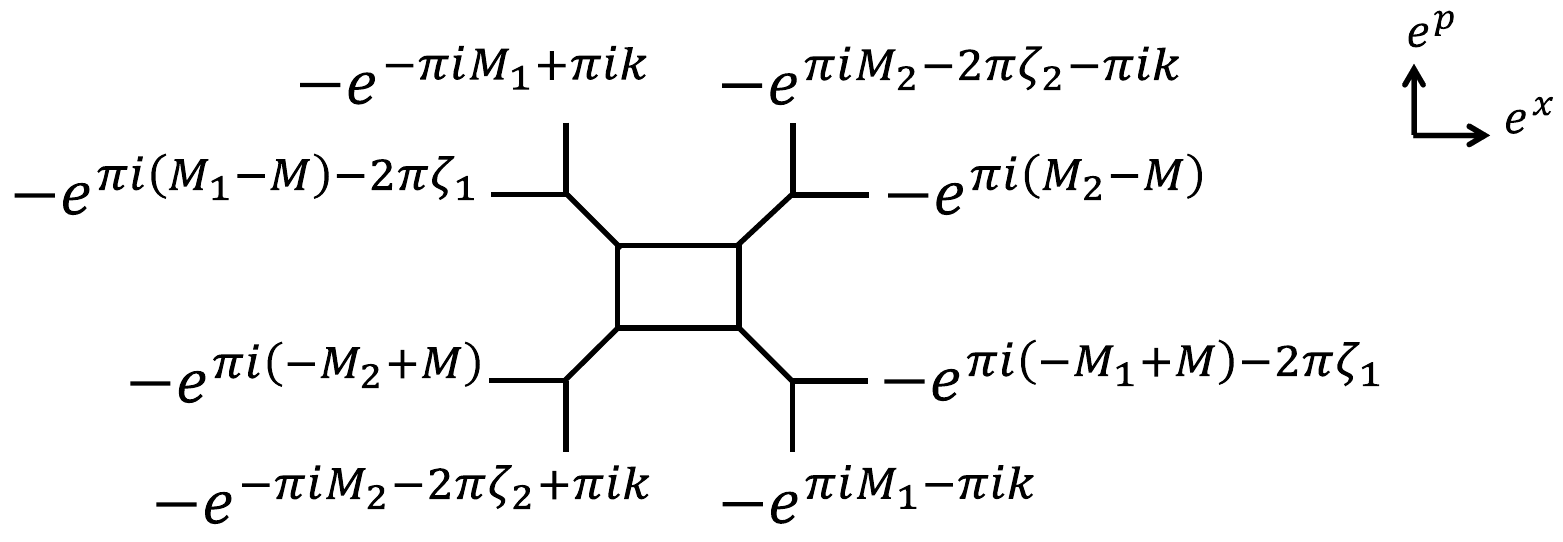}
\par\end{centering}
\caption{The asymptotic behavior of $\left(\widehat{\rho}_{k}^{\text{VI}}\right)^{-1}$, or equivalently \eqref{22modelrhoinverse}. This figure can be regarded as the five-brane web diagram of the 5d $\mathcal{N}=1$ ${\rm SU}\left(2\right)$ gauge theory with $N_{f}=4$ in Fig.\ref{fig_asymptoticpoints}. \label{fig:Toric-VI}}
\end{figure}

The consistent ways to take the limit of the parameters in each step of the coalescence \eqref{eq:Coal-pattern} can be pictorially described through the toric diagram. Indeed,
by Painlev\'e/gauge theory correspondence \cite{Bonelli:2016qwg} we know that each step of the coalescence corresponds to the decoupling limit of a
fundamental hypermultiplet 
in the five dimensional ${\cal N}=1$ gauge theory.
Since the mass parameters are encoded in the Seiberg-Witten curve, which we identify with ${\widehat\rho}_k^{-1}$, as the positions of the asymptotic regions at $x,p=\pm\infty$, i.e.~the external legs in
Fig.\ref{fig_asymptoticpoints} (see also Fig.\ref{fig:Toric-VI}),
the mass decoupling limit corresponds to sending one pair of horizontal and vertical legs to infinity (or to zero) while keeping the other legs fixed.

\begin{figure}
\begin{centering}
\includegraphics[scale=0.35]{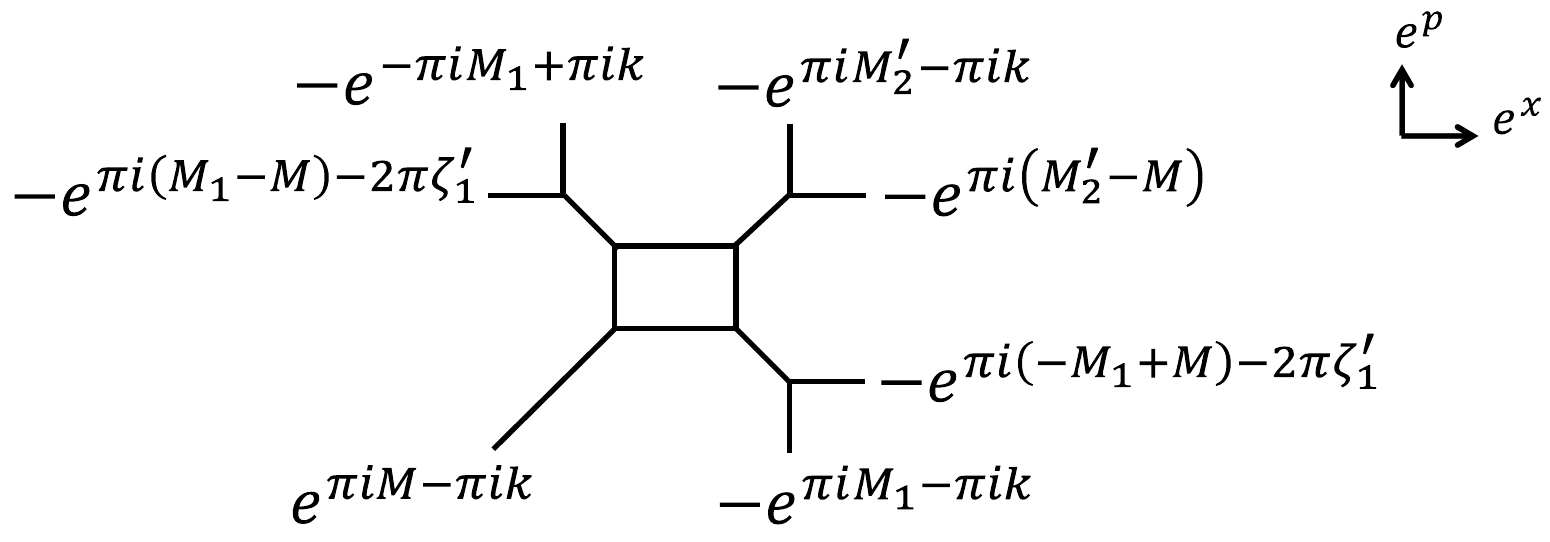}
\par\end{centering}
\caption{The asymptotic behavior of \eqref{eq:QCV-Def}. This figure can be regarded as the five-brane web diagram of the 5d $\mathcal{N}=1$ ${\rm SU}\left(2\right)$ gauge theory with $N_{f}=3$. This diagram is the degeneration of the diagram in Fig.\ref{fig:Toric-VI}. \label{fig:Toric-V}}
\end{figure}

The first coalescence from $\mathfrak{q}$-Painlev\'e $\text{VI}$ to $\mathfrak{q}$-Painlev\'e $\text{V}$ is realized by sending the bottom-left pair of external legs to zero.
This can be achieved by introducing the following new parameters
\begin{equation}
iM_{2}=iM_{2}'+2\Lambda,\quad\zeta_{1}=\zeta_{1}'-\Lambda,\quad\zeta_{2}=\Lambda,\label{eq:paraVI-Def}
\end{equation}
and taking the limit $\Lambda\rightarrow\infty$ while keeping the other parameters fixed, and at the same time shifting ${\widehat p}$ as ${\widehat p}\rightarrow {\widehat p}+2\pi\zeta_2$ so that the locations of the other external legs in \eqref{220115mtildeandttilde1}-\eqref{220115mtildeandttilde} are kept finite:
\begin{align}
&{\widetilde m}_1^{\text{VI}}=-e^{\pi i\left(M_2-M\right)}
&&\rightarrow\quad {\widetilde m}_1^{\text{V}}=e^{-2\pi\Lambda}{\widetilde m}_1^{\text{VI}}=-e^{\pi i\left(M_2'-M\right)},\\
&{\widetilde m}_2^{\text{VI}}=-e^{\pi i\left(-M_1+\right)-2\pi\zeta_1}
&&\rightarrow \quad {\widetilde m}_2^{\text{V}}=e^{-2\pi\Lambda}{\widetilde m}_2^{\text{VI}}=-e^{\pi i\left(-M_1+M\right)-2\pi\zeta_1'},\\
&{\widetilde m}_3^{\text{VI}}=-e^{\pi i\left(M_1-M\right)-2\pi\zeta_1}
&&\rightarrow \quad {\widetilde m}_3^{\text{V}}=e^{-2\pi\Lambda}{\widetilde m}_3^{\text{VI}}=-e^{\pi i\left(M_1-M\right)-2\pi\zeta_1'},\\
&{\widetilde m}_4^{\text{VI}}=-e^{\pi i\left(-M_2+M\right)}
&&\rightarrow \quad {\widetilde m}_4^{\text{V}}=e^{-2\pi\Lambda}{\widetilde m}_4^{\text{VI}}=-e^{\pi i\left(-M_2'+M\right)-4\pi\Lambda}=0,\\
&{\widetilde t}_1^{\text{VI}}=-e^{\pi iM_2-2\pi\zeta_2-\pi ik}
&&\rightarrow \quad {\widetilde t}_1^{\text{V}}={\widetilde t}_1^{\text{VI}}=-e^{\pi iM_2'-\pi ik},\\
&{\widetilde t}_2^{\text{VI}}=-e^{\pi iM_1-\pi ik}
&&\rightarrow \quad {\widetilde t}_2^{\text{V}}={\widetilde t}_2^{\text{VI}}=-e^{\pi iM_1-\pi ik},\\
&{\widetilde t}_3^{\text{VI}}=-e^{-\pi iM_1+\pi ik}
&&\rightarrow \quad {\widetilde t}_3^{\text{V}}={\widetilde t}_3^{\text{VI}}=-e^{-\pi iM_1+\pi ik},\\
&{\widetilde t}_4^{\text{VI}}=-e^{-\pi iM_2-2\pi\zeta_2+\pi ik}
&&\rightarrow \quad {\widetilde t}_4^{\text{V}}={\widetilde t}_4^{\text{VI}}=-e^{-\pi iM_2'+\pi ik-4\pi\Lambda}=0.
\end{align}
See Fig.\ref{fig:Toric-V}.

Note that the ratio $\ell_4={\widetilde m}_4^{\text{V}}\left({\widetilde t}_4^{\text{V}}\right)^{-1}=e^{\pi iM-\pi ik}$ is kept finite under the limit $\Lambda\rightarrow\infty$.
Under this procedure the quantum curve $\left({\widehat\rho}_k^{\text{VI}}\right)^{-1}$ transforms as
\begin{align}
&\left({\widehat\rho}_k^{\text{VI}}\left(M_1,M_2,M,\zeta_1,\zeta_2\right)\right)^{-1}\nonumber \\
&\rightarrow \left({\widehat\rho}_k^{\text{V}}\left(M_1,M_2',M,\zeta_1'\right)\right)^{-1}\equiv \lim_{\Lambda\rightarrow\infty}
e^{-\pi\zeta_{2}}\left.\left(\widehat{\rho}_{k}^{\text{VI}}\left(M_{1},M_{2},M,\zeta_{1},\zeta_{2}\right)\right)^{-1}\right|_{\widehat{p}\rightarrow\widehat{p}+2\pi\zeta_{2}}\nonumber \\
 & =\lim_{\Lambda\rightarrow\infty}\left[e^{\frac{\pi i(-M_{1}+M_{2}')}{2}+\pi\zeta_{1}'}e^{-{\widehat{x}}+{\widehat{p}}}+[e^{\frac{\pi i(-M_{1}-M_{2}')}{2}+\pi\zeta_{1}'+\pi ik}+e^{\frac{\pi i(M_{1}+M_{2}')}{2}+\pi\zeta_{1}'-\pi ik}]e^{{\widehat{p}}}+e^{\frac{\pi i(M_{1}-M_{2}')}{2}+\pi\zeta_{1}'}e^{{\widehat{x}}+{\widehat{p}}}\right.\nonumber \\
 & \quad+[e^{\frac{\pi i(-M_{1}-M_{2}'+2M)}{2}+\pi\zeta_{1}'-4\pi\Lambda}+e^{\frac{\pi i(M_{1}+M_{2}'-2M)}{2}-\pi\zeta_{1}'}]e^{-{\widehat{x}}}+E\nonumber \\
 & \quad+[e^{\frac{\pi i(-M_{1}-M_{2}'+2M)}{2}-\pi\zeta_{1}'}+e^{\frac{\pi i(M_{1}+M_{2}'-2M)}{2}+\pi\zeta_{1}'}]e^{{\widehat{x}}}\nonumber \\
 & \quad+e^{\frac{\pi i(M_{1}-M_{2}')}{2}-\pi\zeta_{1}'-4\pi\Lambda}e^{-{\widehat{x}}-{\widehat{p}}}+[e^{\frac{\pi i(-M_{1}-M_{2}')}{2}-\pi\zeta_{1}'-4\pi\Lambda+\pi ik}+e^{\frac{\pi i(M_{1}+M_{2}')}{2}-\pi\zeta_{1}'-\pi ik}]e^{-{\widehat{p}}}\nonumber \\
 & \left.\quad+e^{\frac{\pi i(-M_{1}+M_{2}')}{2}-\pi\zeta_{1}'}e^{{\widehat{x}}-{\widehat{p}}}\right]\nonumber \\
 & =e^{\frac{\pi i(-M_{1}+M_{2}')}{2}+\pi\zeta_{1}'}e^{-{\widehat{x}}+{\widehat{p}}}+[e^{\frac{\pi i(-M_{1}-M_{2}')}{2}+\pi\zeta_{1}'+\pi ik}+e^{\frac{\pi i(M_{1}+M_{2}')}{2}+\pi\zeta_{1}'-\pi ik}]e^{{\widehat{p}}}+e^{\frac{\pi i(M_{1}-M_{2}')}{2}+\pi\zeta_{1}'}e^{{\widehat{x}}+{\widehat{p}}}\nonumber \\
 & \quad+e^{\frac{\pi i(M_{1}+M_{2}'-2M)}{2}-\pi\zeta_{1}'}e^{-{\widehat{x}}}+E'+[e^{\frac{\pi i(-M_{1}-M_{2}'+2M)}{2}-\pi\zeta_{1}'}+e^{\frac{\pi i(M_{1}+M_{2}'-2M)}{2}+\pi\zeta_{1}'}]e^{{\widehat{x}}}\nonumber \\
 & \quad+e^{\frac{\pi i(M_{1}+M_{2}')}{2}-\pi\zeta_{1}'-\pi ik}e^{-{\widehat{p}}}+e^{\frac{\pi i(-M_{1}+M_{2}')}{2}-\pi\zeta_{1}'}e^{{\widehat{x}}-{\widehat{p}}},\label{eq:QCV-Def}
\end{align}
where $E$ is given by \eqref{eq:ConstTerm-Def} and $E'$ is given by
\begin{equation}
E'=e^{\frac{\pi i(-M_{1}+M_{2}')}{2}-\pi\zeta_{1}'+\pi ia_{1}M}+e^{\frac{\pi i(-M_{1}+M_{2}')}{2}+\pi\zeta_{1}'+\pi ia_{2}M}+e^{\frac{\pi i(M_{1}-M_{2}')}{2}-\pi\zeta_{1}'+\pi ia_{3}M}.
\end{equation}
Here we have rescaled $\left({\widehat\rho}_k^{\text{VI}}\right)^{-1}$
by a factor $e^{-\pi\zeta_2}$ so that the coefficients of limiting curve remain finite.\footnote{
In terms of the grand partition function, which appears in the right hand side of \eqref{eq:tau-GPF}, this operation corresponds to rescale $\kappa$ as $e^{\pi\zeta_{2}}\kappa$.
}
The resulting operator $\left({\widehat\rho}_k^{\text{V}}\right)^{-1}$ can be regarded as the quantum mirror curve of $\mathfrak{q}$-Painlev\'e V.

We now consider the same limit for the matrix model \eqref{eq:ZVI-Def}, or equivalently the left-hand side of the conjecture \eqref{eq:MM-QC-VI}.\footnote{
Matching the similarity transformation and the shift guarantees the equality between the density matrix and the quantum curve at operator level. We explain this point at the end of this section.
}
First, corresponding to the shift ${\widehat p}\rightarrow {\widehat p}+2\pi\zeta_2$ in the quantum curve we perform the following similarity transformation
\begin{equation}
\ket{\mu}\bra{\mu}\rightarrow e^{\frac{i\zeta_{2}}{k}\widehat{x}}\ket{\mu}\bra{\mu}e^{-\frac{i\zeta_{2}}{k}\widehat{x}},\quad\ket{\nu}\bra{\nu}\rightarrow e^{\frac{i\zeta_{2}}{k}\widehat{x}}\ket{\nu}\bra{\nu}e^{-\frac{i\zeta_{2}}{k}\widehat{x}}.\label{eq:ZVI-Sim1}
\end{equation}
Corresponding to the overall rescaling of ${\widehat\rho}_k^{-1}$ by $e^{-\pi\zeta_2}$ we also multiply both sides of \eqref{eq:MM-QC-VI} by $e^{\pi\zeta_{2}N}$.
This changes $\widehat{D}_{2}^{\text{VI}}$ to $e^{\pi\zeta_{2}}\widehat{D}_{2}^{\text{VI}}$. All in all, $\widehat{D}_{i}^{\text{V}}$ and $\widehat{d}_{i}^{\text{V}}$ are changed into
\begin{align}
e^{-\frac{i\zeta_{2}}{k}\widehat{x}}\widehat{D}_{1}^{\text{VI}}e^{\frac{i\zeta_{2}}{k}\widehat{x}} & =e^{-\frac{i\zeta_{1}'}{k}\widehat{x}}e^{\frac{k-M_{1}}{2k}\widehat{x}}\frac{\Phi_{ b}\left(\frac{\widehat{x}}{2\pi b}-\frac{iM_{1}}{2 b}+\frac{i}{2} b\right)}{\Phi_{ b}\left(\frac{\widehat{x}}{2\pi b}+\frac{iM_{1}}{2 b}-\frac{i}{2} b\right)}\frac{1}{2\cosh\frac{\widehat{p}-i\pi M}{2}}e^{\frac{i\zeta_{1}'}{k}\widehat{x}}e^{\frac{M_{1}}{2k}\widehat{x}}\frac{\Phi_{ b}\left(\frac{\widehat{x}}{2\pi b}+\frac{iM_{1}}{2 b}\right)}{\Phi_{ b}\left(\frac{\widehat{x}}{2\pi b}-\frac{iM_{1}}{2 b}\right)},\\
\widehat{d}_{1}^{\text{VI}}e^{\frac{i\zeta_{2}}{k}\widehat{x}} & =e^{\frac{i\zeta_{1}'}{k}\widehat{x}}e^{\frac{M_{1}}{2k}\widehat{x}}\frac{\Phi_{ b}\left(\frac{\widehat{x}}{2\pi b}+\frac{iM_{1}}{2 b}\right)}{\Phi_{ b}\left(\frac{\widehat{x}}{2\pi b}-\frac{iM_{1}}{2 b}\right)},\\
e^{\pi\zeta_{2}}e^{-\frac{i\zeta_{2}}{k}\widehat{x}}\widehat{D}_{2}^{\text{VI}}e^{\frac{i\zeta_{2}}{k}\widehat{x}} & =e^{2\pi\Lambda}e^{\frac{M_{2}'-4i\Lambda}{2k}\widehat{x}}\frac{\Phi_{ b}\left(\frac{\widehat{x}}{2\pi b}+\frac{iM_{2}'+4\Lambda}{2 b}\right)}{\Phi_{ b}\left(\frac{\widehat{x}}{2\pi b}-\frac{iM_{2}'}{2 b}\right)}\frac{1}{2\cosh\frac{\widehat{p}+i\pi M}{2}}e^{\frac{k-M_{2}'+4i\Lambda}{2k}\widehat{x}}\frac{\Phi_{ b}\left(\frac{\widehat{x}}{2\pi b}-\frac{iM_{2}'}{2 b}+\frac{i}{2} b\right)}{\Phi_{ b}\left(\frac{\widehat{x}}{2\pi b}+\frac{iM_{2}'+4\Lambda}{2 b}-\frac{i}{2} b\right)},\\
e^{-\frac{i\zeta_{2}}{k}\widehat{x}}\widehat{d}_{2}^{\text{VI}} & =e^{\frac{\pi}{k}\Lambda\left(M_{2}'-2i\Lambda\right)}e^{\frac{M_{2}'-4i\Lambda}{2k}\widehat{x}}\frac{\Phi_{ b}\left(\frac{\widehat{x}}{2\pi b}+\frac{iM_{2}'+4\Lambda}{2 b}\right)}{\Phi_{ b}\left(\frac{\widehat{x}}{2\pi b}-\frac{iM_{2}'}{2 b}\right)},
\end{align}
where we used the new parameterization \eqref{eq:paraVI-Def} for the right hand side. In this expression, we can take the $\Lambda\rightarrow\infty$ limit. 
In this limit, the quantum dilogarithm function behaves as \eqref{eq:DilogAsym}.
In the third line, the factors depending on $\Lambda$ and the divergent part of the asymptotic value of the quantum dilogarithm cancel out.
On the other hand, in the fourth line, an overall factor
\begin{equation}
e^{\frac{i\pi M}{4k}\left(\left(iM_{2}'+2\Lambda\right)^{2}+4\Lambda^{2}\right)+\frac{i\pi M}{12}\left(k+k^{-1}\right)},
\end{equation}
appears.
However, the same factor also appears in the right hand side of \eqref{eq:MM-QC-VI} since this factor is independent of $N$. 
Therefore, we can get rid of it when we take the limit:
\begin{align}
 & \lim_{\Lambda\rightarrow\infty}e^{-\frac{i\pi M}{4k}\left(\left(iM_{2}'+2\Lambda\right)^{2}+4\Lambda^{2}\right)-\frac{i\pi M}{12}\left(k+k^{-1}\right)}e^{\pi\zeta_{2}N}Z_{k}^{\text{VI}}\left(N;M_{1},M_{2},M,\zeta_{1},\zeta_{2}\right) =Z_{k}^{\text{V}}\left(N;M_{1},M_{2}',M,\zeta_{1}'\right),\label{eq:ZVI-to-ZV}
\end{align}
where we defined
\begin{align}
  Z_{k}^{\text{V}}\left(N;M_{1},M_{2}',M,\zeta_{1}'\right)
 & =\frac{1}{N!\left(N+M\right)!}\int_{-\infty}^\infty \prod_{n=1}^{N}\frac{d\mu_{n}}{2\pi}\prod_{n=1}^{N+M}\frac{d\nu_{n}}{2\pi}\det\left(\begin{array}{c}
\left[\braket{\mu_{m}|\widehat{D}_{1}^{\text{V}}|\nu_{n}}\right]_{m,n}^{N\times\left(N+M\right)}\\
\left[\bbraket{t_{0,M,r}|\widehat{d}_{1}^{\text{V}}|\nu_{n}}\right]_{r,n}^{M\times\left(N+M\right)}
\end{array}\right)\nonumber \\
 & \quad\times\det\left(\begin{array}{cc}
\left[\braket{\nu_{m}|\widehat{D}_{2}^{\text{V}}|\mu_{n}}\right]_{m,n}^{\left(N+M\right)\times N} & \left[\brakket{\nu_{m}|\widehat{d}_{2}^{\text{V}}|-t_{0,M,r}}\right]_{m,r}^{\left(N+M\right)\times M}\end{array}\right),\label{eq:ZV-Def}
\end{align}
with $\widehat{D}_{i}^{\text{V}}$ and $\widehat{d}_{i}^{\text{V}}$ defined as
\begin{align}
\widehat{D}_{1}^{\text{V}} & =e^{-\frac{i\zeta_{1}'}{k}\widehat{x}}e^{\frac{k-M_{1}}{2k}\widehat{x}}\frac{\Phi_{ b}\left(\frac{\widehat{x}}{2\pi b}-\frac{iM_{1}}{2 b}+\frac{i}{2} b\right)}{\Phi_{ b}\left(\frac{\widehat{x}}{2\pi b}+\frac{iM_{1}}{2 b}-\frac{i}{2} b\right)}\frac{1}{2\cosh\frac{\widehat{p}-i\pi M}{2}}e^{\frac{i\zeta_{1}'}{k}\widehat{x}}e^{\frac{M_{1}}{2k}\widehat{x}}\frac{\Phi_{ b}\left(\frac{\widehat{x}}{2\pi b}+\frac{iM_{1}}{2 b}\right)}{\Phi_{ b}\left(\frac{\widehat{x}}{2\pi b}-\frac{iM_{1}}{2 b}\right)},\\
\widehat{d}_{1}^{\text{V}}  & =e^{\frac{i\zeta_{1}'}{k}\widehat{x}}e^{\frac{M_{1}}{2k}\widehat{x}}\frac{\Phi_{ b}\left(\frac{\widehat{x}}{2\pi b}+\frac{iM_{1}}{2 b}\right)}{\Phi_{ b}\left(\frac{\widehat{x}}{2\pi b}-\frac{iM_{1}}{2 b}\right)},\\
\widehat{D}_{2}^{\text{V}} & =e^{\frac{i\pi}{4}k-\frac{i\pi}{2} M_{2}'}\frac{e^{\frac{i}{2\hbar}\widehat{x}^{2}}}{\Phi_{ b}\left(\frac{\widehat{x}}{2\pi b}-\frac{iM_{2}'}{2 b}\right)}\frac{1}{2\cosh\frac{\widehat{p}+i\pi M}{2}}\frac{\Phi_{ b}\left(\frac{\widehat{x}}{2\pi b}-\frac{iM_{2}'}{2 b}+\frac{i}{2} b\right)}{e^{\frac{i}{2\hbar}\widehat{x}^{2}}},\quad
\widehat{d}_{2}^{\text{V}}  =\frac{e^{\frac{i}{2\hbar}\widehat{x}^{2}}}{\Phi_{ b}\left(\frac{\widehat{x}}{2\pi b}-\frac{iM_{2}'}{2 b}\right)}.
\end{align}
The grand partition function associated to $Z_{k}^{\text{V}}$ is expected to be the matrix model representation of the $\mathfrak{q}$-Painlev\'e V tau function.
We remark that, as was the case for $Z_k^{\text{VI}}$ \eqref{220116_ZkVIproductform}, $Z_{k}^{\text{V}}$ can also be expressed without using the operator formalism by using \eqref{eq:op-Form1} and \eqref{eq:CauchyDet1},\eqref{eq:CauchyDet} as
\begin{align}
Z_{k}^{\text{V}}\left(N;M_{1},M_{2}',M,\zeta_{1}'\right) & =\frac{e^{\frac{i\pi}{4}kN-\frac{i\pi}{2}M_{2}'N}}{N!\left(N+M\right)!}
\int_{-\infty}^\infty \prod_{n=1}^{N}\frac{d\mu_{n}}{2\pi k}\prod_{n=1}^{N+M}\frac{d\nu_{n}}{2\pi k}\nonumber \\
 & \quad\times\prod_{n=1}^{N}e^{\left(-\frac{i\zeta_{1}'}{k}+\frac{k-M_{1}}{2k}\right)\mu_{n}-\frac{i}{4\pi k}\mu_n^2}
 \frac{\Phi_{b}\left(\frac{\mu_{n}}{2\pi b}-\frac{iM_{1}}{2b}+\frac{i}{2}b\right)
 \Phi_{b}\left(\frac{\mu_{n}}{2\pi b}-\frac{iM_{2}'}{2b}+\frac{i}{2}b\right)}
 {\Phi_{b}\left(\frac{\mu_{n}}{2\pi b}+\frac{iM_{1}}{2b}-\frac{i}{2}b\right)}\nonumber \\
 & \quad\times\prod_{n=1}^{N+M}e^{\left(\frac{i\zeta_{1}'}{k}+\frac{M_1}{2k}\right)\nu_n+\frac{i}{4\pi k}\nu_n^2}
 \frac{\Phi_{b}\left(\frac{\nu_{n}}{2\pi b}+\frac{iM_{1}}{2b}\right)
}{\Phi_{b}\left(\frac{\nu_{n}}{2\pi b}-\frac{iM_{1}}{2b}\right)
\Phi_{b}\left(\frac{\nu_{n}}{2\pi b}-\frac{iM_{2}'}{2b}\right)}\nonumber \\
 & \quad\times\left(\frac{\prod_{m<m'}^{N}2\sinh\frac{\mu_{m}-\mu_{m'}}{2k}\prod_{n<n'}^{N+M}2\sinh\frac{\nu_{n}-\nu_{n'}}{2k}}{\prod_{m=1}^{N}\prod_{n=1}^{N+M}2\cosh\frac{\mu_{m}-\nu_{n}}{2k}}\right)^{2}.
\end{align}
Combining \eqref{eq:MM-QC-VI}, \eqref{eq:QCV-Def} and \eqref{eq:ZVI-to-ZV}, we find that the coalescence limit $\text{VI}\rightarrow \text{V}$ \eqref{eq:paraVI-Def} reduces the conjecture \eqref{eq:MM-QC-VI} into the following
\begin{align}
 & Z_{k}^{\text{V}}\left(N;M_{1},M_{2}',M,\zeta_{1}'\right)\nonumber \\
 & =Z_{k}^{\text{V}}\left(0;M_{1},M_{2}',M,\zeta_{1}'\right)\frac{1}{N!}\int_{-\infty}^\infty \prod_{n=1}^{N}d\mu_{n}\det\left(\left[\braket{\mu_{m}|\widehat{\rho}_{k}^{\text{V}}\left(M_{1},M_{2}',M,\zeta_{1}'\right)|\mu_{n}}\right]_{m,n}^{N\times N}\right).\label{eq:MM-QC-V}
\end{align}

\begin{figure}
\begin{centering}
\includegraphics[scale=0.35]{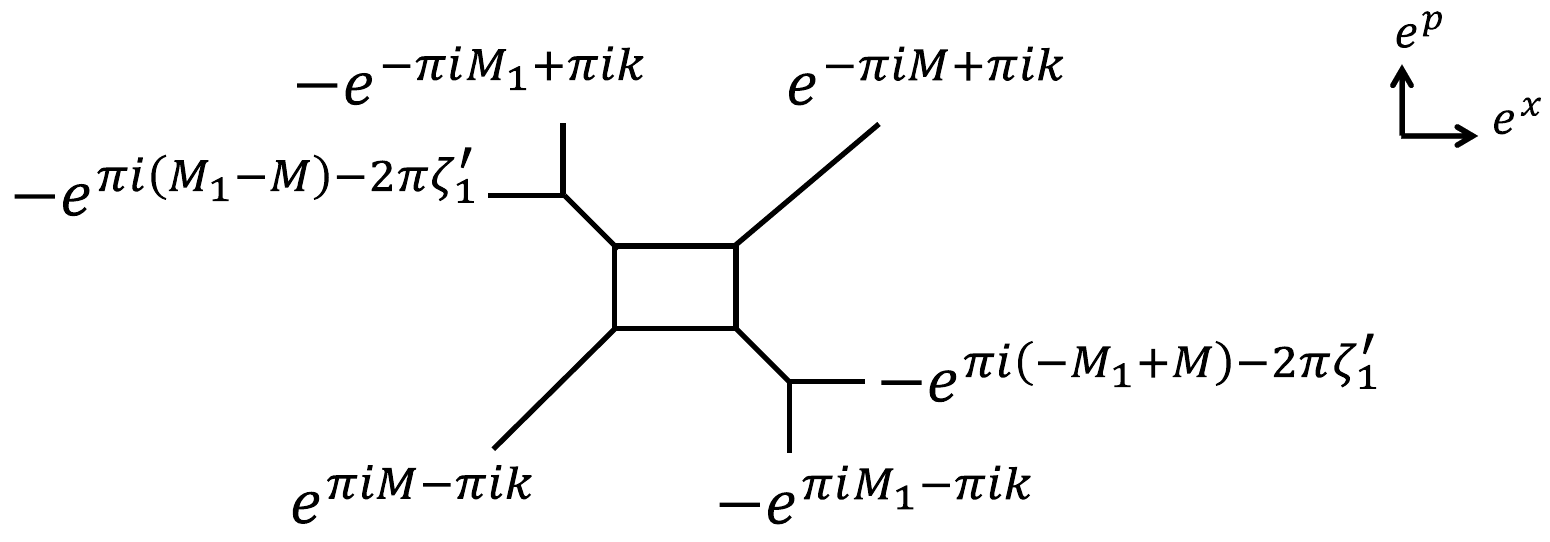}
\par\end{centering}
\caption{The asymptotic behavior of \eqref{eq:QCIII1-Def}. This figure can be regarded as the five-brane web diagram of the 5d $\mathcal{N}=1$ ${\rm SU}\left(2\right)$ gauge theory with $N_{f}=2$. This diagram is the degeneration of the diagram in Fig.\ref{fig:Toric-V}. \label{fig:Toric-III1}}
\end{figure}

To implement the remaining steps of the coalescence, we can simply repeat the same procedure.
First, we consider the coalescence from $\mathfrak{q}$-Painlev\'e $\text{V}$ to $\mathfrak{q}$-Painlev\'e $\text{III}_1$, where we send the top-right legs to infinity.
This is achieved by taking
\begin{align}
iM_2'\rightarrow \infty,
\end{align}
under which the positions of the external legs become
\begin{align}
&{\widetilde m}_1=-e^{-\pi iM+\pi iM_2'}=\infty,
\quad{\widetilde m}_2=-e^{\pi i\left(-M_1+M\right)-2\pi\zeta_1'},
\quad{\widetilde m}_3=-e^{\pi i\left(M_1-M\right)-2\pi\zeta_1'},\\
&{\widetilde t}_1=-e^{-\pi ik+\pi iM_2'}=\infty,
\quad{\widetilde t}_2=-e^{\pi iM_1-\pi ik},
\quad{\widetilde t}_3=-e^{-\pi iM_1+\pi ik},\\
&\ell_4=\frac{{\widetilde m}_4}{t_4}=e^{\pi iM-\pi ik},
\end{align}
with $\ell_1={\widetilde m}_1{\widetilde t}_1^{-1}=e^{-\pi iM+\pi ik}$ finite; see Fig.\ref{fig:Toric-III1}.

The quantum curve, with an appropriate overall rescaling, becomes
\begin{align}
 & \lim_{iM_{2}'\rightarrow\infty}e^{-\frac{1}{2}i\pi M_{2}'}\left(\widehat{\rho}_{k}^{\text{V}}\left(M_{1},M_{2}',M,\zeta_{1}'\right)\right)^{-1}=\left(\widehat{\rho}_{k}^{\text{\ensuremath{\text{III}_{1}}}}\left(M_{1},M,\zeta_{1}'\right)\right)^{-1},
\end{align}
where we defined
\begin{align}
 & \left(\widehat{\rho}_{k}^{\text{\ensuremath{\text{III}_{1}}}}\left(M_{1},M,\zeta_{1}'\right)\right)^{-1}\nonumber \\
 & =e^{-\frac{\pi iM_{1}}{2}+\pi\zeta_{1}'}e^{-{\widehat{x}}+{\widehat{p}}}+e^{\frac{\pi iM_{1}}{2}+\pi\zeta_{1}'-\pi ik}e^{{\widehat{p}}}\nonumber \\
 & \quad+e^{\frac{\pi i(M_{1}-2M)}{2}-\pi\zeta_{1}'}e^{-{\widehat{x}}}+[e^{-\frac{\pi iM_{1}}{2}-\pi\zeta_{1}'+\pi ia_{1}M}+e^{-\frac{\pi iM_{1}}{2}+\pi\zeta_{1}'+\pi ia_{2}M}]+e^{\frac{\pi i(M_{1}-2M)}{2}+\pi\zeta_{1}'}e^{{\widehat{x}}}\nonumber \\
 & \quad+e^{\frac{\pi iM_{1}}{2}-\pi\zeta_{1}'-\pi ik}e^{-{\widehat{p}}}+e^{-\frac{\pi iM_{1}}{2}-\pi\zeta_{1}'}e^{{\widehat{x}}-{\widehat{p}}}.\label{eq:QCIII1-Def}
\end{align}
This operator can be regarded as the mirror curve corresponding to $\mathfrak{q}$-Painlev\'e $\text{III}_{1}$.
Correspondingly, on the matrix model side we first multiply by an overall factor $e^{\frac{1}{2}i\pi M_{2}'N}$ and then take the $iM_{2}'\rightarrow\infty$ limit, to obtain
\begin{align}
 & \lim_{iM_{2}'\rightarrow\infty}e^{\frac{1}{2}i\pi M_{2}'N}Z_{k}^{\text{V}}\left(N;M_{1},M_{2}',M,\zeta_{1}'\right)=Z_{k}^{\text{\ensuremath{\text{III}_{1}}}}\left(N;M_{1},M,\zeta_{1}'\right),
\end{align}
where we defined
\begin{align}
 & Z_{k}^{\text{\ensuremath{\text{III}_{1}}}}\left(N;M_{1},M,\zeta_{1}'\right)\nonumber \\
 & =\frac{1}{N!\left(N+M\right)!}\int_{-\infty}^\infty \prod_{n=1}^{N}\frac{d\mu_{n}}{2\pi}\prod_{n=1}^{N+M}\frac{d\nu_{n}}{2\pi}\det\left(\begin{array}{c}
\left[\braket{\mu_{m}|\widehat{D}_{1}^{\text{III}_{1}}|\nu_{n}}\right]_{m,n}^{N\times\left(N+M\right)}\\
\left[\bbraket{t_{0,M,r}|\widehat{d}_{1}^{\text{III}_{1}}|\nu_{n}}\right]_{r,n}^{M\times\left(N+M\right)}
\end{array}\right)\nonumber \\
 & \quad\times\det\left(\begin{array}{cc}
\left[\braket{\nu_{m}|e^{\frac{i\pi}{4}k}e^{\frac{i}{2\hbar}\widehat{x}^{2}}\frac{1}{2\cosh\frac{\widehat{p}+i\pi M}{2}}e^{-\frac{i}{2\hbar}\widehat{x}^{2}}|\mu_{n}}\right]_{m,n}^{\left(N+M\right)\times N} & \left[\brakket{\nu_{m}|e^{\frac{i}{2\hbar}\widehat{x}^{2}}|-t_{0,M,r}}\right]_{m,r}^{\left(N+M\right)\times M}\end{array}\right).\label{eq:ZIII1-Def}
\end{align}
In this limit, $\widehat{D}_{1}^{\text{V}}$ and $\widehat{d}_{1}^{\text{V}}$ are not affected:
\begin{align}
 & \widehat{D}_{1}^{\text{III}_{1}}=\widehat{D}_{1}^{\text{V}},\quad\widehat{d}_{1}^{\text{III}_{1}}=\widehat{d}_{1}^{\text{V}}.
\end{align}
$Z_{k}^{\text{\ensuremath{\text{III}_{1}}}}$ can be regarded as the matrix model corresponding to $\mathfrak{q}$-Painlev\'e $\text{III}_{1}$.
Again we can also write the integrand of $Z_k^{\text{III}_1}$ in the product form:
\begin{align}
&Z_{k}^{\text{III}_1}\left(N;M_{1},M,\zeta_{1}'\right)\nonumber \\
&=\frac{e^{\frac{i\pi}{4}kN}}{N!\left(N+M\right)!}
\int_{-\infty}^\infty \prod_{n=1}^{N}\frac{d\mu_{n}}{2\pi k}\prod_{n=1}^{N+M}\frac{d\nu_{n}}{2\pi k}
\prod_{n=1}^{N}e^{\left(-\frac{i\zeta_{1}'}{k}+\frac{k-M_{1}}{2k}\right)\mu_{n}-\frac{i}{4\pi k}\mu_n^2}
 \frac{\Phi_{b}\left(\frac{\mu_{n}}{2\pi b}-\frac{iM_{1}}{2b}+\frac{i}{2}b\right)}
 {\Phi_{b}\left(\frac{\mu_{n}}{2\pi b}+\frac{iM_{1}}{2b}-\frac{i}{2}b\right)}\nonumber \\
&\quad \times \prod_{n=1}^{N+M}e^{\left(\frac{i\zeta_{1}'}{k}+\frac{M_1}{2k}\right)\nu_n+\frac{i}{4\pi k}\nu_n^2}
 \frac{\Phi_{b}\left(\frac{\nu_{n}}{2\pi b}+\frac{iM_{1}}{2b}\right)
}{\Phi_{b}\left(\frac{\nu_{n}}{2\pi b}-\frac{iM_{1}}{2b}\right)}
\left(\frac{\prod_{m<m'}^{N}2\sinh\frac{\mu_{m}-\mu_{m'}}{2k}\prod_{n<n'}^{N+M}2\sinh\frac{\nu_{n}-\nu_{n'}}{2k}}{\prod_{m=1}^{N}\prod_{n=1}^{N+M}2\cosh\frac{\mu_{m}-\nu_{n}}{2k}}\right)^{2}.
\end{align}
Combining the above results, the conjecture \eqref{eq:MM-QC-VI} is now reduced to the following
\begin{align}
 & Z_{k}^{\text{\ensuremath{\text{III}_{1}}}}\left(N;M_{1},M,\zeta_{1}'\right)\nonumber \\
 & =Z_{k}^{\text{\ensuremath{\text{III}_{1}}}}\left(0;M_{1},M,\zeta_{1}'\right)\frac{1}{N!}\int_{-\infty}^\infty \prod_{n=1}^{N}d\mu_{n}\det\left(\left[\braket{\mu_{m}|\widehat{\rho}_{k}^{\text{\ensuremath{\text{III}_{1}}}}\left(M_{1},M,\zeta_{1}'\right)|\mu_{n}}\right]_{m,n}^{N\times N}\right).\label{eq:MM-QC-III1}
\end{align}

\begin{figure}
\begin{centering}
\includegraphics[scale=0.35]{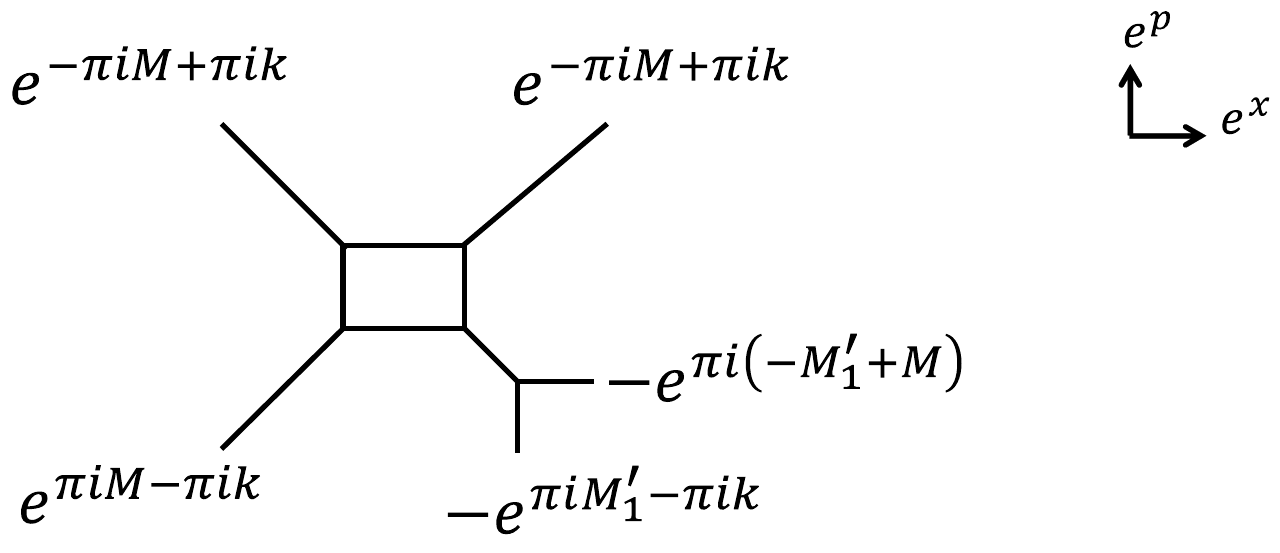}
\par\end{centering}
\caption{The asymptotic behavior of \eqref{eq:QCIII2-Def}. This figure can be regarded as the five-brane web diagram of the 5d $\mathcal{N}=1$ ${\rm SU}\left(2\right)$ Yang-Mills theory with $N_{f}=1$. This diagram is degenerated one from the diagram in figure \ref{fig:Toric-III1}. \label{fig:Toric-III2}}
\end{figure}

We now consider the coalescence from $\mathfrak{q}$-Painlev\'e $\text{III}_1$ to $\mathfrak{q}$-Painlev\'e $\text{III}_2$.
We send the top-left pair of external legs to ${\widetilde m}_3\rightarrow \infty$ and ${\widetilde t}_3\rightarrow 0$, by first shifting $x,p$ as ${\widehat x}\rightarrow {\widehat x}-\pi\zeta_1'$, ${\widehat p}\rightarrow {\widehat p}-\pi\zeta_1'$ and then taking $\Lambda\rightarrow \infty$ with the following reparameterization
\begin{equation}
iM_{1}=iM_{1}'+\Lambda,\quad\zeta_{1}'=-\Lambda.
\label{220116PIII1toIII2reparametrization}
\end{equation}
Under this procedure we are left with
\begin{align}
&{\widetilde m}_2=-e^{\pi i\left(-M_1'+M\right)},\\
&{\widetilde t}_2=-e^{\pi iM_1'-\pi ik},\\
&\ell_1=\frac{{\widetilde m}_1}{{\widetilde t}_1}=e^{-\pi iM+\pi ik},
\quad\ell_4=\frac{{\widetilde m}_4}{{\widetilde t}_4}=e^{\pi iM-\pi ik},
\quad\ell_3={\widetilde m}_3{\widetilde t}_3=e^{-\pi iM+\pi ik},
\end{align}
see Fig.\ref{fig:Toric-III2}.
The quantum curve becomes
\begin{align}
 & \lim_{\Lambda\rightarrow\infty}e^{\frac{1}{2}\pi\zeta_{1}'}\left.\left(\widehat{\rho}_{k}^{\text{\ensuremath{\text{III}_{1}}}}\left(M_{1},M,\zeta_{1}'\right)\right)^{-1}\right|_{\widehat{x}\rightarrow\widehat{x}-\pi\zeta_{1}',\quad\widehat{p}\rightarrow\widehat{p}-\pi\zeta_{1}'}=\left(\widehat{\rho}_{k}^{\text{\ensuremath{\text{III}_{2}}}}\left(M_{1}',M\right)\right)^{-1},
\end{align}
where
\begin{align}
\left(\widehat{\rho}_{k}^{\text{\ensuremath{\text{III}_{2}}}}\left(M_{1}',M\right)\right)^{-1} & =e^{\frac{\pi iM_{1}'}{2}-\pi ik}e^{{\widehat{p}}}\nonumber \\
 & \quad+e^{\frac{\pi i(M_{1}'-2M)}{2}}e^{-{\widehat{x}}}+e^{-\frac{\pi iM_{1}'}{2}+\pi ia_{1}M}+e^{\frac{\pi i(M_{1}'-2M)}{2}}e^{{\widehat{x}}}\nonumber \\
 & \quad+e^{\frac{\pi iM_{1}'}{2}-\pi ik}e^{-{\widehat{p}}}+e^{-\frac{\pi iM_{1}'}{2}}e^{{\widehat{x}}-{\widehat{p}}}.\label{eq:QCIII2-Def}
\end{align}
This operator can be regarded as the mirror curve corresponding to $\mathfrak{q}$-Painlev\'e $\text{III}_{2}$.
On the matrix model side, we first multiply $e^{-\frac{1}{2}\pi\zeta_{1}'N}$, perform the similarity transformation
\begin{equation}
\ket{\lambda}\bra{\lambda}\rightarrow e^{-\frac{i\zeta_{1}'}{2k}\widehat{x}}e^{\frac{i\zeta_{1}'}{2k}\widehat{p}}\ket{\lambda}\bra{\lambda}e^{-\frac{i\zeta_{1}'}{2k}\widehat{p}}e^{\frac{i\zeta_{1}'}{2k}\widehat{x}},
\end{equation}
both for $\ket{\mu}\bra{\mu}$ and for $\ket{\nu}\bra{\nu}$,
and then take the limit $\Lambda\rightarrow\infty$ with the reparametrization \eqref{220116PIII1toIII2reparametrization}.
We end up with
\begin{align}
 & \lim_{\Lambda\rightarrow\infty}e^{-\frac{i\pi M}{4k}\left(iM_{1}'+\Lambda\right)^{2}-\frac{i\pi M}{12}\left(k+k^{-1}\right)}e^{-\frac{1}{2}\pi\zeta_{1}'N}Z_{k}^{\text{\ensuremath{\text{III}_{1}}}}\left(N;M_{1},M,\zeta_{1}'\right)=Z_{k}^{\text{\ensuremath{\text{III}_{2}}}}\left(N;M_{1}',M\right),
\end{align}
where
\begin{align}
 & Z_{k}^{\text{\ensuremath{\text{III}_{2}}}}\left(N;M_{1}',M\right)\nonumber \\
 & =\frac{1}{N!\left(N+M\right)!}\int_{-\infty}^\infty \prod_{n=1}^{N}\frac{d\mu_{n}}{2\pi}\prod_{n=1}^{N+M}\frac{d\nu_{n}}{2\pi}\det\left(\begin{array}{c}
\left[\braket{\mu_{m}|\widehat{D}_{1}^{\text{III}_{2}}|\nu_{n}}\right]_{m,n}^{N\times\left(N+M\right)}\\
\left[\bbraket{t_{0,M,r}|\widehat{d}_{1}^{\text{III}_{2}}|\nu_{n}}\right]_{r,n}^{M\times\left(N+M\right)}
\end{array}\right)\nonumber \\
 & \quad\times\det\left(\begin{array}{cc}
\left[\braket{\nu_{m}|e^{\frac{i\pi}{4}k}e^{\frac{i}{2\hbar}\widehat{x}^{2}}\frac{1}{2\cosh\frac{\widehat{p}+i\pi M}{2}}e^{-\frac{i}{2\hbar}\widehat{x}^{2}}|\mu_{n}}\right]_{m,n}^{\left(N+M\right)\times N} & \left[\brakket{\nu_{m}|e^{\frac{i}{2\hbar}\widehat{x}^{2}}|-t_{0,M,r}}\right]_{m,r}^{\left(N+M\right)\times M}\end{array}\right).\label{eq:ZIII2-Def}
\end{align}
$\widehat{D}_{1}^{\text{III}_{1}}$ and $\widehat{d}_{1}^{\text{III}_{1}}$ are changed into
\begin{align}
\widehat{D}_{1}^{\text{III}_{2}} & =e^{\frac{i\pi}{4}k}e^{-\frac{1}{2}i\pi M_{1}'}\frac{\Phi_{ b}\left(\frac{\widehat{x}}{2\pi b}-\frac{iM_{1}'}{2 b}+\frac{i}{2} b\right)}{e^{\frac{i}{2\hbar}\widehat{x}^{2}}}\frac{1}{2\cosh\frac{\widehat{p}-i\pi M}{2}}\frac{e^{\frac{i}{2\hbar}\widehat{x}^{2}}}{\Phi_{ b}\left(\frac{\widehat{x}}{2\pi b}-\frac{iM_{1}'}{2 b}\right)},\\
\widehat{d}_{1}^{\text{III}_{2}} & =\frac{e^{\frac{i}{2\hbar}\widehat{x}^{2}}}{\Phi_{ b}\left(\frac{\widehat{x}}{2\pi b}-\frac{iM_{1}'}{2 b}\right)}.
\end{align}
In the product form we have
\begin{align}
&Z_{k}^{\text{III}_2}\left(N;M_{1},M\right)\nonumber \\
&=\frac{e^{\frac{i\pi}{2}kN-\frac{\pi iM_1'N}{2}}}{N!\left(N+M\right)!}
\int_{-\infty}^\infty \prod_{n=1}^{N}\frac{d\mu_{n}}{2\pi k}\prod_{n=1}^{N+M}\frac{d\nu_{n}}{2\pi k}
\prod_{n=1}^{N}e^{-\frac{i}{2\pi k}\mu_n^2}
\Phi_{b}\left(\frac{\mu_{n}}{2\pi b}-\frac{iM_{1}'}{2b}+\frac{i}{2}b\right)
\prod_{n=1}^{N+M}e^{\frac{i}{2\pi k}\nu_n^2}
\frac{1}{\Phi_{b}\left(\frac{\nu_{n}}{2\pi b}-\frac{iM_{1}'}{2b}\right)}\nonumber \\
&\quad \times \left(\frac{\prod_{m<m'}^{N}2\sinh\frac{\mu_{m}-\mu_{m'}}{2k}\prod_{n<n'}^{N+M}2\sinh\frac{\nu_{n}-\nu_{n'}}{2k}}{\prod_{m=1}^{N}\prod_{n=1}^{N+M}2\cosh\frac{\mu_{m}-\nu_{n}}{2k}}\right)^{2}.
\end{align}
$Z_{k}^{\text{\ensuremath{\text{III}_{2}}}}$ can be regarded as the matrix model corresponding to $\mathfrak{q}$-Painlev\'e $\text{III}_{2}$.
Combining the above results, we obtain
\begin{align}
 & Z_{k}^{\text{\ensuremath{\text{III}_{2}}}}\left(N;M_{1}',M\right)=Z_{k}^{\text{\ensuremath{\text{III}_{2}}}}\left(0;M_{1}',M\right)\frac{1}{N!}\int_{-\infty}^\infty \prod_{n=1}^{N}d\mu_{n}\det\left(\left[\braket{\mu_{m}|\widehat{\rho}_{k}^{\text{\ensuremath{\text{III}_{2}}}}\left(M_{1}',M\right)|\mu_{n}}\right]_{m,n}^{N\times N}\right).\label{eq:MM-QC-III2}
\end{align}

\begin{figure}
\begin{centering}
\includegraphics[scale=0.35]{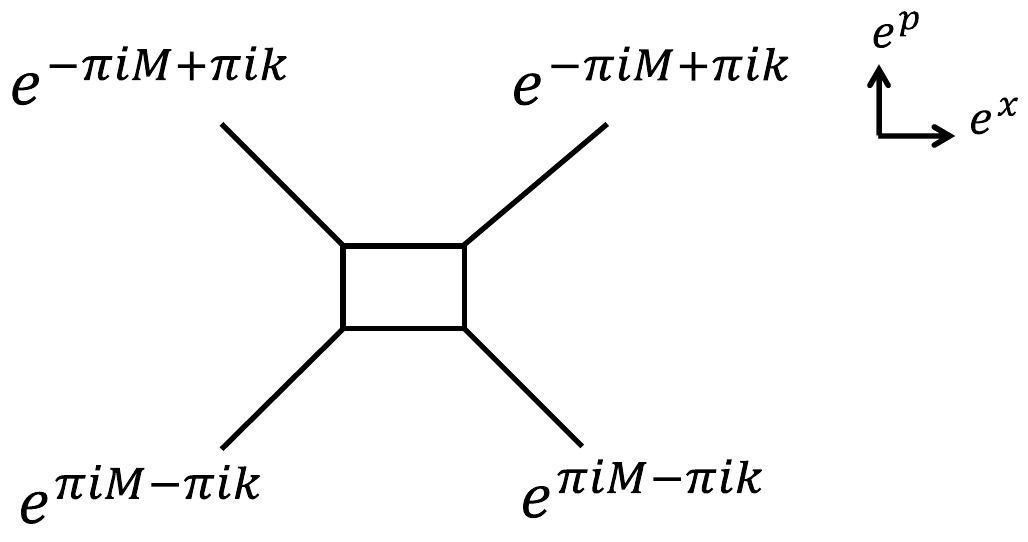}
\par\end{centering}
\caption{The asymptotic behavior of \eqref{eq:QCIII3-Def}. This figure can be regarded as the five-brane web diagram of the 5d $\mathcal{N}=1$ ${\rm SU}\left(2\right)$ Yang-Mills theory. This diagram is 
the degeneration of the diagram in Fig.\ref{fig:Toric-III2}. \label{fig:Toric-III3}}
\end{figure}

Finally, we consider the coalescence from $\mathfrak{q}$-Painlev\'e $\text{III}_2$ to $\mathfrak{q}$-Painlev\'e $\text{III}_3$.
We can send the bottom-right legs to ${\widetilde m}_2\rightarrow 0$ and ${\widetilde t}_2\rightarrow \infty$ by taking the $iM_{1}'\rightarrow\infty$ limit, which leaves $\ell_2={\widetilde m}_2{\widetilde t}_2=e^{\pi iM-\pi ik}$ finite; see Fig.\ref{fig:Toric-III3}.

The quantum curve becomes
\begin{align}
 & \lim_{iM_{1}'\rightarrow\infty}e^{-\frac{1}{2}i\pi M_{1}'}\left(\widehat{\rho}_{k}^{\text{\ensuremath{\text{III}_{2}}}}\left(M_{1}',M\right)\right)^{-1}=\left(\widehat{\rho}_{k}^{\text{\ensuremath{\text{III}_{3}}}}\left(M\right)\right)^{-1},
\end{align}
where
\begin{align}
\left(\widehat{\rho}_{k}^{\text{\ensuremath{\text{III}_{3}}}}\left(M\right)\right)^{-1} & =e^{-\pi iM}e^{{\widehat{x}}}+e^{-\pi iM}e^{-{\widehat{x}}}+e^{-\pi ik}e^{{\widehat{p}}}+e^{-\pi ik}e^{-{\widehat{p}}}.\label{eq:QCIII3-Def}
\end{align}
This operator can be regarded as the mirror curve corresponding to $\mathfrak{q}$-Painlev\'e $\text{III}_{3}$.
On the matrix model side we multiply the overall factor $e^{\frac{1}{2}i\pi M_{1}'N}$ and take the limit $iM_1'\rightarrow\infty$ to obtain
\begin{align}
 & \lim_{iM_{1}'\rightarrow\infty}e^{\frac{1}{2}i\pi M_{1}'N}Z_{k}^{\text{\ensuremath{\text{III}_{2}}}}\left(N;M_{1}',M\right)=Z_{k}^{\text{\ensuremath{\text{III}_{3}}}}\left(N;M\right),
\end{align}
where
\begin{align}
 & Z_{k}^{\text{\ensuremath{\text{III}_{3}}}}\left(N;M\right)\nonumber \\
 & =\frac{1}{N!\left(N+M\right)!}\int_{-\infty}^\infty \prod_{n=1}^{N}\frac{d\mu_{n}}{2\pi}\prod_{n=1}^{N+M}\frac{d\nu_{n}}{2\pi}\det\left(\begin{array}{c}
\left[\braket{\mu_{m}|e^{\frac{i\pi}{4}k}e^{-\frac{i}{2\hbar}\widehat{x}^{2}}\frac{1}{2\cosh\frac{\widehat{p}-i\pi M}{2}}e^{\frac{i}{2\hbar}\widehat{x}^{2}}|\nu_{n}}\right]_{m,n}^{N\times\left(N+M\right)}\\
\left[\bbraket{t_{0,M,r}|e^{\frac{i}{2\hbar}\widehat{x}^{2}}|\nu_{n}}\right]_{r,n}^{M\times\left(N+M\right)}
\end{array}\right)\nonumber \\
 & \quad\times\det\left(\begin{array}{cc}
\left[\braket{\nu_{m}|e^{\frac{i\pi}{4}k}e^{\frac{i}{2\hbar}\widehat{x}^{2}}\frac{1}{2\cosh\frac{\widehat{p}+i\pi M}{2}}e^{-\frac{i}{2\hbar}\widehat{x}^{2}}|\mu_{n}}\right]_{m,n}^{\left(N+M\right)\times N} & \left[\brakket{\nu_{m}|e^{\frac{i}{2\hbar}\widehat{x}^{2}}|-t_{0,M,r}}\right]_{m,r}^{\left(N+M\right)\times M}\end{array}\right).\label{eq:ZIII3-Def}
\end{align}
As we mentioned in the beginning of this section, this matrix model coincides with the partition function of ABJM theory with the gauge group $\text{U}(N)_{2k}\times \text{U}(N+M)_{-2k}$.
This integral can be regarded as the matrix model corresponding to $\mathfrak{q}$-Painlev\'e $\text{III}_{3}$.
Combining the above results, we obtain
\begin{align}
 & Z_{k}^{\text{\ensuremath{\text{III}_{3}}}}\left(N;M\right)=Z_{k}^{\text{\ensuremath{\text{III}_{3}}}}\left(0;M\right)\frac{1}{N!}\int_{-\infty}^\infty \prod_{n=1}^{N}\frac{d\mu_{n}}{2\pi}\det\left(\left[\braket{\mu_{m}|\widehat{\rho}_{k}^{\text{\ensuremath{\text{III}_{3}}}}\left(M\right)|\mu_{n}}\right]_{m,n}^{N\times N}\right).\label{eq:MM-QC-III3}
\end{align}

In this section we derived the relations between the matrix models and the quantum curves in \eqref{eq:MM-QC-VI}, \eqref{eq:MM-QC-V}, \eqref{eq:MM-QC-III1}, \eqref{eq:MM-QC-III2} and \eqref{eq:MM-QC-III3}. We expect that the grand canonical partition function of these matrix models
provide a conjectural Fredholm determinant expression of the $\tau$-functions for the relevant 
$\mathfrak{q}$-Painlev\'e equations, at least at the specific values of the moduli that we are analysing.

The result \eqref{eq:MM-QC-III3} we get is consistent with previous findings on
$\mathfrak{q}$-P$\text{III}_3$ equation. Indeed \eqref{eq:QCIII3-Def} is the quantum curve associated with local $\mathbb{P}^{1}\times\mathbb{P}^{1}$. The spectral determinant of this quantum curve was already known to satisfy the bilinear equation of the $\mathfrak{q}$-Painlev\'e $\text{III}_{3}$ \cite{Bonelli:2017gdk}. On the other hand, the matrix model \eqref{eq:ZIII3-Def} we find  $Z_{k}^{\text{\ensuremath{\text{III}_{3}}}}\left(N;M\right)$ is the one associated with ABJM theory, as expected from
the results in \cite{Bonelli:2017gdk}.
The identification of the matrix models goes along the same lines we followed in the beginning of the section, \eqref{eq:ZVI-Def} and \eqref{220116_ZkVIproductform}.
We get
\begin{align}
Z_{k}^{\text{\ensuremath{\text{III}_{3}}}}\left(N;M\right) & =\frac{e^{\frac{i\pi}{2}kN}}{N!\left(N+M\right)!}\int_{-\infty}^\infty \prod_{n=1}^{N}\frac{d\mu_{n}}{2\pi}\prod_{n=1}^{N+M}\frac{d\nu_{n}}{2\pi}e^{\frac{ik}{2\pi}\sum_{n=1}^{N}\mu_{n}^{2}-\frac{ik}{2\pi}\sum_{n=1}^{N+M}\nu_{n}^{2}}\nonumber \\
 & \quad\times\frac{\prod_{n<n'}^{N}\left(2\sinh\frac{\mu_{n}-\mu_{n'}}{2}\right)^{2}}{\prod_{m=1}^{N}\prod_{n=1}^{N+M}2\cosh\frac{\mu_{m}-\nu_{n}}{2}}\frac{\prod_{n<n'}^{N+M}\left(2\sinh\frac{\nu_{n}-\nu_{n'}}{2}\right)^{2}}{\prod_{m=1}^{N+M}\prod_{n=1}^{N}2\cosh\frac{\nu_{m}-\mu_{n}}{2}},
\label{220119_ABJMKWY}
\end{align}
which is the matrix model associated to the ABJM theory with ${\rm U}\left(N\right)_{2k}\times{\rm U}\left(N+M\right)_{-2k}$ gauge group. Notice that the Chern-Simons level gets renormalised along the flow to
$k^{{\rm ABJM}}= 2k$. Let us also observe that in the relation between ABJM theory and, $\mathfrak{q}$-Painlev\'e $\text{III}_{3}$, the rank parameter $M$ corresponds to the time variable, while this is not the case for the $\mathfrak{q}$-Painlev\'e $\text{VI}$ equation we start from. 

For the particular case \eqref{eq:MM-QC-III3} we can actually prove our conjecture \eqref{eq:MM-QC-VI}.
Indeed in this case an explicit calculation of the inverse of the spectral density matrix 
$\left(\widehat{\rho}_{k}^{\text{\ensuremath{\text{III}_{3}}}}\right)^{-1}$
 can be performed by showing that this is indeed the quantum Seiberg-Witten curve for the local $\mathbb{P}^1\times\mathbb{P}^1$ geometry $\widehat{\mathcal{O}}_{{\mathbb P}^1\times{\mathbb P}^1}$.
By using \eqref{eq:Hyper-QC1},\eqref{eq:Hyper-QC}, we can show that
\cite{Kashaev:2015wia}
\begin{align}
 \widehat{\rho}_{k}^{\text{\ensuremath{\text{III}_{3}}}}\left(M\right)=
 \widehat{\mathcal{O}}_{{\mathbb P}^1\times{\mathbb P}^1}^{-1}=
 e^{\frac{i\pi}{2}k}\frac{\prod_{j}^{M}2\sinh\frac{\widehat{x}+t_{0,M,j}}{4k}}{2\cosh\frac{\widehat{x}}{2}}\frac{1}{2\cosh\frac{\widehat{p}}{2}}\frac{1}{\prod_{j}^{M}2\cosh\frac{\widehat{x}+t_{0,M,j}}{4k}},
\end{align}
so that \eqref{eq:MM-QC-III3} becomes
\begin{align}
Z_{k}^{\text{\ensuremath{\text{III}_{3}}}}\left(N;M\right) & =\frac{i^{\frac{1}{2}M^{2}}e^{i\theta_{2k}\left(M,0\right)}}{N!}Z_{2k,M}^{\left(\mathrm{CS}\right)}\int_{-\infty}^\infty \prod_{n=1}^{N}\frac{d\mu_{n}}{2\pi}\nonumber \\
 & \quad\times\det\left(\left[\braket{\mu_{m}|e^{\frac{i\pi}{2}k}\frac{1}{\prod_{j}^{M}2\cosh\frac{\widehat{x}+t_{0,M,j}}{4k}}\frac{1}{2\cosh\frac{\widehat{p}}{2}}\frac{\prod_{j}^{M}2\sinh\frac{\widehat{x}+t_{0,M,j}}{4k}}{2\cosh\frac{\widehat{x}}{2}}|\mu_{n}}\right]_{m,n}^{N\times N}\right).\label{eq:ZIII3-FGF}
\end{align}
The above can be shown to coincide with \eqref{220119_ABJMKWY} \cite{Honda:2014npa}.
 This is a non-trivial check of the conjecture \eqref{22modelrhoinverse}. On the other hand, \eqref{eq:MM-QC-V}, \eqref{eq:MM-QC-III1} and \eqref{eq:MM-QC-III2} provide new conjectural relations between matrix models and quantum curves.
Let us also stress that the normalisation we choose to define 
$Z_{k}^{\text{VI}}$ in \eqref{eq:ZVI-Def} 
is consistent with the one found in \cite{Bonelli:2017gdk}.

Finally, let us notice that 
the conjectured relations \eqref{eq:MM-QC-V}, \eqref{eq:MM-QC-III1} and \eqref{eq:MM-QC-III2} can actually be proved in the $M=0$ case by making use of the results of
appendix \ref{sec:FGF}. In these cases, the matrix models can be written in the form of \eqref{eq:FGF-M0-Form} by using gluing formula \eqref{eq:Det-Glue},
and the inverse of the density matrix coincides with the quantum Seiberg-Witten curve. Explicitly, \eqref{eq:DensM-M0} and \eqref{eq:QC-M0-1} are equal as explained in appendix \ref{sec:FGF}. 
For $M=0$, we can show that the other conjectured relations can be proved at the operator level in a similar way.  
For example, to show \eqref{eq:MM-QC-V}, it is enough to prove
\begin{align}
 & \left[e^{-\frac{i\zeta_{1}'}{k}\widehat{x}}e^{\frac{k-M_{1}}{2k}\widehat{x}}\frac{\Phi_{ b}\left(\frac{\widehat{x}}{2\pi b}-\frac{iM_{1}}{2 b}+\frac{i}{2} b\right)}{\Phi_{ b}\left(\frac{\widehat{x}}{2\pi b}+\frac{iM_{1}}{2 b}-\frac{i}{2} b\right)}\frac{1}{2\cosh\frac{\widehat{p}}{2}}e^{\frac{i\zeta_{1}'}{k}\widehat{x}}e^{\frac{M_{1}}{2k}\widehat{x}}\frac{\Phi_{ b}\left(\frac{\widehat{x}}{2\pi b}+\frac{iM_{1}}{2 b}\right)}{\Phi_{ b}\left(\frac{\widehat{x}}{2\pi b}-\frac{iM_{1}}{2 b}\right)}\right.\nonumber \\
 & \quad\left.\times e^{\frac{i\pi}{4}k}e^{-\frac{1}{2}i\pi M_{2}'}\frac{e^{\frac{i}{2\hbar}\widehat{x}^{2}}}{\Phi_{ b}\left(\frac{\widehat{x}}{2\pi b}-\frac{iM_{2}'}{2 b}\right)}\frac{1}{2\cosh\frac{\widehat{p}}{2}}\frac{\Phi_{ b}\left(\frac{\widehat{x}}{2\pi b}-\frac{iM_{2}'}{2 b}+\frac{i}{2} b\right)}{e^{\frac{i}{2\hbar}\widehat{x}^{2}}}\right]^{-1}\nonumber \\
 & =\left[\left(e^{-\frac{1}{2}i\pi M_{2}'}e^{\frac{1}{2}\widehat{x}}+e^{\frac{1}{2}i\pi M_{2}'}e^{-\frac{1}{2}\widehat{x}}\right)e^{\frac{1}{2}\widehat{p}}+e^{\frac{1}{2}i\pi M_{2}'}e^{\frac{1}{2}\widehat{x}}e^{-\frac{1}{2}\widehat{p}}\right]\nonumber \\
 & \quad\times\left[e^{\pi\zeta_{1}'}e^{\frac{1}{2}\widehat{p}}\left(e^{\frac{1}{2}i\pi M_{1}}e^{\frac{1}{2}\widehat{x}}+e^{-\frac{1}{2}i\pi M_{1}}e^{-\frac{1}{2}\widehat{x}}\right)+e^{-\pi\zeta_{1}'}e^{-\frac{1}{2}\widehat{p}}\left(e^{-\frac{1}{2}i\pi M_{1}}e^{\frac{1}{2}\widehat{x}}+e^{\frac{1}{2}i\pi M_{1}}e^{-\frac{1}{2}\widehat{x}}\right)\right].
\end{align}
The right hand side is the quantum curve \eqref{eq:QCV-Def} with $M=0$. This identity can be proved by using \eqref{eq:Hyper-QC01},\eqref{eq:Hyper-QC0} and
\begin{align}
  e^{-\frac{i\pi}{4}k}e^{\frac{1}{2}i\pi M_{2}'}\frac{e^{\frac{i}{2\hbar}\widehat{x}^{2}}}{\Phi_{ b}\left(\frac{\widehat{x}}{2\pi b}-\frac{iM_{2}'}{2 b}+\frac{i}{2} b\right)}e^{\pm\frac{1}{2}\widehat{p}}\frac{\Phi_{ b}\left(\frac{\widehat{x}}{2\pi b}-\frac{iM_{2}'}{2 b}\right)}{e^{\frac{i}{2\hbar}\widehat{x}^{2}}} 
  &=e^{-\frac{i\pi}{4}k}e^{\frac{1}{2}i\pi M_{2}'}\frac{\Phi_{ b}\left(\frac{\widehat{x}}{2\pi b}-\frac{iM_{2}'}{2 b}\mp\frac{i}{2} b\right)}{\Phi_{ b}\left(\frac{\widehat{x}}{2\pi b}-\frac{iM_{2}'}{2 b}+\frac{i}{2} b\right)}e^{\pm\frac{1}{2}\left(\widehat{p}-\widehat{x}\right)}\nonumber \\
 & =\begin{cases}
\left(e^{-\frac{1}{2}i\pi M_{2}'}e^{\frac{1}{2}\widehat{x}}+e^{\frac{1}{2}i\pi M_{2}'}e^{-\frac{1}{2}\widehat{x}}\right)e^{\frac{1}{2}\widehat{p}}\\
e^{\frac{1}{2}i\pi M_{2}'}e^{\frac{1}{2}\widehat{x}}e^{-\frac{1}{2}\widehat{p}}
\end{cases}.
\end{align}
We can also prove \eqref{eq:MM-QC-III1} and \eqref{eq:MM-QC-III2} at the operator level when $M=0$.

\section{Coalescence limits:  $\mathfrak{q}$-difference equations} \label{sei}

	Having discussed the coalescence limits at the level of the matrix models, we now turn to the $\mathfrak{q}$-difference equations their grand canonical partition function conjecturally satisfy when identified with $\mathfrak{q}$-Painlev\'e $\tau$-functions.
	In this section we consider the coalescence limit of the $\tau$-functions in the short-time expansion as in \eqref{tau}. This is the large-radius expansion of the topological string whose gauge theory counterpart is the 5d
	$\mathcal{N}=1$ $\text{SU}(2)$ with four to zero flavors in the electric frame of its Coulomb branch.
	Although natural from the perspective of gauge theory, this limit is not ideal from the point of view of the matrix model, which we expect to be describing the dual magnetic frame
	\cite{Bonelli:2016idi,Bonelli:2017ptp}. The main issue is the rescaling of the parameter $s$. Based on the TS/ST correspondence, we expect the spectral determinant to be calculating the $\tau$-function 
	\eqref{tau}
	at $s=1$ at any step in the coalescence series.
	This can be consistently implemented in a given set of
	identifications of parameters \eqref{treventi} 
	by choosing a suitable set of Weyl transformations $w$
	so that the $s$ parameter doesn't flow along the coalescence.

	In the following we first briefly recall the standard flow as in \cite{Matsuhira:2018qtx}, then we discuss the alternative flow in a concrete choice of
	parameterization. 
	
	\subsection{The two types of flow}\label{61}
	In 5 dimensions, we find there are two different ways to decouple a massive hypermultiplet. This section will serve as an introduction to details of both at the level of $\tau$-functions. With that in mind, let us write a generic 5d Nekrasov-Okounkov
$\tau$-function, focusing on one particular hypermultiplet of mass $\theta$ whose decoupling will be described in the two schemes:
	\begin{align}
	    \tau\left(\theta,\vec\theta_{\text{rest}};s,\sigma,t\right)&=\sum_{n\in\mathbb{Z}}s^n t^{\left(\sigma+n\right)^2}
	C\left(\theta,\vec\theta_{\text{rest}};\sigma+n\right)
	Z\left(\theta,\vec\theta_{\text{rest}};\sigma+n,t\right),\\
	C\left(\theta,\vec\theta_{\text{rest}};\sigma\right)&=\left(\prod_{\epsilon=\pm} G_{\mathfrak{q}}(1-\theta +\epsilon\sigma) \right)C_{\text{rest}}\left(\vec\theta_{\text{rest}};\sigma\right)
	,\\
	Z\left(\theta,\vec\theta_{\text{rest}};\sigma,t\right)&=\sum_{\lambda_+,\lambda_-}t^{|\lambda_+|+|\lambda_-|}
	\prod_{\epsilon=\pm}
	N_{\phi,\lambda_{\epsilon}}\left(\mathfrak{q}^{-\theta+\epsilon\sigma}\right)
	z_{\text{rest};\lambda_+,\lambda_-}\left(\vec\theta_{\text{rest}};\sigma\right).
	\end{align}
	Here, $C_{\text{rest}}$ and $z_{\text{rest};\lambda_+,\lambda_-}$ are respectively 
	the one-loop and instanton terms describing the contributions of the rest of the theory.
	
    Let us first recall the standard holomorphic decoupling limit of a massive hypermultiplet
\begin{equation}
\mathfrak{q}^{-\theta}\to\infty,\quad t\to 0,\quad t_1=t
\mathfrak{q}^{-\theta}\text{ finite}.
\end{equation}
    We will realize this limit by setting $\theta=\vartheta + i\Lambda$, $t= t_1 \mathfrak{q}^{\vartheta+i\Lambda}$.
    In our case, $\mathfrak{q}=e^{2\pi i/k}$ with $k>0$, so $\mathfrak{q}^{i\Lambda} =e^{-\frac{2\pi i}{k}\Lambda}\to 0$
    as $\Lambda\to\infty$. Up to some difference of imaginary units, this is the case worked out in \cite{Matsuhira:2018qtx}, the main points of the calculation whereof we recall. To begin with, consider the instanton counting part $Z\left(\theta,\vec\theta_{\text{rest}};\sigma,t\right)$. Since
    \begin{equation}
	N_{\lambda,\mu}(u)=\prod\limits_{c\in\lambda}\left(1-\mathfrak{q}^{-l_\lambda(c)-a_\mu(c)-1} u\right)\prod\limits_{c\in\mu}\left(1-\mathfrak{q}^{l_\mu(c)+a_\lambda(c)+1} u\right),
	\end{equation}
	we have
\begin{equation}
\mathfrak{q}^{|\lambda|(\vartheta+i\Lambda)}N_{\phi,\lambda}(u \mathfrak{q}^{-\vartheta-i\Lambda})=\prod\limits_{c\in\mu}\left(\mathfrak{q}^{\vartheta+i\Lambda}-\mathfrak{q}^{l_\mu(c)+a_\lambda(c)+1} u\right)\to f_{\lambda}(u^{-1}),
\end{equation}
	where 
$f_{\lambda}(u)=\prod\limits_{c\in\lambda}\left(-\mathfrak{q}^{l_\lambda(c)+a_\varnothing(c)+1} u^{-1}\right)$. 
This term is nothing but the five dimensional Chern-Simons coupling. 
	
	Slightly tougher is the calculation of the one-loop part, and here also the $n$-dependence of the summands has to be taken into account. The functional relations of Appendix \ref{appA} enable us to write for a shift $n\in\mathbb{Z}$
	\begin{align}\label{Gq_expansion}
	G_\mathfrak{q}\left(1+x \pm\left(\sigma+n\right)\right) = G_\mathfrak{q}\left(1+x \pm\sigma\right) \left[\frac{ \Gamma_\mathfrak{q}(x+\sigma)}{
	\Gamma_\mathfrak{q}(x-\sigma)
	}\right]^n \prod_{j=0}^{|n|-1}[x+\text{sgn}(n)\sigma]\prod_{k=1}^j[x\pm(\sigma+k)],
	\end{align}
	a relation the reader should make a mental note of, as
 we will
	return to it again from a different perspective. Presently though, we have $x=-\vartheta-i\Lambda$ and want the limit $\Lambda\to\infty$. Recalling the definition of $\mathfrak{q}$-numbers, we have for any $\alpha\in\mathbb{C}$
	\begin{equation}
	    [\alpha-i\Lambda] =\frac{1-\mathfrak{q}^{\alpha-i\Lambda}}{1-\mathfrak{q}} \sim \frac{\mathfrak{q}^{\alpha-i\Lambda}}{\mathfrak{q}-1},
	\end{equation}
	as $\Lambda\to\infty$. Before returning to \eqref{Gq_expansion}, we find it useful to rewrite
	\begin{equation}
	  \frac{ \Gamma_\mathfrak{q}(x+\sigma)}{
	\Gamma_\mathfrak{q}(x-\sigma)
	} =  \frac{ \Gamma_\mathfrak{q}(x+\sigma)\Gamma_\mathfrak{q}(1+x-\sigma)}{
	\Gamma_\mathfrak{q}(1+x+\sigma)\Gamma_\mathfrak{q}(x-\sigma)
	}\cdot \frac{ \Gamma_\mathfrak{q}(1+x+\sigma)}{
	\Gamma_\mathfrak{q}(1+x-\sigma)
	}
	=\frac{1-\mathfrak{q}^{x-\sigma}}{1-\mathfrak{q}^{x+\sigma}}\frac{ \Gamma_\mathfrak{q}(1+x+\sigma)}{
	\Gamma_\mathfrak{q}(1+x-\sigma)
	}\sim
	\mathfrak{q}^{-2\sigma}\frac{ \Gamma_\mathfrak{q}(1+x+\sigma)}{
	\Gamma_\mathfrak{q}(1+x-\sigma)
	},
	\end{equation}
	and define 
	\begin{equation}\label{68}
	    \check{s} = s (\mathfrak{q}-1)^{-2\sigma} q^{2\sigma(\vartheta+i\Lambda-\frac{1}{2})}
	    \frac{ \Gamma_\mathfrak{q}(1-\vartheta-i\Lambda+\sigma)}{
	\Gamma_\mathfrak{q}(1-\vartheta-i\Lambda-\sigma)}.
	\end{equation}
	Collecting it all and returning to \eqref{Gq_expansion} we have asymptotically 
	\begin{align}
	G_\mathfrak{q}\left(1-\vartheta-i\Lambda \pm\left(\sigma+n\right)\right) &\sim G_\mathfrak{q}\left(1-\vartheta-i\Lambda \pm\sigma\right) \left[\frac{ \Gamma_\mathfrak{q}(-\vartheta-i\Lambda+\sigma)}{
	\Gamma_\mathfrak{q}(-\vartheta-i\Lambda-\sigma)
	}\right]^n (\mathfrak{q}-1)^{-n^2} q^{n^2(-\vartheta-i\Lambda)+n\sigma} \nonumber \\
	&=G_\mathfrak{q}\left(1-\vartheta-i\Lambda \pm\sigma\right) \left(\frac{\check{s}}{s}\right)^n(\mathfrak{q}-1)^{-(\sigma+n)^2+\sigma^2} \left[q^{-\vartheta-i\Lambda}\right]^{(\sigma+n)^2-\sigma^2}\nonumber\\
&=X^{(1)}(\theta)^{-1} \left(\frac{\check{s}}{s}\right)^n(\mathfrak{q}-1)^{-(\sigma+n)^2} \left[\frac{t_1}{t}\right]^{(\sigma+n)^2},
	\end{align}
where in the last line we defined the $n$-independent factor
	\begin{equation}
X^{(1)}\left(\theta;\sigma\right)^{-1}= (\mathfrak{q}-1)^{\sigma^2}\mathfrak{q}^{\theta \sigma^2}G_\mathfrak{q}\left(1-\vartheta-i\Lambda \pm\sigma\right).
	\end{equation}
	It's now clear that as $\Lambda\to\infty$,
	\begin{equation}
\tau\left(\theta,\vec{\theta}_{\text{rest}};s,\sigma,t\right)\to X^{(1)}\left(\theta;\sigma\right)^{-1}
\tau^{\text{el}}\left(\vec{\theta}_{\text{rest}};\check{s},\sigma,t_1\right),\label{matilde}
	\end{equation}
	where 
	\begin{align}
	    \tau^{\text{el}}\left(\vec{\theta}_{\text{rest}};s,\sigma,t\right)&=\sum_{n\in\mathbb{Z}}s^n t^{\left(\sigma+n\right)^2}
	C^{\text{el}}\left(\vec\theta_{\text{rest}};\sigma+n\right)
	Z^{\text{el}}\left(\vec\theta_{\text{rest}};\sigma+n,t\right),\\
	C^{\text{el}}\left(\vec\theta_{\text{rest}};\sigma\right)&=(\mathfrak{q}-1)^{-\sigma^2}C_{\text{rest}}\left(\vec\theta_{\text{rest}};\sigma\right)
	,\\
	Z^{\text{el}}\left(\vec\theta_{\text{rest}};\sigma,t\right)&=\sum_{\lambda_+,\lambda_-}t^{|\lambda_+|+|\lambda_-|}
	\prod_{\epsilon=\pm}
	f_{\lambda_{\epsilon}}\left(\mathfrak{q}^{\epsilon\sigma}\right)
	z_{\text{rest};\lambda_+,\lambda_-}\left(\vec\theta_{\text{rest}};\sigma\right),
	\end{align}
and the superscript "el" stands for the fact that the holomorphic decoupling limit is studied in the electric frame of the gauge theory in five dimensions.
	Being the relevant
	$\mathfrak{q}$-Painlev\'e equation homogeneous,
	the multiplicative redefinition in 
	\eqref{matilde} preserves it.
	Notice that the holomorphic decoupling limit implies a redefinition of the $s$ parameter as in \eqref{68}.
	
	A different kind of decoupling limit has to be defined if one 
	studies the flow of solutions of the 
	$\mathfrak{q}$-Painlev\'e equation
	with fixed initial condition
	$s=1$ before and after the 
	coalescence.
	We now turn to discuss a class of decoupling limits suitable in the sense above. Consider the limit
	\begin{equation}
	\label{ml}
\mathfrak{q}^{-\theta}\to0,
	\end{equation}
	while keeping all other parameters fixed. In the following, we put $\theta=\vartheta-i\Lambda$ and consider the $\Lambda\to\infty$ limit. 
	Consider again the instanton counting part $Z\left(\theta,\vec\theta_{\text{rest}};\sigma,t\right)$. Since
    \begin{equation}
	N_{\lambda,\mu}(0)=\prod\limits_{c\in\lambda}\left(1-\mathfrak{q}^{-l_\lambda(c)-a_\mu(c)-1} \cdot 0\right)\prod\limits_{c\in\mu}\left(1-\mathfrak{q}^{l_\mu(c)+a_\lambda(c)+1} \cdot 0\right)=1,
	\end{equation}
we are
	immediately left with just the $z_{\text{rest};\lambda_+,\lambda_-}$ part in each term of the sum.
	We now turn to the one-loop part, referring back to \eqref{Gq_expansion}, this time with $x=-\vartheta+i\Lambda$. Consider first the $\mathfrak{q}$-numbers:
	\begin{equation}
	    [\alpha+i\Lambda] =\frac{1-\mathfrak{q}^{\alpha+i\Lambda}}{1-\mathfrak{q}} \to \frac{1}{1-\mathfrak{q}}.
	\end{equation}
	Therefore, the entire nested product in  \eqref{Gq_expansion} gives $(1-\mathfrak{q})^{-n^2}$. Next consider the $\mathfrak{q}$-Gamma functions. Using their definition, we can write
	\begin{equation}
	    \frac{\Gamma_{\mathfrak{q}}(x+\sigma)}{\Gamma_{\mathfrak{q}}(x-\sigma)} = (1-\mathfrak{q})^{-2\sigma}\frac{\left(q^{x-\sigma};q\right)_{\infty}}{\left(q^{x+\sigma};q\right)_{\infty}}\to(1-\mathfrak{q})^{-2\sigma},
	\end{equation}
since  $\mathfrak{q}^{x}=\mathfrak{q}^{-\vartheta-i\Lambda}\to0$ and $\left(0;\mathfrak{q}\right)_{\infty}=1$. All in all, we have
	\begin{equation}
	G_\mathfrak{q}\left(1-\vartheta+i\Lambda \pm\left(\sigma+n\right)\right) \sim G_\mathfrak{q}\left(1-\vartheta+i\Lambda \pm\sigma\right) (1-\mathfrak{q})^{-(\sigma+n)^2+\sigma^2}.
	\end{equation}
	Therefore, letting 
	\begin{equation}
X^{(2)}\left(\theta;\sigma\right)^{-1}=(1-\mathfrak{q})^{\sigma^2}G_\mathfrak{q}\left(1-\theta \pm\sigma\right),
	\end{equation} 
	we have as $\Lambda\to\infty$,
	\begin{equation}
\tau\left(\theta,\vec{\theta}_{\text{rest}};s,\sigma,t\right)\to X^{(2)}\left(\theta;\sigma\right)^{-1} \tau^{\text{magn}}\left(\vec{\theta}_{\text{rest}};s,\sigma,t\right)
	\end{equation}
	where 
	\begin{align}
	    \tau^{\text{magn}}\left(\vec{\theta}_{\text{rest}};s,\sigma,t\right)&=\sum_{n\in\mathbb{Z}}s^n t^{\left(\sigma+n\right)^2}
	C^{\text{magn}}\left(\vec\theta_{\text{rest}};\sigma+n\right)
	Z^{\text{magn}}\left(\vec\theta_{\text{rest}};\sigma+n,t\right),\\
	C^{\text{magn}}\left(\vec\theta_{\text{rest}};\sigma\right)&=(1-\mathfrak{q})^{-\sigma^2}C_{\text{rest}}\left(\vec\theta_{\text{rest}};\sigma\right)
	,\\
 Z^{\text{magn}}\left(\vec\theta_{\text{rest}};\sigma,t\right)&=\sum_{\lambda_+,\lambda_-}t^{|\lambda_+|+|\lambda_-|}
	z_{\text{rest};\lambda_+,\lambda_-}\left(\vec\theta_{\text{rest}};\sigma\right).
	\end{align}
	The superscript "magn" indicates that the limit is suitable to preserve the matrix model realisation of the $\tau$-function, which conjecturally describes the gauge theory in the magnetic  
	frame.
	
	\subsection{Coalescence limits in the electric frame}
	
	Having illustrated the calculation of coalescence limits at a generic step, we turn to concrete realizations of our dictionary. We first show that
	the dictionary \eqref{identificationwithoutWeyl}
	reproduces the coalescence limits in the electric frame as described in \cite{Matsuhira:2018qtx}. 
	Formula \eqref{identificationwithoutWeyl} reads in components as
		\begin{align}
		&\theta_0=\frac{M_2-M_1-2i\zeta_1}{4},\quad
		\theta_1=\frac{2M-M_1-M_2+2i\zeta_2}{4},\quad
		\theta_t=\frac{2M-M_1-M_2-2i\zeta_2}{4},\\
		&\theta_\infty=\frac{M_1-M_2-2i\zeta_1}{4},\quad
		t=\mathfrak{q}^{k+M-M_1-M_2},\quad \mathfrak{q}=e^{\frac{2\pi i}{k}}.
		\label{NoWeyl_dictionary}
	\end{align} 
	Let us now examine the holomorphic decoupling limits in the matrix model parameters.
	First, the redefinition $M_2=M_2'-2i\Lambda$, $\zeta_1 =\zeta_1'-\Lambda$, $\zeta_2 =\Lambda$ leads to 
	\begin{equation}
	    \mathfrak{q}^{-\theta_1-\theta_\infty}\to\infty,\quad t\to 0,\quad t_1= t \mathfrak{q}^{-\theta_1-\theta_\infty} = \mathfrak{q}^{\frac{M-2M_1-M_2'+i\zeta_1'}{2}},\quad\Lambda\to\infty,
	\end{equation}
	which corresponds to section 3 of \cite{Matsuhira:2018qtx}. Next, it can be checked that setting $M_2'=-i\Lambda$ leads to 
	\begin{equation}
	    \mathfrak{q}^{-\theta_t+\theta_0}\to\infty,\quad t_1\to 0,\quad t_2= t_1 \mathfrak{q}^{-\theta_t+\theta_0} = \mathfrak{q}^{-M_1},\quad\Lambda\to\infty,
	\end{equation}
	which corresponds to section 4 of \cite{Matsuhira:2018qtx}, up to a redefinition $\theta_0\leftrightarrow-\theta_0$, which is a symmetry of the original $\mathfrak{q}-$PVI tau function. Next, it can be checked that $M_1=M_1'-i\Lambda$, $\zeta_1'=-\Lambda$ leads to 
	\begin{equation}
	    \mathfrak{q}^{-\theta_t-\theta_0}\to\infty,\quad t_2\to 0,\quad t_3= t_2 \mathfrak{q}^{-\theta_t-\theta_0} = \mathfrak{q}^{-\frac{M+M_1'}{2}},\quad\Lambda\to\infty,
	\end{equation}
	which corresponds to section 5 of \cite{Matsuhira:2018qtx}. Finally, one can check that $M_1'=-i\Lambda$ leads to 
	\begin{equation}
	    \mathfrak{q}^{-\theta_1+\theta_\infty}\to\infty,\quad t_3\to 0,\quad t_4= t_3 \mathfrak{q}^{-\theta_1+\theta_\infty} = \mathfrak{q}^{-M},\infty,
	\end{equation}
	which corresponds to section 6 of \cite{Matsuhira:2018qtx}. 
	
	\subsection{Coalescence limits in the magnetic frame}
	
	In the following subsections we focus on the coalescence limits suitable for the magnetic frame, where the $\tau$-functions at $s=1$ are identified with the
    grand
	partition functions of the suitable quiver Chern-Simons matrix models.
	This can be achieved by modifying the dictionary \eqref{identificationwithoutWeyl} as in \eqref{treventi} by a specific class of Weyl transformations. By inspection, one finds that there are $48=4!\times 2$ possibilities corresponding to 
    the permutations of four masses and the inversion $t\rightarrow t^{-1}$.

We now 
	describe in detail a specific parameter identification among them and describe the corresponding coalescence limits. At each step, we give the sets of shifted $\tau$ functions and their bilinear relations.
	These depend on
	the specific choice of coalescence.
	All the choices, however, end on the identical $\mathfrak{q}$-PIII$_3$ equation.
    For the other choices, see Appendix \ref{220208app_weyltransformations}.

	\subsubsection{$\mathfrak{q}$-PVI $\to$ $\mathfrak{q}$-PV}
	We consider the dictionary:
	\begin{align}
	\begin{pmatrix}
	\theta_0\\
	\theta_1\\
	\theta_t\\
	\theta_\infty\\
	\frac{\log t}{\log \mathfrak{q}}
	\end{pmatrix}
	=\frac{1}{4}\begin{pmatrix}
	-1& 1& 0&2& 0\\
	-1&-1& 2&0&-2\\
	-1&-1& 2&0&2\\
	1&-1& 0&2& 0\\
	-4&-4& 4&0& 0
	\end{pmatrix}
	\tilde{w}
	\begin{pmatrix}
	M_1-k\\
	M_2-k\\
	M-k\\
	-i\zeta_1\\
	-i\zeta_2
	\end{pmatrix},
	\end{align}
	with the Weyl group element
	\begin{align}
	\tilde{w}=s_2 s_1 s_3 s_4 s_2 s_3 =
	\begin{pmatrix}
     \frac{1}{2} & \frac{1}{2} & -1 & 0 & 1 \\
     \frac{1}{2} & \frac{1}{2} & -1 & 0 & -1 \\
     1 & 1 & -1 & 0 & 0 \\
     0 & 0 & 0 & 1 & 0 \\
     -\frac{1}{2} & \frac{1}{2} & 0 & 0 & 0 \\
    \end{pmatrix},
	\end{align}
	which leads to the simple identification 
		\begin{align}
		&\theta_0=-\frac{i}{2}\left(\zeta_1-\zeta_2\right),\quad
		\theta_1=\frac{M_1-k}{2},\quad
		\theta_t=\frac{M_2-k}{2},\quad
		\theta_\infty=-\frac{i}{2}\left(\zeta_1+\zeta_2\right),\quad
		t=\mathfrak{q}^{M-k}=e^{\frac{2\pi iM}{k}},\\ &\mathfrak{q}=e^{\frac{2\pi i}{k}}.
		\label{43limit_parameters}
	\end{align}

	We proceed with the limit by setting $M_2=M_2'-2i\Lambda$, $\zeta_1 =\zeta_1'-\Lambda$, $\zeta_2 =\Lambda$. We have
	\begin{equation}
	q^{-\theta_t+\theta_0}=e^{\frac{\pi i (k-M_2'-i\zeta_1')}{k}}e^{\frac{4\pi\Lambda}{k}}\to 0, \quad \Lambda\to\infty.
	\end{equation} 
	Further, let $\theta_{\star}=\theta_0+\theta_t=\frac{M_2'-i\zeta_1'-k}{2}$ and define the $\mathfrak{q}$-PV tau function as
	\begin{align}
	\tau^{\text{V}}\left(\theta_1,\theta_{\star},\theta_{\infty};s,\sigma,t\right)&=\sum_{n\in\mathbb{Z}}s^nt^{\left(\sigma+n\right)^2}
	C^{\text{V}}\left(\theta_1,\theta_{\star},\theta_{\infty};\sigma+n\right)
	Z^{\text{V}}\left(\theta_1,\theta_{\star},\theta_{\infty};\sigma+n,t\right),\\
	C^{\text{V}}\left(\theta_1,\theta_{\star},\theta_{\infty};\sigma\right)&=(1-\mathfrak{q})^{-\sigma^2}\prod_{\epsilon,\epsilon'=\pm}
	G_{\mathfrak{q}}\left(1+\epsilon\theta_\infty-\theta_1+\epsilon'\sigma\right)\prod_{\epsilon=\pm}\frac{G_{\mathfrak{q}}\left(1-\theta_{\star}+\epsilon\sigma\right)}{G_{\mathfrak{q}}\left(1+2\epsilon\sigma\right)}
	,\\
	Z^{\text{V}}\left(\theta_1,\theta_{\star},\theta_{\infty};\sigma,t\right)&=\sum_{\lambda_+,\lambda_-}t^{|\lambda_+|+|\lambda_-|}\frac{
	\prod_{\epsilon,\epsilon'=\pm}
	N_{\phi,\lambda_{\epsilon'}}\left(\mathfrak{q}^{\epsilon\theta_{\infty}-\theta_1-\epsilon'\sigma}\right)
	\prod_{\epsilon=\pm}
	N_{\lambda_{\epsilon},\phi}\left(\mathfrak{q}^{\epsilon\sigma-\theta_{\star}}\right)}{\prod_{\epsilon,\epsilon'}N_{\lambda_\epsilon,\lambda_{\epsilon'}}\left(\mathfrak{q}^{\left(\epsilon-\epsilon'\right)\sigma}\right)}.
	\end{align}
	Let us consider the limit for the $\tau$ functions. First, define the scaling prefactors
	\begin{align}
	X_{1,2}&=X\left(\theta_0,\theta_1,\theta_t,\theta_{\infty}\pm\frac{1}{2},\sigma,t\right),&
	X_{3,4}&=X\left(\theta_0\pm\frac{1}{2},\theta_1,\theta_t,\theta_{\infty},\sigma\pm\frac{1}{2},t\right),\\
	X_{5,6}&=X\left(\theta_0,\theta_1\mp\frac{1}{2},\theta_t,\theta_{\infty},\sigma,t\right),&
	X_{7,8}&=X\left(\theta_0,\theta_1,\theta_t\mp\frac{1}{2},\theta_{\infty},\sigma\pm\frac{1}{2},t\right),
	\end{align}
with
\begin{equation}
	X(\theta_0,\theta_1,\theta_t,\theta_{\infty},\sigma,t)=t^{-\theta_t^2-\theta_0^2}(1-\mathfrak{q})^{\sigma^2}\prod_{\epsilon=\pm}G_{\mathfrak{q}}\left(1-\theta_t+\theta_0+\epsilon\sigma\right)^{-1}.
	\end{equation}
	Here, the first or second index will correspond to the choice of the upper or lower sign, respectively. 
	Next, define the shifted $\mathfrak{q}-$PV tau functions \begin{align}
	\tau_{1,2}^{\text{V}}&=\tau^{\text{V}}\left(\theta_1,\theta_\star,\theta_\infty\pm\frac{1}{2};s,\sigma,t\right),  &       
	\tau_{3,4}^{\text{V}}&=\tau^{\text{V}}\left(\theta_1,\theta_\star\pm\frac{1}{2},\theta_\infty;s,\sigma\pm\frac{1}{2},t\right),\\
	\tau_{5,6}^{\text{V}}&=\tau^{\text{V}}\left(\theta_1\mp\frac{1}{2},\theta_\star,\theta_\infty;s,\sigma,t\right).
	\end{align}
	Then, our discussion in subsection \ref{61} implies that
	\begin{align}
	    C_{i}\tau_i^{\text{VI}}(\theta_0,\theta_1,\theta_t,\theta_{\infty};s,\sigma,t) &\to \tau^{\text{V}}_i(\theta_1,\theta_\star,\theta_{\infty};s,\sigma,t),\quad i = 1,2,3,4,5,6\\
	    C_{7}\tau_7^{\text{VI}}(\theta_0,\theta_1,\theta_t,\theta_{\infty};s,\sigma,t) &\to s^{-1}\cdot \tau^{\text{V}}_4(\theta_0,\theta_*,\theta_{\infty};s,\sigma,t),\\
	    C_{8}\tau_8^{\text{VI}}(\theta_0,\theta_1,\theta_t,\theta_{\infty};s,\sigma,t) &\to s\cdot \tau^{\text{V}}_3(\theta_0,\theta_*,\theta_{\infty};s,\sigma,t).
	\end{align}
	For the tau functions as we have defined them, w
	We then obtain the following nontrivial bilinear identities for the $\mathfrak{q}$-PV $\tau$-functions 
	-- omitting the "$\text{V}$" superscript for readability:
	\begin{align}
	\tau_1\tau_2-(1-\mathfrak{q})^{-\frac{1}{2}}t^{\frac{1}{2}}q^{-2\theta_1}\tau_3\tau_4-\left(1-\mathfrak{q}^{-2\theta_1} t\right)\tau_5\tau_6&=0,\\
	\tau_1\tau_2-(1-\mathfrak{q})^{-\frac{1}{2}}t^{\frac{1}{2}}\tau_3\tau_4-\underline{\tau}_5\bar{\tau}_6&=0,\\
	\tau_1\tau_2-(1-\mathfrak{q})^{-\frac{1}{2}}t^{-\frac{1}{2}}\tau_3\tau_4+(1-\mathfrak{q})^{-\frac{1}{2}}t^{-\frac{1}{2}}\left(1-\mathfrak{q}^{-2\theta_1} t\right)\bar{\tau}_3\underline{\tau}_4&=0,\\
	\underline{\tau}_5\tau_6+\mathfrak{q}^{-\frac{1}{4}-\theta_1-\theta_\infty}(1-\mathfrak{q})^{-\frac{1}{2}}t^{\frac{1}{2}}\tau_3\underline{\tau}_4-\underline{\tau}_1\tau_2&=0,\\
	\underline{\tau}_5\tau_6+\mathfrak{q}^{-\frac{1}{4}-\theta_1+\theta_\infty}(1-\mathfrak{q})^{-\frac{1}{2}}t^{\frac{1}{2}}\underline{\tau}_3\tau_4-\tau_1\underline{\tau}_2&=0,\\
	\underline{\tau}_5\tau_6+\mathfrak{q}^{\frac{1}{2}+\theta_{\star}}(1-\mathfrak{q})^{-\frac{1}{2}}t^{-\frac{1}{2}}\left(\tau_3\underline{\tau_4}-\underline{\tau}_3\tau_4\right)&=0.
	\label{qPVbilineareq}
	\end{align}
Note that \eqref{qPVIbilineareq4} and \eqref{qPVIbilineareq} became trivial.
	
\subsubsection{ $\mathfrak{q}$-PV $\to$ $\mathfrak{q}$-PIII$_1$}
	The next limit amounts to $i M_2'\to \infty$ which means \begin{equation}
	\mathfrak{q}^{-\theta_\star} = \mathfrak{q}^{\frac{k+i \zeta_1'}{2}}e^{-\pi\Lambda/k}\to0, \quad \Lambda\to\infty.
	\end{equation}
	Let us define the appropriate $\tau$ function as
	\begin{align}
	\tau^{\text{III}_1}\left(\theta_1,\theta_\infty;s,\sigma,t\right)&=\sum_{n\in\mathbb{Z}}s^n t^{\left(\sigma+n\right)^2}
	C^{\text{III}_1}\left(\theta_1,\theta_\infty;\sigma+n\right)
	Z^{\text{III}_1}\left(\theta_1,\theta_\infty;\sigma+n,t\right),\\
	C^{\text{III}_1}\left(\theta_1,\theta_\infty;\sigma\right)&=(1-\mathfrak{q})^{-2\sigma^2} \frac{\prod_{\epsilon,\epsilon'=\pm}
	G_{\mathfrak{q}}\left(1-\theta_1+\epsilon\theta_\infty + \epsilon'\sigma\right)}{
	\prod_{\epsilon=\pm}
	G_{\mathfrak{q}}\left(1+2\epsilon\sigma\right)
	}
	,\\
	Z^{\text{III}_1}\left(\theta_1,\theta_\infty;\sigma,t\right)&=\sum_{\lambda_+,\lambda_-}t^{|\lambda_+|+|\lambda_-|}
	\prod_{\epsilon,\epsilon'=\pm}
	\frac{
	N_{\phi,\lambda_{\epsilon}}\left(\mathfrak{q}^{-\theta_1+\epsilon\theta_\infty-\epsilon'\sigma}\right)}{N_{\lambda_\epsilon,\lambda_{\epsilon'}}\left(\mathfrak{q}^{\left(\epsilon-\epsilon'\right)\sigma}\right)}.
	\end{align}
	Calculations are completely analogous to the previous section. Because of $\mathfrak{q}^{-\theta_*}\to 0$, it's easy to see that $Z^{V}\left(\theta_0,\theta_*,\theta_t;\sigma,t\right)\to Z^{\text{III}_1}\left(\theta_0,\theta_t;\sigma,t\right)$. We define the scaling prefactors
\begin{align}
X_{1,2}(\theta_1,\theta_\star,\theta_\infty,\sigma)&= X\left(\theta_1,\theta_\star,\theta_\infty\pm\frac{1}{2},\sigma\right),\\
X_{3,4}(\theta_1,\theta_\star,\theta_\infty,\sigma)&= X\left(\theta_1,\theta_\star\pm\frac{1}{2},\theta_\infty,\sigma\pm\frac{1}{2}\right),\\
X_{5,6}(\theta_1,\theta_\star,\theta_\infty,\sigma)&= X\left(\theta_1\mp\frac{1}{2},\theta_\star,\theta_\infty,\sigma\right),
\end{align}
with
\begin{equation}
X(\theta_1,\theta_\star,\theta_\infty,\sigma)=(1-\mathfrak{q})^{-\sigma^2}\prod_{\epsilon=\pm}G_{\mathfrak{q}}\left(1-\theta_\star+\epsilon\sigma\right)^{-1},
\end{equation}
	and the shifted $\mathfrak{q}$-PIII$_1$ tau functions
	\begin{align}
	\tau_{1,2}^{\text{III}_1}&=\tau^{\text{III}_1}\left(\theta_1,\theta_\infty\pm\frac{1}{2};s,\sigma,t\right), & &\\
	\tau_{3}^{\text{III}_1}&=s^{\frac{1}{2}}\tau^{\text{III}_1}\left(\theta_1,\theta_\infty;s,\sigma+\frac{1}{2},t\right),  &       
	\tau_{4,5}^{\text{III}_1}&=\tau^{\text{III}_1}\left(\theta_1\mp\frac{1}{2},\theta_\infty;s,\sigma,t\right) ,	
	\end{align}
	so that in the limit we find
	\begin{align}
	    C_{1,2} \tau_{1,2}^{\text{V}}(\theta_1,\theta_\star,\theta_\infty;s,\sigma,t) &\to \tau_{1,2}^{\text{III}_1}(\theta_1,\theta_\infty;s,\sigma,t),\\
	    C_{3,4} \tau_{3,4}^{\text{V}}(\theta_1,\theta_\star,\theta_\infty;s,\sigma,t) &\to s^{\mp\frac{1}{2}}\tau_3^{\text{III}_1}(\theta_1,\theta_\infty;s,\sigma,t),\\
	    C_{5,6} \tau_{5,6}^{\text{V}}(\theta_1,\theta_\star,\theta_\infty;s,\sigma,t) &\to \tau_{4,5}^{\text{III}_1}(\theta_1,\theta_\infty;s,\sigma,t).
	\end{align}
	The resulting bilinear equations are the following:
	\begin{align}
	\tau_1\tau_2 + (\mathfrak{q}-1)^{-1}\mathfrak{q}^{-2\theta_1}t^{\frac{1}{2}}\tau_3^2-\left(1-\mathfrak{q}^{-2\theta_1}t\right)\tau_4\tau_5&=0,\\						\tau_1\tau_2 + (\mathfrak{q}-1)^{-1}t^{\frac{1}{2}}\tau_3^2-\underline{\tau}_4\bar{\tau}_5&=0,\\
	\tau_1\tau_2 + (\mathfrak{q}-1)^{-1}t^{-\frac{1}{2}}\tau_3^2-(\mathfrak{q}-1)^{-1}t^{-\frac{1}{2}}\left(1-\mathfrak{q}^{-2\theta_1}t\right)\underline{\tau}_3\bar{\tau}_3&=0,\\
	\underline{\tau}_4\tau_5 - (\mathfrak{q}-1)^{-1}\mathfrak{q}^{-\frac{1}{4}-\theta_1-\theta_\infty}t^{\frac{1}{2}}\underline{\tau}_3\tau_3-\underline{\tau}_1\tau_2&=0,\\
	\underline{\tau}_4\tau_5 - (\mathfrak{q}-1)^{-1}\mathfrak{q}^{-\frac{1}{4}-\theta_1+\theta_\infty}t^{\frac{1}{2}}\underline{\tau}_3\tau_3-\underline{\tau}_2\tau_1&=0.
	\label{qPIII1bilineareq}
	\end{align}

\subsubsection{ $\mathfrak{q}$-PIII$_1$ $\to$ $\mathfrak{q}$-PIII$_2$}

    We now consider the limit $M_1\to M_1'-i\Lambda, \zeta_{1}'=-\Lambda $ which in our dictionary implies 
	\begin{equation}
     \mathfrak{q}^{-\theta_1+\theta_\infty} = \mathfrak{q}^{\frac{k-M_1'}{2}}e^{-2\pi\Lambda/k}\to\infty.
	\end{equation}
	Let us denote by $\theta_{*}=\theta_1+\theta_\infty$ and define the appropriate $\tau$ function as
	\begin{align}
	\tau^{\text{III}_2}\left(\theta_{*};s,\sigma,t\right)&=\sum_{n\in\mathbb{Z}}s^nt^{\left(\sigma+n\right)^2}
	C^{\text{III}_2}\left(\theta_{*};\sigma+n\right)
	Z^{\text{III}_2}\left(\theta_{*};\sigma+n,t\right),\\
	C^{\text{III}_2}\left(\theta_{*};\sigma\right)&=(1-\mathfrak{q})^{-3\sigma^2}\prod_{\epsilon=\pm}\frac{			
	G_{\mathfrak{q}}\left(1-\theta_{*}+\epsilon\sigma\right)							}{
	G_{\mathfrak{q}}\left(1+2\epsilon\sigma\right)						},\\
	Z^{\text{III}_2}\left(\theta_{*};\sigma,t\right)&=\sum_{\lambda_+,\lambda_-}t^{|\lambda_+|+|\lambda_-|}
	\frac{
	\prod_{\epsilon=\pm} N_{\phi,\lambda_{\epsilon}}\left(\mathfrak{q}^{-\theta_*-\epsilon\sigma}\right)}{
	\prod_{\epsilon,\epsilon'=\pm}
	N_{\lambda_\epsilon,\lambda_{\epsilon'}}\left(\mathfrak{q}^{\left(\epsilon-\epsilon'\right)\sigma}\right)}.
	\end{align}
	We define the scaling prefactors
	\begin{align}
	X_{1,2}(\theta_1,\theta_\infty,\sigma)&=X\left(\theta_1,\theta_\infty\pm\frac{1}{2},\sigma\right),\\
	X_{3}(\theta_1,\theta_\infty,\sigma)&=X\left(\theta_1,\theta_\infty,\sigma+\frac{1}{2}\right),\\
    X_{4,5}(\theta_1,\theta_\infty,\sigma)&=X\left(\theta_1\mp\frac{1}{2},\theta_\infty,\sigma\right),
	\end{align}
	with
	\begin{equation}
	X(\theta_1,\theta_\infty,\sigma)=(1-\mathfrak{q})^{-\sigma^2}\prod_{\epsilon=\pm}G_{\mathfrak{q}}\left(1-\theta_1+\theta_\infty+\epsilon\sigma\right)^{-1},
	\end{equation}
	and the shifted $\tau$ functions:
	\begin{align}
	\tau^{\text{III}_2}_{1,2}&=\tau^{\text{III}_2}\left(\theta_{*}\pm\frac{1}{2};s,\sigma,t\right),\\
	\tau^{\text{III}_2}_{3}&=s^{\frac{1}{2}}\cdot\tau^{\text{III}_2}\left(\theta_{*};s,\sigma+\frac{1}{2},t\right) .
	\end{align}
	Then we have as $\Lambda\to\infty$,
	\begin{align}
	    C_{1,5}\tau_{1,5}^{\text{III}_1}(\theta_1,\theta_\infty;s,\sigma,t)&\to \tau^{\text{III}_2}_1(\theta_{*};s,\sigma,t),\\
	    C_{2,4}\tau_{2,4}^{\text{III}_1}(\theta_1,\theta_\infty;s,\sigma,t)&\to \tau^{\text{III}_2}_2(\theta_{*};s,\sigma,t),\\
	    C_{3}\tau_{3}^{\text{III}_1}(\theta_1,\theta_\infty;s,\sigma,t)&\to \tau^{\text{III}_2}_3(\theta_{*};s,\sigma,t).
	\end{align}
	The resulting bilinear equations are
		\begin{align}
	\tau_1\tau_2 - \left(1-\mathfrak{q}\right)^{-\frac{1}{2}}t^{\frac{1}{2}}\tau_3^2 - \bar{\tau}_1\underline{\tau}_2
	&= 0,\\
	\tau_1\tau_2 - \left(1-\mathfrak{q}\right)^{-\frac{1}{2}}t^{-\frac{1}{2}}\left(\tau_3^2 - \underline{\tau}_3\bar{\tau}_3\right)
	&= 0,\\
	\underline{\tau}_1\tau_2-\tau_1\underline{\tau}_2 - \left(1-\mathfrak{q}\right)^{-\frac{1}{2}}t^{\frac{1}{2}}\mathfrak{q}^{-\frac{1}{4}-\theta_*}\underline{\tau}_3\tau_3
	&=0.
	\end{align}
	
\subsubsection{ $\mathfrak{q}$-PIII$_2$ $\to$ $\mathfrak{q}$-PIII$_3$}
	Finally we have $M_1'=-i\Lambda,\Lambda\to \infty$, which corresponds to
	\begin{equation}
\mathfrak{q}^{-\theta_*} = - e^{-\pi\Lambda/k}\to 0.
	\end{equation}
	Let us define the appropriate $\tau$ function
	\begin{align}
	\tau^{\text{III}_3}\left(s,\sigma,t\right)&=\sum_{n\in\mathbb{Z}}s^nt^{\left(\sigma+n\right)^2}
	C^{\text{III}_3}\left(\sigma+n\right)
	Z^{\text{III}_3}\left(\sigma+n,t\right),\\
	C^{\text{III}_3}\left(\sigma\right)&=(1-\mathfrak{q})^{-4\sigma^2} \prod_{\epsilon=\pm}\frac{1}
	{G_{\mathfrak{q}}\left(1+2\epsilon\sigma\right)}
	,\\
	Z^{\text{III}_3}\left(\theta_0,\theta_t;\sigma,t\right)&=\sum_{\lambda_+,\lambda_-}t^{|\lambda_+|+|\lambda_-|}
	\frac{
	1}{
	\prod_{\epsilon,\epsilon'=\pm}
	N_{\lambda_\epsilon,\lambda_{\epsilon'}}\left(\mathfrak{q}^{\left(\epsilon-\epsilon'\right)\sigma}\right)}.
	\end{align}
	To perform the limit, we first define the scaling prefactors  
	\begin{equation}
	X_{1,2}(\theta_{*},\sigma)=X\left(\theta_{*}\pm\frac{1}{2},\sigma\right),  \quad
	X_3(\theta_{*},\sigma)=X\left(\theta_{\star},\sigma+\frac{1}{2}\right) 
	\end{equation} 
	with
	\begin{equation}
	X(\theta_{*},\sigma)=(1-\mathfrak{q})^{-\sigma^2}\prod_{\epsilon=\pm}G_{\mathfrak{q}}\left(1-\theta_{\star}+\epsilon\sigma\right)^{-1},
	\end{equation}
	and the shifted $\tau$ functions		
	\begin{equation}
	\tau_1^{\text{III}_3} = \tau^{\text{III}_3}(s,\sigma,t), \quad  \tau_2^{\text{III}_3} = s^{\frac{1}{2}}\tau^{\text{III}_3}\left(s,\sigma+\frac{1}{2},t\right).
	\end{equation}
	Then we have as $\Lambda\to\infty$,
	\begin{align}
	    C_{1,2}\tau_{1,2}^{\text{III}_2}(\theta_{*};s,\sigma,t)&\to \tau^{\text{III}_3}_1(s,\sigma,t),\\
	    C_3\tau_3^{\text{III}_2}(\theta_{*};s,\sigma,t)&\to \tau^{\text{III}_2}_2 (s,\sigma,t),
	\end{align}
	and, as expected, we find these to satisfy the bilinear equations of $\mathfrak{q}$-Painlev\'e III$_3$, 
	\begin{align}
	\tau_1^2 - \underline{\tau}_1\bar{\tau}_1 -t^{\frac{1}{2}}\tau_2^2&=0,\\
	\tau_1^2 -t^{-\frac{1}{2}}\left(\tau_2^2- \underline{\tau}_2\bar{\tau}_2 \right)&=0.				\end{align}

\section{Discussion and open questions}
\label{sec_discussion}

In this paper we propose that the grand partition function of the four node circular quiver superconformal Chern-Simons theory, see Fig.\ref{fig_IIBbranesetup}, solves the $\mathfrak{q}$-deformed Painlev\'e VI equation in $\tau$-form \cite{2001CMaPh.220..165S,2006CMaPh.262..595T}.
We show that this theory describes the full moduli space of
$\text{SU}(2)$ gauge theory on $\mathbb{R}^4\times S^1$ with $N_f=4$ by extending the previous findings in \cite{Moriyama:2017gye,Moriyama:2017nbw,Kubo:2018cqw,Kubo:2019ejc}. Let us notice that a solution to the above $\mathfrak{q}$-difference system was previously proposed in terms of the Nekrasov-Okounkov partition function of the
gauge theory in \cite{Jimbo:2017ael}.
While this solution 
is valid in a short time expansion, which is perturbative in the gauge coupling,
the Fredholm determinant realisation arising from the quiver Chern-Simons theory can be naturally expanded in the different regime of small $\kappa$, corresponding to the magnetic phase of the gauge theory. We provide several explicit checks
of this proposal at low orders in $\kappa$.
Our result therefore allows to study the five dimensional gauge theory in terms of a matrix model in a regime which is otherwise difficult to access.
Moreover, the explicit results obtained in this paper, motivated by TS/ST correspondence, give a stronger check of the latter and enlarge the set of examples 
where its rigorous realisation is verified.

Let us list in the following several questions 
left open by our analysis
that it would be interesting to further investigate.

\begin{itemize}
\item 
The matrix models discussed in this paper can be used to 
study systematically the dual prepotential of five  dimensional $\text{SU}(2)$ gauge theories with $N_f\le 4$.
The four dimensional limit can be also studied by introducing a suitable dual scaling along the lines of \cite{Bonelli:2016idi}. 

\item
Generalise our matrix model to the case $M\neq 0$ in a 
representation which can be analytically continued in $M$ as well as in $M_1,M_2$.

\item We provided analytic evidence of our conjecture by explicit checks at low orders in $\kappa$ and numerical checks at fixed moduli. It would be great to be able to provide an analytic proof, either by induction in the power of $\kappa$ or by suitable Ward identities on the matrix model itself.

\item The $\mathfrak{q}$-Painlev\'e VI $\tau$-functions given as the Nekrasov-Okounkov partition function \eqref{tau} and the one proposed in this paper
as the grand partition function of Chern-Simons quiver \eqref{eq:tau-GPF} should be matched by fixing the ambiguity of the $C$ coefficients and relating the parameters $\kappa$ and $\sigma$.
The latter are linked through the quantum mirror map whose explicit expression is proposed in \cite{Furukawa:2019sfy}, which
one could check by comparing the two expressions for the $\tau$-function.

\item The identification of the $\tau$-function with the spectral determinant of the quantum operator implies that the analysis of the zeroes of the first solves the quantum spectrum of the latter \cite{Bonelli:2016idi,Bershtein:2021uts}.
Therefore, the results obtained in this paper provide a method to quantize the integrable spin chain systems associated to 5d gauge theories with $N_f\le 4$. It would be interesting to
pursue this direction and compare the results with the ones 
that can be derived from the Nekrasov-Shatashvili quantization method \cite{Nekrasov:2009rc}. 

\item
Conversely, by extending the relevant $\mathfrak{q}$-difference equations to higher rank simple gauge groups, it would be possible to 
use them as a tool to compute the multi-instantons expansions
of matter gauge theories in five dimensions, by extending the approach 
elaborated in \cite{Bonelli:2021rrg} for the four dimensional case.

\item
$\mathfrak{q}$-P$\text{III}_3$ equation has been shown to be related to five dimensional Nakajima-Yoshioka blowup equation 
\cite{Nakajima:2005fg} for pure Super Yang-Mills
\cite{Bershtein:2018zcz,Bershtein:2016aef}.
It would be interesting to extend such an analysis to the
$\mathfrak{q}$-Painlev\'e equations corresponding to the gauge theory in five dimensions coupled to massive hypermultiplets.

\item 
One can also consider the mass deformed quiver Chern-Simons matter theories  \cite{Hosomichi:2008jb,Gomis:2008vc}.
It was found that the grand partition function of the 
 ABJM theory with ${\cal N}=6$ preserving mass deformation satisfies 
 a modified version of $\mathfrak{q}$-P$\text{III}_3$
 \cite{Nosaka:2020tyv}.
It would be worth investigating whether the grand partition function of the four node quiver theory with mass deformation also obey a modified version of 
$\mathfrak{q}$-P$\text{VI}$.
Conversely, one could also investigate 
the grand partition function of the mass deformed CS quiver theories
by exploiting the $\mathfrak{q}$-difference bilinear equations, this in particular 
concerning 
the novel phase transition which was discovered for the mass deformed ABJM theory in \cite{Nosaka:2015bhf,Nosaka:2016vqf,Honda:2018pqa} and which is expected to exist also for more general quiver Chern-Simons matter theories with mass deformation \cite{Nosaka:2015iiw,Honda:2018pqa,Nosakatoappear}.

\item 
Generalise the 
Chern-Simons matter quiver theory and 
identify its grand canonical partition function with a Fredholm determinant of a suitable quantum operator and a related integrable system.

\item Investigate the insertion of observables and their r\^ole in these correspondences. It would be particularly interesting to find the relevant observable of the three-dimensional Chern-Simons quiver theory allowing to describe the full set of initial conditions of $\mathfrak{q}$-Painlev\'e equations, or from the five dimensional gauge theory viewpoint, the insertion of real co-dimension two defects.
We expect this to provide a description of the wave functions of the associated quantum integrable systems.

\item 
The relation between the $S^3$ quiver matter Chern Simons partition functions and  
the NO partition function of 
5d gauge theories on 
$\mathbb{R}^4\times S^1$ 
based on bilinear $\mathfrak{q}$-Painlev\'e
is suitable to 
be dimensionally lift to 
a relation 
between a quiver supersymmetric gauge theory on $S^3\times S^1$ and 
${\mathcal N}=1$ gauge theory on
$\mathbb{R}^4\times T^2$. 
Let us notice in this perspective 
that the Fermi gas formalism applied to
the 3d partition function was crucial to the study
of
$\mathfrak{q}$-Painlev\'e system we performed.
It is indeed known that 
the Fermi gas formalism extends to the Schur index of a certain class of four dimensional gauge theories \cite{Bourdier:2015sga,Drukker:2015spa,Bourdier:2015wda} where the relevant integrand is as elliptic lift of our \eqref{eq_MM_Def}.
It is therefore natural to expect that these Schur indices could be related to some gauge theories on $\mathbb{R}^4\times T^2$
and to a related cluster integrable systems.

\item A direct link between the three dimensional Chern-Simons quiver theories on $S^3$ and the gauge theories on $\mathbb{R}^4\times S^1$ involved in this game is to our knowledge still missing. It is expected to arise from a chain of
string theory dualities and geometric transitions, that it would be worth exploring also with the aim of a deeper understanding of the TS/ST correspondence.

\end{itemize}

\section*{Acknowledgements}
We
would like to thank
Fabrizio Del Monte, Alba Grassi,
Hirotaka Hayashi, Andrew P.~Kels, Marcos Mari\~no, 
Sanefumi Moriyama, Takahiro Nishinaka and Yasuhiko Yamada for valuable discussions and useful comments.
Part of the results was computed by using the high performance computing facilities provided by SISSA (Ulysses) and by Yukawa Institute for Theoretical Physics (Sushiki server).
The work of NK is supported by Grant-in-Aid for JSPS Fellows No.20J12263.
This research is partially supported by the INFN Research Projects GAST and ST\&FI, by PRIN ``Geometria delle variet\`a algebriche'' and by PRIN ``Non-perturbative Aspects Of Gauge Theories And Strings'' and by GNFM of INdAM.
Data sharing not applicable to this article as no datasets were generated or analysed during the current study.

\appendix

\section{Quantum dilogarithm and other special functions}
\label{appA}
In the following we assume $|\mathfrak{q}|\neq1$. We recall the definition of $\mathfrak{q}-$numbers, 
\begin{equation}
    [u]=\frac{1-\mathfrak{q}^{u}}{1-\mathfrak{q}},
\end{equation}
and the infinite multiple $\mathfrak{q}$-Pochhammer symbol
\begin{equation}
    (z;\mathfrak{q}_1,\dots,\mathfrak{q}_k)_{\infty}=\exp{\left(-\sum\limits_{p=1}^\infty \frac{z^p}{p}\frac{1}{1-\mathfrak{q}_1^p}\cdots\frac{1}{1-\mathfrak{q}_k^p}
    \right)
    }
    \stackrel{|\mathfrak{q}|<1}{=}\prod_{l_1,\dots,l_k=0}^{\infty}\left(1-z \mathfrak{q}_1^{l_1}\cdots \mathfrak{q}_k^{l_k}\right).
\end{equation}
This is defined for $|z|<1$, but can be analytically continued to $z\in\mathbb{C}$ using the latter equality. We extend the definition to an empty symbol $(z;)_\infty:=1-z$. For any $k\geq1$ we then have the relations
\begin{equation}
\frac{( z;\mathfrak{q}_1,\dots,\mathfrak{q}_k)_{\infty}}{(\mathfrak{q}_1 z;\mathfrak{q}_1,\dots,\mathfrak{q}_k)_{\infty}}= ( z;\mathfrak{q}_2,\dots,\mathfrak{q}_k)_{\infty}.
\end{equation}
Let us introduce the $\mathfrak{q}$-Gamma and $\mathfrak{q}$-Barnes G functions
\begin{equation}
    \Gamma_{\mathfrak{q}}(u)=\frac{(\mathfrak{q};\mathfrak{q})_\infty}{(\mathfrak{q}^u;\mathfrak{q})_\infty}(1-\mathfrak{q})^{u-1},\quad G_{\mathfrak{q}}(u)=\frac{(\mathfrak{q}^u;\mathfrak{q},\mathfrak{q})_\infty}{(\mathfrak{q};\mathfrak{q},\mathfrak{q})_\infty}(\mathfrak{q};\mathfrak{q})_\infty^{u-1}(1-\mathfrak{q})^{-\frac{(u-1)(u-2)}{2}},
\end{equation}
where $|\mathfrak{q}|<1$, which satisfy the $\mathfrak{q}$-analogues of the usual properties of Gamma and Barnes G functions,
\begin{equation}
\Gamma_{\mathfrak{q}}(u+1)=[u]\Gamma_{\mathfrak{q}}(u),\quad G_{\mathfrak{q}}(u+1)=\Gamma_{\mathfrak{q}}(u) G_{\mathfrak{q}}(u)
\end{equation}
and are both equal to one at $u=1$ and log-convex \cite{Nishizawa1996}. These can also be easily analytically continued to $|\mathfrak{q}|>1$ using
\begin{equation}
    \Gamma_{\mathfrak{q}}(u)=\mathfrak{q}^{\frac{(u-1)(u-2)}{2}}\Gamma_{\mathfrak{q}^{-1}}(u), \quad G_{\mathfrak{q}}(u) = \mathfrak{q}^{\frac{(u-1)(u-2)(u-3)}{6}}G_{\mathfrak{q}^{-1}}(u).
\end{equation}
Let us finally introduce the quantum dilogarithm function $\Phi_b\left(z\right)$ as \cite{Kashaev:2015wia}
\begin{align}
\Phi_b\left(z\right)=\frac{
\left(e^{2\pi b\left(z+\frac{i}{2}\left(b+\frac{1}{b}\right)\right)};e^{2\pi ib^2}\right)_\infty
}{
\left(e^{\frac{2\pi}{b}\left(z-\frac{i}{2}\left(b+\frac{1}{b}\right)\right)};e^{-\frac{2\pi i}{b^2}}\right)_\infty
}.
\label{quantumdilogdefinition}
\end{align}
Note that $\Phi_b\left(z\right)$ satisfies the following recursive relations
\begin{align}
\frac{\Phi_b\left(z+ib\right)}{\Phi_b\left(z\right)}=\frac{1}{1+e^{\pi ib^2}e^{2\pi bz}},\quad
\frac{\Phi_b\left(z+\frac{i}{b}\right)}{\Phi_b\left(z\right)}=\frac{1}{1+e^{\frac{\pi i}{b^2}}e^{\frac{2\pi z}{b}}},
\label{211107quantumdilogrecursive}
\end{align}
and that its asymptotic behavior is 
\begin{align}
\Phi_{b}\left(z\right)\sim\begin{cases}
\exp\left(i\pi z^{2}+\frac{\pi i}{12}\left(b^{2}+b^{-2}\right)\right) & \left({\rm Re}\left[z\right]\rightarrow\infty\right)\\
1 & \left({\rm Re}\left[z\right]\rightarrow-\infty\right)
\end{cases}.\label{eq:DilogAsym}
\end{align}

\section{Fermi gas formalism\label{sec:FGF}}
In this appendix, we address the study of the matrix model \eqref{eq_MM_Def} in the Fermi gas formalism.
First of all, we introduce our notations for one-dimensional quantum mechanics.
We denote the canonical position operators as ${\widehat x}$ and define the canonical momentum operator ${\widehat p}$ normalized as $[{\widehat x},{\widehat p}]=i\hbar$ with $\hbar=2\pi k$.
We denote a position eigenstate as $|x\rangle$ and a momentum eigenstate as $|p\rangle\!\rangle$, which are normalized as
\begin{align}
&{\widehat x}|x\rangle=x|x\rangle,\quad
\langle x|y\rangle=2\pi\delta\left(x-y\right),\label{eq:QM-Def1}\\
&{\widehat p}|p\rangle\!\rangle=p|p\rangle\!\rangle,\quad
\langle\!\langle p|p'\rangle\!\rangle=2\pi\delta\left(p-p'\right),\quad
\langle x|p\rangle\!\rangle=\frac{1}{\sqrt{k}}e^{\frac{ixp}{2\pi k}}.\label{eq:QM-Def}
\end{align}
For later purposes, we also introduce a symbol $t_{\zeta,n,r}$ defined as
\begin{align}
t_{\zeta,n,r}=2\pi\zeta+2\pi i\left(\frac{n+1}{2}-r\right).\label{eq:t-Def}
\end{align}

Our goal is to derive the expressions for the matrix model \eqref{pippo} and ${\widehat\rho}_k^{-1}$ \eqref{22modelrhoinverse}.
Our derivation extends the one in \cite{Kubo:2019ejc} to the present case with rank deformations and generic Fayet-Iliopoulos parameters. In the following we assume that the FI parameters $\zeta_{i}$ are real, we use the parametrization
\begin{equation}
N_{1}=N+M_{1},\quad N_{2}=N+M,\quad N_{3}=N+M_{2},\quad N_{4}=N,
\end{equation}
and assume that $N$, $M_{1}$, $M_{2}$ and $M$ are non-negative integers as in the main text. We also assume $M_{1}\geq M$ and $M_{2}\geq M$.

The integrand of the matrix model, after shifting all of the integration variables as $\mu\rightarrow\frac{\mu}{k}$, can be divided into two parts as
\begin{equation}
Z_k\left(N;M_1,M_2,M,\zeta_1,\zeta_2\right)
=\frac{1}{N_{2}!N_{4}!}\int_{-\infty}^\infty \prod_{n=1}^{N_{4}}\frac{d\mu_{n}}{2\pi k}\prod_{n=1}^{N_{2}}\frac{d\nu_{n}}{2\pi k}Y_{N_{4},N_{1},N_{2}}\left(0,\zeta_{1};\mu,\nu\right)\left(Y_{N_{4},N_{3},N_{2}}\left(\zeta_{2},0;\mu,\nu\right)\right)^{*},\label{eq:MM-FG-C1}
\end{equation}
where
\begin{align}
 & Y_{N_{4},\tilde{N},N_{2}}\left(\zeta,\zeta';\mu,\nu\right)\nonumber \\
 & =\frac{i^{-\frac{\tilde{N}^{2}}{2}}}{\tilde{N}!}\int_{-\infty}^\infty \prod_{n=1}^{\tilde{N}}\frac{d\lambda_{n}}{2\pi k}e^{\frac{i}{4\pi k}\sum_{n=1}^{\tilde{N}}\lambda_{n}^{2}}e^{-\frac{i\zeta}{k}\left(\sum_{n=1}^{N_{4}}\mu_{n}-\sum_{n=1}^{\tilde{N}}\lambda_{n}\right)-\frac{i\zeta'}{k}\left(\sum_{n=1}^{\tilde{N}}\lambda_{n}-\sum_{n=1}^{N_{2}}\nu_{n}\right)}\nonumber \\
 & \quad\times\frac{\prod_{m<m'}^{N_{4}}2\sinh\frac{\mu_{m}-\mu_{m'}}{2k}\prod_{n<n'}^{\tilde{N}}2\sinh\frac{\lambda_{n}-\lambda_{n'}}{2k}}{\prod_{m=1}^{N_{4}}\prod_{n=1}^{\tilde{N}}2\cosh\frac{\mu_{m}-\lambda_{n}}{2k}}\frac{\prod_{m<m'}^{\tilde{N}}2\sinh\frac{\lambda_{m}-\lambda_{m'}}{2k}\prod_{n<n'}^{N_{2}}2\sinh\frac{\nu_{n}-\nu_{n'}}{2k}}{\prod_{m=1}^{\tilde{N}}\prod_{n=1}^{N_{2}}2\cosh\frac{\lambda_{m}-\nu_{n}}{2k}}.\label{eq:Ydef}
\end{align}

We first focus on $Y_{N_{4},\tilde{N},N_{2}}$ and rewrite it in the operator formalism.
By combining the Cauchy determinant formula and the Vandermonde determinant formula \cite{Matsumoto:2013nya}
\begin{align}
&\frac{\prod_{m<m'}^{K}2\sinh\frac{\mu_{m}-\mu_{m'}}{2k}\prod_{n<n'}^{K+L}2\sinh\frac{\lambda_{n}-\lambda_{n'}}{2k}}{\prod_{m=1}^{K}\prod_{n=1}^{K+L}2\cosh\frac{\mu_{m}-\lambda_{n}}{2k}}
=\det\left(\begin{array}{c}
\left[\left(-1\right)^{L}\frac{e^{\frac{L}{2k}\left(\mu_{m}-\lambda_{n}\right)}}{2\cosh\frac{\mu_{m}-\lambda_{n}}{2k}}\right]_{m,n}^{K\times\left(K+L\right)}\\
\left[e^{\frac{1}{k}\left(\frac{L+1}{2}-r\right)\lambda_{n}}\right]_{r,n}^{L\times\left(K+L\right)}
\end{array}\right),\label{eq:CauchyDet1}\\
&\frac{\prod_{m<m'}^{K+L}2\sinh\frac{\lambda_{m}-\lambda_{m'}}{2k}\prod_{n<n'}^{K}2\sinh\frac{\nu_{n}-\nu_{n'}}{2k}}{\prod_{m=1}^{K+L}\prod_{n=1}^{K}2\cosh\frac{\lambda_{m}-\nu_{n}}{2k}}\nonumber\\
&=\det\left(\begin{array}{cc}
\left[\left(-1\right)^{L}\frac{e^{-\frac{L}{2k}\left(\lambda_{m}-\nu_{n}\right)}}{2\cosh\frac{\lambda_{m}-\nu_{n}}{2k}}\right]_{m,n}^{\left(K+L\right)\times K} & \left[e^{\frac{1}{k}\left(\frac{L+1}{2}-r\right)\lambda_{m}}\right]_{m,r}^{\left(K+L\right)\times L}\end{array}\right),\label{eq:CauchyDet}
\end{align}
we can rewrite the third line of \eqref{eq:Ydef} as the determinant of the product of two matrices. 
The notation for the operator formalism is in \eqref{eq:QM-Def1},\eqref{eq:QM-Def}.
The elements of the matrices can be rewritten in the operator formalism.
We will make use of the following identities
\begin{align}
\left(-1\right)^{L}\frac{e^{\frac{L}{2k}\left(\mu-\lambda\right)}}{2\cosh\frac{\mu-\lambda}{2k}} 
&=\left(-1\right)^{L}\int_{-\infty}^{\infty}\frac{dp}{2\pi} \frac{e^{\frac{i}{2\pi k}\left(p-i\pi L\right)\left(\mu -\lambda\right)}}{2\cosh\frac{p}{2}}\nonumber \\
&=k\braket{\mu|\frac{1}{2\cosh\frac{\widehat{p}-i\pi L}{2}}|\lambda}
+\sum_{r}^{\left\lfloor \frac{L+1}{2}\right\rfloor }\left(-1\right)^{L+r+1}e^{\frac{1}{k}\left(\frac{L+1}{2}-r\right)\left(\mu-\lambda\right)},\label{eq:op-Form11}\\
e^{\frac{1}{k}\sigma\lambda} & =\sqrt{k}\bbraket{2\pi i\sigma|\lambda},\label{eq:op-Form1}
\end{align}
where 
in the second line, as we shifted the integration contour from $\mathbb{R}$ to $\mathbb{R}+i\pi L$, we obtain a summation over
the resulting residues.
Fortunately, the contribution to the determinant of the latter vanishes
being a linear combination of rows.
Indeed it is evident from
 \eqref{eq:CauchyDet1},\eqref{eq:CauchyDet} that the sum of the residues is a linear combination of the lower (or right) elements.

The third line of $Y_{N_{4},\tilde{N},N_{2}}$ is now written in the operator formalism.
To write all of factors in the operator formalism, we multiply the first matrix
by a Fresnel factor $e^{\frac{i}{4\pi k}\sum_{n=1}^{\bar{N}}\lambda_n^2}$.
We also include the first FI factor, depending on $\zeta$, in the first matrix and the second FI factor, depending on $\zeta'$, in the second matrix. 
After performing the similarity transformations
\begin{equation}
e^{-\frac{2\pi i\zeta}{\hbar}\widehat{x}}f\left(\widehat{p}\right)e^{\frac{2\pi i\zeta}{\hbar}\widehat{x}}=f\left(\widehat{p}+2\pi\zeta\right),\quad\bbra pe^{\frac{2\pi i\zeta}{\hbar}\widehat{x}}=\bbra{p-2\pi\zeta},
\end{equation}
we obtain
\begin{align}
&Y_{N_{4},\tilde{N},N_{2}}\left(\zeta,\zeta';\mu,\nu\right)\nonumber \\
&=  \frac{i^{-\frac{\tilde{N}^{2}}{2}}}{\tilde{N}!}\int_{-\infty}^\infty \prod_{n=1}^{\tilde{N}}\frac{d\lambda_{n}}{2\pi k}\det\left(\begin{array}{c}
\left[k\braket{\mu_{m}|\frac{1}{2\cosh\frac{\widehat{p}+2\pi\zeta-i\pi \tilde{M}}{2}}e^{\frac{i}{4\pi k}\widehat{x}^{2}}|\lambda_{n}}\right]_{m,n}^{N_{4}\times \tilde{N}}\\
\left[\sqrt{k}\bbraket{t_{-\zeta,\tilde{M},r}|e^{\frac{i}{4\pi k}\widehat{x}^{2}}|\lambda_{n}}\right]_{r,n}^{\tilde{M}\times \tilde{N}}
\end{array}\right)\nonumber \\
 & \quad\times\det\left(\begin{array}{cc}
\left[k\braket{\lambda_{m}|\frac{1}{2\cosh\frac{\widehat{p}+2\pi\zeta'+i\pi\left(\tilde{N}-N_{2}\right)}{2}}|\nu_{n}}\right]_{m,n}^{_{\times N_{2}}^{\tilde{N}}} & \left[\sqrt{k}\brakket{\lambda_{m}|-t_{\zeta',\tilde{N}-N_{2},r}}\right]_{m,r}^{_{\times\left(\tilde{N}-N_{2}\right)}^{\tilde{N}}}\end{array}\right),
\end{align}
where $\tilde{M}=N_4-\tilde{N}$ and $t_{\zeta,n,r}$ is defined in \eqref{eq:t-Def}.

We now return to the matrix model \eqref{eq:MM-FG-C1}. 
Upon the similarity transformation
\begin{align}
\int_{-\infty}^\infty \prod_{n=1}^{N_{4}}d\mu_{n}\ket{\mu_{n}}\bra{\mu_{n}} & =\int_{-\infty}^\infty \prod_{n=1}^{N_{4}}d\mu_{n}e^{\frac{i}{2\hbar}\widehat{x}^{2}}e^{\frac{i}{2\hbar}\widehat{p}^{2}}\ket{\mu_{n}}\bra{\mu_{n}}e^{-\frac{i}{2\hbar}\widehat{p}^{2}}e^{-\frac{i}{2\hbar}\widehat{x}^{2}},\\
\int_{-\infty}^\infty \prod_{n=1}^{\tilde{N}}d\lambda_{n}\ket{\lambda_{n}}\bra{\lambda_{n}} & =\int_{-\infty}^\infty \prod_{n=1}^{\tilde{N}}d\lambda_{n}e^{\frac{i}{2\hbar}\widehat{p}^{2}}\ket{\lambda_{n}}\bra{\lambda_{n}}e^{-\frac{i}{2\hbar}\widehat{p}^{2}},\\
\int_{-\infty}^\infty \prod_{n=1}^{N_{2}}d\nu_{n}\ket{\nu_{n}}\bra{\nu_{n}} & =\int_{-\infty}^\infty \prod_{n=1}^{N_{2}}d\nu_{n}e^{\frac{i}{2\hbar}\widehat{p}^{2}}\ket{\nu_{n}}\bra{\nu_{n}}e^{-\frac{i}{2\hbar}\widehat{p}^{2}},
\label{220110_similaritytrsf}
\end{align}
and by using the formulae
\begin{equation}
e^{-\frac{i}{2\hbar}\widehat{p}^{2}}e^{-\frac{i}{2\hbar}\widehat{x}^{2}}f\left(\widehat{p}\right)e^{\frac{i}{2\hbar}\widehat{x}^{2}}e^{\frac{i}{2\hbar}\widehat{p}^{2}}=f\left(\widehat{x}\right),\quad
\bbra pe^{\frac{i}{2\hbar}\hat{x}^{2}}e^{\frac{i}{2\hbar}\hat{p}^{2}}=\sqrt{i}e^{-\frac{i}{2\hbar}p^{2}}\bra p,
\end{equation}
we find that the matrix models can be written as
\begin{align}
 & Z_{k}\left(N;M_{1},M_{2},M,\zeta_{1},\zeta_{2}\right)\nonumber \\
 & =\frac{1}{N_{2}!N_{4}!}e^{i\Theta_{k}\left(M_{1},M_{2},M,\zeta_{1},\zeta_{2}\right)}\int_{-\infty}^\infty \prod_{n=1}^{N_{4}}\frac{d\mu_{n}}{2\pi k}\prod_{n=1}^{N_{2}}\frac{d\nu_{n}}{2\pi k}\tilde{Y}_{N_{4},N_{1},N_{2}}\left(0,\zeta_{1};\mu,\nu\right)\left(\tilde{Y}_{N_{4},N_{3},N_{2}}\left(\zeta_{2},0;\mu,\nu\right)\right)^{*},\label{eq:Z-with-Ytilde}
\end{align}
where
\begin{align}
&\tilde{Y}_{N_{4},\tilde{N},N_{2}}\left(\zeta,\zeta';\mu,\nu\right) \nonumber \\
& =\frac{i^{-\frac{\tilde{N}^{2}}{2}+\frac{\tilde{M}}{2}}}{\tilde{N}!}\int_{-\infty}^\infty \prod_{n=1}^{N}\frac{d\lambda_{n}}{2\pi k}\det\left(\begin{array}{c}
\left[k\braket{\mu_{m}|\frac{1}{2\cosh\frac{\widehat{x}+2\pi\zeta-i\pi \tilde{M}}{2}}|\lambda_{n}}\right]_{m,n}^{N_{4}\times \tilde{N}}\\
\left[\sqrt{k}\braket{t_{-\zeta,\tilde{M},r}|\lambda_{n}}\right]_{r,n}^{\tilde{M}\times \tilde{N}}
\end{array}\right)\nonumber \\
 & \quad\times\det\left(\begin{array}{cc}
\left[k\braket{\lambda_{m}|\frac{1}{2\cosh\frac{\widehat{p}+2\pi\zeta'+i\pi\left(\tilde{N}-N_{2}\right)}{2}}|\nu_{n}}\right]_{m,n}^{_{\times N_{2}}^{\tilde{N}}} & \left[\sqrt{k}\brakket{\lambda_{m}|-t_{\zeta',\tilde{N}-N_{2},r}}\right]_{m,r}^{_{\times\left(\tilde{N}-N_{2}\right)}^{\tilde{N}}}\end{array}\right),
\end{align}
and
\begin{align}
\Theta_{k}\left(M_{1},M_{2},M,\zeta_{1},\zeta_{2}\right) & =\theta_{k}\left(M_{1},0\right)+\theta_{k}\left(M_{1}-M,\zeta_{1}\right)-\theta_{k}\left(M_{2},\zeta_{2}\right)-\theta_{k}\left(M_{2}-M,0\right),\\
\theta_{k}\left(M,\zeta\right) & =\frac{\pi}{k}\left[\frac{1}{12}\left(M^{3}-M\right)-M\zeta^{2}\right].\label{eq:Theta-Def}
\end{align}
$\tilde{Y}_{N_{4},\tilde{N},N_{2}}$ can be computed as follows. Since the second determinant is an anti-symmetric function of $\lambda_{m}$, we can simplify the first determinant by using
\begin{align}
\frac{1}{N!}\int_{-\infty}^\infty  d^{N}\lambda\det\left(\left[g_{m}\left(\lambda_{n}\right)\right]_{m,n}^{N\times N}\right)f\left(\lambda_{1},\lambda_{2},\ldots,\lambda_{N}\right)=\int_{-\infty}^\infty  d^{N}\lambda\prod_{n}^{N}g_{n}\left(\lambda_{n}\right)f\left(\lambda_{1},\lambda_{2},\ldots,\lambda_{N}\right),\label{eq:Diagonalize}
\end{align}
which holds for any anti-symmetric function $f\left(\bm{\lambda}\right)$. 
We decompose the other determinant by using \eqref{eq:op-Form11},\eqref{eq:op-Form1} and \eqref{eq:CauchyDet1},\eqref{eq:CauchyDet} backwards. 
Now we can perform the integration by using the delta functions coming form the inner products of the position operators. 
After a short computation, we obtain
\begin{align}
\tilde{Y}_{N_{4},\tilde{N},N_{2}}\left(\zeta,\zeta';\mu,\nu\right)= & i^{-\frac{\tilde{N}^{2}}{2}+\frac{\tilde{M}^{2}}{2}}e^{\frac{2\pi i\zeta\zeta'\tilde{M}}{k}}e^{-\frac{i\zeta'}{k}\left(\sum_{n=1}^{N_{4}}\mu_{n}-\sum_{n=1}^{N_{2}}\nu_{n}\right)}Z_{k}^{\left(\cs\right)}\left(\tilde{M}\right)\nonumber \\
 & \times\prod_{n=1}^{N_{4}}\frac{\prod_{r=1}^{\tilde{M}}2\sinh\frac{\mu_{n}+t_{\zeta,\tilde{M},r}}{2k}}{2\cosh\frac{\mu_{n}+2\pi\zeta-i\pi \tilde{M}}{2}}\prod_{n=1}^{N_{2}}\frac{1}{\prod_{r=1}^{\tilde{M}}2\cosh\frac{\nu_{n}+t_{\zeta,\tilde{M},r}}{2k}}\nonumber \\
 & \times\frac{\prod_{m<m'}^{N_{4}}2\sinh\frac{\mu_{m}-\mu_{m'}}{2k}\prod_{n<n'}^{N_{2}}2\sinh\frac{\nu_{n}-\nu_{n'}}{2k}}{\prod_{m=1}^{N_{4}}\prod_{n=1}^{N_{2}}2\cosh\frac{\mu_{m}-\nu_{n}}{2k}},
\end{align}
where
\begin{align}
Z_{k}^{\left(\cs\right)}\left(L\right) & =\frac{1}{k^{\frac{L}{2}}}\prod_{j<j'}^{L}2\sin\frac{\pi}{k}\left(j'-j\right),\label{eq:zCS-Def}
\end{align}
is the partition function of ${\rm U}\left(L\right)_{k}$ pure Chern-Simons theory. 
We again use the determinant formula \eqref{eq:CauchyDet1},\eqref{eq:CauchyDet} and the operator formula \eqref{eq:op-Form11},\eqref{eq:op-Form1} for the factor at the third line, and we include the FI factors and the factors in the second line into the matrix.
As a result, we obtain
\begin{align}
 \tilde{Y}_{N_{4},\tilde{N},N_{2}}\left(\zeta,\zeta';\mu,\nu\right) &=i^{-\frac{N_{4}^{2}}{2}}e^{\frac{2\pi i\zeta\zeta'\tilde{M}}{k}}Z_{k}^{\left(\cs\right)}\left(\tilde{M}\right)\nonumber \\
&\quad \times \det\left(\begin{array}{c}
\left[k\braket{\mu_{m}|S_{\tilde{M}}\left(\widehat{x}+2\pi\zeta\right)\frac{1}{2\cosh\frac{\widehat{p}+2\pi\zeta'-i\pi M}{2}}C_{\tilde{M}}\left(\widehat{x}+2\pi\zeta\right)|\nu_{n}}\right]_{m,n}^{N_{4}\times N_{2}}\\
\left[\sqrt{k}\bbraket{t_{-\zeta',M,r}|C_{\tilde{M}}\left(\widehat{x}+2\pi\zeta\right)|\nu_{n}}\right]_{r,n}^{M\times N_{2}}
\end{array}\right),
\label{220112_Ytildefinal}
\end{align}
where
\begin{align}
S_{L}\left(x\right)  =i^{L}\frac{\prod_{r=1}^{L}2\sinh\frac{x-2\pi i\left(\frac{L+1}{2}-r\right)}{2k}}{2\cosh\frac{x+i\pi L}{2}},\quad
C_{L}\left(x\right)  =\frac{1}{\prod_{r=1}^{L}2\cosh\frac{x-2\pi i\left(\frac{L+1}{2}-r\right)}{2k}}.\label{eq:FGF-SandC}
\end{align}

By using the recursive formula for the quantum dilogarithm functions \eqref{211107quantumdilogrecursive}, \eqref{eq:FGF-SandC} can be written in terms of the quantum dilogarithm as ($b=\sqrt{k}$)
\begin{align}
S_{L}\left(x\right)=e^{\frac{k-L}{2k}x}\frac{\Phi_{ b}\left(\frac{x}{2\pi b}-\frac{iL}{2b}+\frac{i}{2} b\right)}{\Phi_{ b}\left(\frac{x}{2\pi b}+\frac{iL}{2b}-\frac{i}{2} b\right)},\quad C_{L}\left(x\right)=e^{\frac{L}{2k}x}\frac{\Phi_{ b}\left(\frac{x}{2\pi b}+\frac{iL}{2 b}\right)}{\Phi_{ b}\left(\frac{x}{2\pi b}-\frac{iL}{2 b}\right)}.\label{eq:SandC-QDilog}
\end{align}
By substituting \eqref{220112_Ytildefinal} into \eqref{eq:Z-with-Ytilde}, we finally arrive at
\begin{align}
 & Z_k\left(N;M_{1},M_{2},M,\zeta_{1},\zeta_{2}\right)\nonumber \\
 & =
\frac{e^{i\Theta_{k}\left(M_{1},M_{2},M,\zeta_{1},\zeta_{2}\right)}Z_{k}^{\left(\cs\right)}\left(M_{1}\right)Z_{k}^{\left(\cs\right)}\left(M_{2}\right)}{N!\left(N+M\right)!}
\int_{-\infty}^\infty \prod_{n=1}^{N}\frac{d\mu_{n}}{2\pi}\prod_{n=1}^{N+M}\frac{d\nu_{n}}{2\pi}\nonumber \\
&\quad \times \det\left(\begin{array}{c}
\left[\braket{\mu_{m}|\widehat{D}_{1}^{\text{VI}}|\nu_{n}}\right]_{m,n}^{N\times\left(N+M\right)}\\
\left[\bbraket{t_{0,M,r}|\widehat{d}_{1}^{\text{VI}}|\nu_{n}}\right]_{r,n}^{M\times\left(N+M\right)}
\end{array}\right)\nonumber \\
 & \quad\times\det\left(\begin{array}{cc}
\left[\braket{\nu_{m}|\widehat{D}_{2}^{\text{VI}}|\mu_{n}}\right]_{m,n}^{\left(N+M\right)\times N} & \left[\brakket{\nu_{m}|\widehat{d}_{2}^{\text{VI}}|-t_{0,M,r}}\right]_{m,r}^{\left(N+M\right)\times M}\end{array}\right),
\label{eq:FGF-Res-Gen}
\end{align}
where
\begin{align}
\widehat{D}_{1}^{\text{VI}} & =e^{-\frac{i\zeta_{1}}{k}\widehat{x}}S_{M_1}\left({\widehat x}\right)\frac{1}{2\cosh\frac{\widehat{p}-i\pi M}{2}}e^{\frac{i\zeta_{1}}{k}\widehat{x}}C_{M_1}\left({\widehat x}\right),\\
\widehat{d}_{1}^{\text{VI}} & =e^{\frac{i\zeta_{1}}{k}\widehat{x}}C_{M_1}\left({\widehat x}\right),\\
\widehat{D}_{2}^{\text{VI}} & =C_{M_2}\left({\widehat x}+2\pi\zeta_2\right)\frac{1}{2\cosh\frac{\widehat{p}+\pi iM}{2}}S_{M_2}\left({\widehat x}+2\pi\zeta_2\right),\\
\widehat{d}_{2}^{\text{VI}} & =C_{M_2}\left({\widehat x}+2\pi\zeta_2\right).
\end{align}

Let us perform a short digression on the formulas which can be used to rephrase our final result \eqref{eq:FGF-Res-Gen} in a more concise way, though we do not use them in the main text.
Note that by using the formula
\begin{equation}
\frac{1}{N!}\int_{-\infty}^\infty  d^{N}\nu\det\left(\left[f_{m}\left(\nu_{n}\right)\right]_{m,n}^{N\times N}\right)\det\left(\left[g_{n}\left(\nu_{m}\right)\right]_{m,n}^{N\times N}\right)=\det\left(\left[\int_{-\infty}^\infty  d\nu f_{m}\left(\nu\right)g_{n}\left(\nu\right)\right]_{m,n}^{N\times N}\right),\label{eq:Det-Glue}
\end{equation}
the partition function for $M>0$ \eqref{eq:FGF-Res-Gen}, which is written as a $\left(2N+M\right)$ dimensional integral, can be further reduced to a $N$ dimensional integral:
\begin{align}
 & Z_k\left(N;M_{1},M_{2},M,\zeta_{1},\zeta_{2}\right)\nonumber \\
 & =
\frac{e^{i\Theta_{k}\left(M_{1},M_{2},M,\zeta_{1},\zeta_{2}\right)}Z_{k}^{\left(\cs\right)}\left(M_{1}\right)Z_{k}^{\left(\cs\right)}\left(M_{2}\right)}{N!}
\int_{-\infty}^\infty \prod_{n=1}^{N}\frac{d\mu_{n}}{2\pi}\nonumber \\
&\quad \times \det\begin{pmatrix}
\left[\braket{\mu_{m}|\widehat{D}_{1}^{\text{VI}}\widehat{D}_{2}^{\text{VI}}|\mu_{n}}\right]_{m,n}^{N\times N}
&\left[\brakket{\mu_{m}|\widehat{D}_{1}^{\text{VI}}\widehat{d}_{2}^{\text{VI}}|-t_{0,M,s}}\right]_{m,s}^{N\times M}\\
\left[\bbraket{t_{0,M,r}|\widehat{d}_{1}^{\text{VI}}\widehat{D}_{2}^{\text{VI}}|\mu_{n}}\right]_{r,n}^{M\times N}
&
\left[\bbrakket{t_{0,M,r}|\widehat{d}_{1}^{\text{VI}}\widehat{d}_{2}^{\text{VI}}|-t_{0,M,s}}\right]_{r,s}^{M\times M}
\end{pmatrix},
\label{220112ZkVIfourblocks}
\end{align}
which implies that the grand partition function \eqref{211223Fredholm} can be written as \cite{Matsumoto:2013nya}
\begin{align}
&\Xi_k\left(\kappa;M_1,M_2,M,\zeta_1,\zeta_2\right)
=\sum_{N=0}^\infty \kappa^N\frac{Z_k\left(N;M_1,M_2,M,\zeta_1,\zeta_2\right)}{Z_k\left(0;M_1,M_2,M,\zeta_1,\zeta_2\right)}\nonumber \\
&=
\frac{
\text{Det}\left(1+\kappa {\widehat D}_1^{\text{VI}}{\widehat D}_2^{\text{VI}}\right)
\det_{r,s}\left[\bbrakket{t_{0,M,r}|{\widehat d}^{\text{VI}}_1\frac{1}{1+\kappa{\widehat D}_2^{\text{VI}}{\widehat D}_1^{\text{VI}}}{\widehat d}^{\text{VI}}_2|-t_{0,M,s}}\right]
}{
\det_{r,s}\left[\bbrakket{t_{0,M,r}|{\widehat d}^{\text{VI}}_1{\widehat d}^{\text{VI}}_2|-t_{0,M,s}}\right]
}.
\end{align}

\subsection{$M=0$ case}

When $M=0$, the matrix model \eqref{220112ZkVIfourblocks} simplifies to
\begin{align}
Z_k\left(N;M_1,M_2,0,\zeta_1,\zeta_2\right)
= &  e^{i\Theta_{k}\left(M_{1},M_{2},0,\zeta_{1},\zeta_{2}\right)}Z_{k}^{\left(\cs\right)}\left(M_{1}\right)Z_{k}^{\left(\cs\right)}\left(M_{2}\right)\nonumber \\
 & \times\frac{1}{N!}\int_{-\infty}^\infty \prod_{n=1}^{N}\frac{d\mu_{n}}{2\pi}\det\left(\left[\braket{\mu_{m}|\widehat{\rho}_{k}\left(M_{1},M_{2},0,\zeta_{1},\zeta_{2}\right)|\mu_{n}}\right]_{m,n}^{N\times N}\right),\label{eq:FGF-Res-M0}
\end{align}
where
\begin{align}
\widehat{\rho}_{k}\left(M_{1},M_{2},0,\zeta_{1},\zeta_{2}\right)
&=\left.{\widehat D}_1^{\text{VI}}
{\widehat D}_2^{\text{VI}}
\right|_{M=0}\nonumber \\
&=
S_{M_{1}}\left(\widehat{x}\right)\frac{1}{2\cosh\frac{\widehat{p}+2\pi\zeta_{1}}{2}}C_{M_{1}}\left(\widehat{x}\right)C_{M_{2}}\left(\widehat{x}+2\pi\zeta_{2}\right)\frac{1}{2\cosh\frac{\widehat{p}}{2}}S_{M_{2}}\left(\widehat{x}+2\pi\zeta_{2}\right). \label{eq:DensM-M0}
\end{align}
This is the same as \eqref{211015rhoM1M2integerM30}.
For this expression, we can relate the density matrix to the quantum curve \cite{Kashaev:2015wia,Zakany:2018dio}. The important relations are
\begin{align}
C_{L}^{-1}\left(\widehat{x}\right)e^{\pm\frac{1}{2}\widehat{p}}S_{L}^{-1}\left(\widehat{x}\right) & =e^{\mp\frac{1}{2}i\pi L}e^{\pm\frac{1}{2}\widehat{p}}\frac{\Phi_{ b}\left(\frac{\widehat{x}}{2\pi b}-\frac{iL}{2 b}\pm\frac{i}{2} b\right)}{\Phi_{ b}\left(\frac{\widehat{x}}{2\pi b}+\frac{iL}{2 b}\pm\frac{i}{2} b\right)}\frac{\Phi_{ b}\left(\frac{\widehat{x}}{2\pi b}+\frac{iL}{2b}-\frac{i}{2} b\right)}{\Phi_{ b}\left(\frac{\widehat{x}}{2\pi b}-\frac{iL}{2b}+\frac{i}{2} b\right)}e^{-\frac{1}{2}\widehat{x}}\nonumber \\
 & =e^{\pm\frac{1}{2}\widehat{p}}\left(e^{\pm\frac{1}{2}i\pi L}e^{\frac{1}{2}\widehat{x}}+e^{\mp\frac{1}{2}i\pi L}e^{-\frac{1}{2}\widehat{x}}\right),\label{eq:Hyper-QC01}\\
S_{L}^{-1}\left(\widehat{x}\right)e^{\pm\frac{1}{2}\widehat{p}}C_{L}^{-1}\left(\widehat{x}\right) & =e^{\pm\frac{1}{2}i\pi L}e^{-\frac{1}{2}\widehat{x}}\frac{\Phi_{ b}\left(\frac{\widehat{x}}{2\pi b}+\frac{iL}{2 b}-\frac{i}{2} b\right)}{\Phi_{ b}\left(\frac{\widehat{x}}{2\pi b}-\frac{iL}{2 b}+\frac{i}{2} b\right)}\frac{\Phi_{ b}\left(\frac{\widehat{x}}{2\pi b}-\frac{iL}{2 b}\mp\frac{i}{2} b\right)}{\Phi_{ b}\left(\frac{\widehat{x}}{2\pi b}+\frac{iL}{2 b}\mp\frac{i}{2} b\right)}e^{\pm\frac{1}{2}\widehat{p}}\nonumber \\
 & =\left(e^{\mp\frac{1}{2}i\pi L}e^{\frac{1}{2}\widehat{x}}+e^{\pm\frac{1}{2}i\pi L}e^{-\frac{1}{2}\widehat{x}}\right)e^{\pm\frac{1}{2}\widehat{p}},\label{eq:Hyper-QC0}
\end{align}
where we used the Baker--Campbell--Hausdorff formula $e^{\alpha\widehat{x}}e^{\beta\widehat{p}}=e^{2\pi i\alpha \beta k}e^{\beta\widehat{p}}e^{\alpha\widehat{x}}$ and \eqref{211107quantumdilogrecursive}. By using these relations, we obtain
\begin{align}
S_{L}\left(\widehat{x}\right)\frac{1}{2\cosh\frac{\widehat{p}}{2}}C_{L}\left(\widehat{x}\right)&=\left[e^{\frac{1}{2}\widehat{p}}\left(e^{\frac{1}{2}i\pi L}e^{\frac{1}{2}\widehat{x}}+e^{-\frac{1}{2}i\pi L}e^{-\frac{1}{2}\widehat{x}}\right)+e^{-\frac{1}{2}\widehat{p}}\left(e^{-\frac{1}{2}i\pi L}e^{\frac{1}{2}\widehat{x}}+e^{\frac{1}{2}i\pi L}e^{-\frac{1}{2}\widehat{x}}\right)\right]^{-1},\label{eq:Hyper-QC1}\\
C_{L}\left(\widehat{x}\right)\frac{1}{2\cosh\frac{\widehat{p}}{2}}S_{L}\left(\widehat{x}\right)&=\left[\left(e^{-\frac{1}{2}i\pi L}e^{\frac{1}{2}\widehat{x}}+e^{\frac{1}{2}i\pi L}e^{-\frac{1}{2}\widehat{x}}\right)e^{\frac{1}{2}\widehat{p}}+\left(e^{\frac{1}{2}i\pi L}e^{\frac{1}{2}\widehat{x}}+e^{-\frac{1}{2}i\pi L}e^{-\frac{1}{2}\widehat{x}}\right)e^{-\frac{1}{2}\widehat{p}}\right]^{-1}.\label{eq:Hyper-QC}
\end{align}
The inverse of the density matrix is the product of the above two quantum curves.
Therefore, we finally obtain
\begin{align}
&\widehat{\rho}_{k}^{-1}\left(M_{1},M_{2},0,\zeta_{1},\zeta_{2}\right)\nonumber \\
&=\left[\left(e^{-\frac{1}{2}i\pi M_2+\pi \zeta_2}e^{\frac{1}{2}\widehat{x}}+e^{\frac{1}{2}i\pi M_2-\pi \zeta_2}e^{-\frac{1}{2}\widehat{x}}\right)e^{\frac{1}{2}\widehat{p}}+\left(e^{\frac{1}{2}i\pi M_2+\pi \zeta_2}e^{\frac{1}{2}\widehat{x}}+e^{-\frac{1}{2}i\pi M_2-\pi \zeta_2}e^{-\frac{1}{2}\widehat{x}}\right)e^{-\frac{1}{2}\widehat{p}}\right]
\nonumber \\
&\quad \times 
\left[e^{\pi \zeta_1}e^{\frac{1}{2}\widehat{p}}\left(e^{\frac{1}{2}i\pi M_1}e^{\frac{1}{2}\widehat{x}}+e^{-\frac{1}{2}i\pi M_1}e^{-\frac{1}{2}\widehat{x}}\right)+e^{-\pi \zeta_1}e^{-\frac{1}{2}\widehat{p}}\left(e^{-\frac{1}{2}i\pi M_1}e^{\frac{1}{2}\widehat{x}}+e^{\frac{1}{2}i\pi M_1}e^{-\frac{1}{2}\widehat{x}}\right)\right]
\nonumber \\
&=e^{\frac{\pi i(M_{1}-M_{2})}{2}+\pi(\zeta_{1}+\zeta_{2})}e^{{\widehat{x}}+{\widehat{p}}}+[e^{\frac{\pi i(-M_{1}-M_{2})}{2}+\pi(\zeta_{1}+\zeta_{2})+\pi ik}+e^{\frac{\pi i(M_{1}+M_{2})}{2}+\pi(\zeta_{1}-\zeta_{2})-\pi ik}]e^{{\widehat{p}}}\nonumber\\
&\quad+e^{\frac{\pi i(-M_{1}+M_{2})}{2}+\pi(\zeta_{1}-\zeta_{2})}e^{-{\widehat{x}}+{\widehat{p}}}\nonumber\\
&\quad+[e^{\frac{\pi i(-M_{1}-M_{2})}{2}+\pi(-\zeta_{1}+\zeta_{2})}+e^{\frac{\pi i(M_{1}+M_{2})}{2}+\pi(\zeta_{1}+\zeta_{2})}]e^{{\widehat{x}}}\nonumber\\
&\quad+e^{\frac{\pi i(-M_{1}+M_{2})}{2}+\pi(-\zeta_{1}-\zeta_{2})}+e^{\frac{\pi i(-M_{1}+M_{2})}{2}+\pi(\zeta_{1}+\zeta_{2})}+e^{\frac{\pi i(M_{1}-M_{2})}{2}+\pi(-\zeta_{1}+\zeta_{2})}+e^{\frac{\pi i(M_{1}-M_{2})}{2}+\pi(\zeta_{1}-\zeta_{2})}\nonumber\\
&\quad+[e^{\frac{\pi i(-M_{1}-M_{2})}{2}+\pi(\zeta_{1}-\zeta_{2})}+e^{\frac{\pi i(M_{1}+M_{2})}{2}+\pi(-\zeta_{1}-\zeta_{2})}]e^{-{\widehat{x}}}\nonumber\\
&\quad+e^{\frac{\pi i(-M_{1}+M_{2})}{2}+\pi(-\zeta_{1}+\zeta_{2})}e^{{\widehat{x}}-{\widehat{p}}}+[e^{\frac{\pi i(-M_{1}-M_{2})}{2}+\pi(-\zeta_{1}-\zeta_{2})+\pi ik}+e^{\frac{\pi i(M_{1}+M_{2})}{2}+\pi(-\zeta_{1}+\zeta_{2})-\pi ik}]e^{-{\widehat{p}}}\nonumber\\
&\quad+e^{\frac{\pi i(M_{1}-M_{2})}{2}+\pi(-\zeta_{1}-\zeta_{2})}e^{-{\widehat{x}}-{\widehat{p}}}.\label{eq:QC-M0-1}
\end{align}
This is the quantum curve associated to the (2,2) model for $M=0$.

\section{Proof of \eqref{Weyl_directlyshownonrhohat4}: ${\widehat\rho}_k\left(M_1,M_2,0,\zeta_1,\zeta_2\right)\sim {\widehat\rho}_k\left(M_1,M_2,0,\zeta_2,\zeta_1\right)$ }
\label{sec_proofofzeta1zeta2symmetry}
First we notice the following identity
\begin{align}
&e^{\frac{M}{2k}{\widehat x}}
\frac{\Phi_b\left(\frac{{\widehat x}}{2\pi b}+\frac{iM}{2b}\right)}{\Phi_b\left(\frac{{\widehat x}}{2\pi b}-\frac{iM}{2b}\right)}
\frac{1}{2\cosh\frac{{\widehat p}}{2}}
e^{\left(\frac{1}{2}-\frac{M}{2k}\right){\widehat x}}
\frac{\Phi_b\left(\frac{{\widehat x}}{2\pi b}-\frac{iM}{2b}+\frac{ib}{2}\right)}{\Phi_b\left(\frac{{\widehat x}}{2\pi b}+\frac{iM}{2b}-\frac{ib}{2}\right)}\nonumber \\
&\quad =
e^{\left(\frac{1}{2}-\frac{M}{2k}\right){\widehat p}}
\frac{\Phi_b\left(\frac{{\widehat p}}{2\pi b}-\frac{iM}{2b}+\frac{ib}{2}\right)}{\Phi_b\left(\frac{{\widehat p}}{2\pi b}+\frac{iM}{2b}-\frac{ib}{2}\right)}
\frac{1}{2\cosh\frac{{\widehat x}}{2}}
e^{\frac{M}{2k}{\widehat p}}
\frac{\Phi_b\left(\frac{{\widehat p}}{2\pi b}+\frac{iM}{2b}\right)}{\Phi_b\left(\frac{{\widehat p}}{2\pi b}-\frac{iM}{2b}\right)}.
\label{211107quantumdilogidentity}
\end{align}
This identity can be proved by calculating the inverse of both sides.
The inverse of the left-hand side can be calculated as
\begin{align}
&\left(
e^{\frac{M}{2k}{\widehat x}}
\frac{\Phi_b\left(\frac{{\widehat x}}{2\pi b}+\frac{iM}{2b}\right)}{\Phi_b\left(\frac{{\widehat x}}{2\pi b}-\frac{iM}{2b}\right)}
\frac{1}{2\cosh\frac{{\widehat p}}{2}}
e^{\left(\frac{1}{2}-\frac{M}{2k}\right){\widehat x}}
\frac{\Phi_b\left(\frac{{\widehat x}}{2\pi b}-\frac{iM}{2b}+\frac{ib}{2}\right)}{\Phi_b\left(\frac{{\widehat x}}{2\pi b}+\frac{iM}{2b}-\frac{ib}{2}\right)}
\right)^{-1}\nonumber \\
&=
\frac{\Phi_b\left(\frac{{\widehat x}}{2\pi b}+\frac{iM}{2b}-\frac{ib}{2}\right)}{\Phi_b\left(\frac{{\widehat x}}{2\pi b}-\frac{iM}{2b}+\frac{ib}{2}\right)}
e^{\left(-\frac{1}{2}+\frac{M}{2k}\right){\widehat x}}
\left(e^{\frac{{\widehat p}}{2}}+e^{-\frac{{\widehat p}}{2}}\right)
\frac{\Phi_b\left(\frac{{\widehat x}}{2\pi b}-\frac{iM}{2b}\right)}{\Phi_b\left(\frac{{\widehat x}}{2\pi b}+\frac{iM}{2b}\right)}
e^{-\frac{M}{2k}{\widehat x}}.
\end{align}
By moving $e^{\pm\frac{{\widehat p}}{2}}$ to the right by using the formula $e^{\pm\frac{{\widehat p}}{2}}f\left({\widehat x}\right)e^{\mp\frac{{\widehat p}}{2}}=f\left({\widehat x}\mp \pi ik\right)$, we are left with ratios of $\Phi_b$ which can be simplified by the recursive formula \eqref{211107quantumdilogrecursive}.
We finally obtain
\begin{align}
&\left(
e^{\frac{M}{2k}{\widehat x}}
\frac{\Phi_b\left(\frac{{\widehat x}}{2\pi b}+\frac{iM}{2b}\right)}{\Phi_b\left(\frac{{\widehat x}}{2\pi b}-\frac{iM}{2b}\right)}
\frac{1}{2\cosh\frac{{\widehat p}}{2}}
e^{\left(\frac{1}{2}-\frac{M}{2k}\right){\widehat x}}
\frac{\Phi_b\left(\frac{{\widehat x}}{2\pi b}-\frac{iM}{2b}+\frac{ib}{2}\right)}{\Phi_b\left(\frac{{\widehat x}}{2\pi b}+\frac{iM}{2b}-\frac{ib}{2}\right)}
\right)^{-1}\nonumber \\
&=
\left(e^{-\frac{\pi iM}{2}}e^{\frac{{\widehat x}}{2}}+e^{\frac{\pi iM}{2}}e^{-\frac{{\widehat x}}{2}}\right)e^{\frac{{\widehat p}}{2}}
+\left(e^{\frac{\pi iM}{2}}e^{\frac{{\widehat x}}{2}}+e^{-\frac{\pi iM}{2}}e^{-\frac{{\widehat x}}{2}}\right)e^{-\frac{{\widehat p}}{2}}.
\label{inverseoflhs}
\end{align}
In the same way the inverse of the right-hand side of \eqref{211107quantumdilogidentity} can be simplified as
\begin{align}
&\left(e^{\left(\frac{1}{2}-\frac{M}{2k}\right){\widehat p}}
\frac{\Phi_b\left(\frac{{\widehat p}}{2\pi b}-\frac{iM}{2b}+\frac{ib}{2}\right)}{\Phi_b\left(\frac{{\widehat p}}{2\pi b}+\frac{iM}{2b}-\frac{ib}{2}\right)}
\frac{1}{2\cosh\frac{{\widehat x}}{2}}
e^{\frac{M}{2k}{\widehat p}}
\frac{\Phi_b\left(\frac{{\widehat p}}{2\pi b}+\frac{iM}{2b}\right)}{\Phi_b\left(\frac{{\widehat p}}{2\pi b}-\frac{iM}{2b}\right)}
\right)^{-1}\nonumber \\
&=
e^{\frac{{\widehat x}}{2}}\left(e^{-\frac{\pi iM}{2}}e^{\frac{{\widehat p}}{2}}+e^{\frac{\pi iM}{2}}e^{-\frac{{\widehat p}}{2}}\right)
+e^{-\frac{{\widehat x}}{2}}\left(e^{\frac{\pi iM}{2}}e^{\frac{{\widehat p}}{2}}+e^{-\frac{\pi iM}{2}}e^{-\frac{{\widehat p}}{2}}\right).
\end{align}
This is identical to the inverse \eqref{inverseoflhs} of the left-hand side of \eqref{211107quantumdilogidentity}.

By using the identity \eqref{211107quantumdilogidentity} we can show \eqref{Weyl_directlyshownonrhohat4}, namely ${\widehat\rho}_k\left(M_1,M_2,0,\zeta_1,\zeta_2\right)\sim {\widehat\rho}_k\left(M_1,M_2,0,\zeta_2,\zeta_1\right)$,  as follows.
First we reorder the terms in ${\widehat\rho}$ \eqref{rhowritteninPhib2} cyclically
\begin{align}
{\widehat\rho}_k\left(M_1,M_2,0,\zeta_1,\zeta_2\right)&\sim
e^{\pi \zeta_2}
e^{\left(-\frac{i\zeta_1}{k}+1-\frac{M_1+M_2}{2k}\right){\widehat x}}
e^{\frac{i\zeta_2}{k}{\widehat p}}
\frac{
\Phi_b\left(\frac{{\widehat x}}{2\pi b}-\frac{iM_2}{2b}+\frac{ib}{2}\right)
}{
\Phi_b\left(\frac{{\widehat x}}{2\pi b}+\frac{iM_2}{2b}-\frac{ib}{2}\right)}
\frac{1}{2\cosh\frac{{\widehat p}}{2}}
\frac{
\Phi_b\left(\frac{{\widehat x}}{2\pi b}+\frac{iM_2}{2b}\right)
}{
\Phi_b\left(\frac{{\widehat x}}{2\pi b}-\frac{iM_2}{2b}\right)
}\nonumber \\
&\quad \times e^{-\frac{i\zeta_2}{k}{\widehat p}}
e^{\left(\frac{i\zeta_1}{k}+\frac{M_1+M_2}{2k}\right){\widehat x}}
\frac{
\Phi_b\left(\frac{{\widehat x}}{2\pi b}+\frac{iM_1}{2b}\right)
}{
\Phi_b\left(\frac{{\widehat x}}{2\pi b}+-\frac{iM_1}{2b}\right)}
\frac{1}{2\cosh\frac{{\widehat p}}{2}}
\frac{
\Phi_b\left(\frac{{\widehat x}}{2\pi b}-\frac{iM_1}{2b}+\frac{ib}{2}\right)
}{
\Phi_b\left(\frac{{\widehat x}}{2\pi b}+\frac{iM_1}{2b}-\frac{ib}{2}\right)
}.
\end{align}
Now to each line we can apply the identity \eqref{211107quantumdilogidentity}.
After combining the exponential factors together and cyclically reordering the terms, we obtain
\begin{align}
&{\widehat\rho}_k\left(M_1,M_2,0,\zeta_1,\zeta_2\right)\nonumber \\
&\sim e^{\pi\zeta_2}
e^{\left(-\frac{i\zeta_1}{k}+1-\frac{M_1+M_2}{2k}\right){\widehat x}}
e^{\frac{i\zeta_2}{k}{\widehat p}}
e^{\left(-\frac{1}{2}+\frac{M_2}{2k}\right){\widehat x}}
e^{\frac{M_2}{2k}{\widehat p}}
\frac{
\Phi_b\left(\frac{{\widehat p}}{2\pi b}+\frac{iM_2}{2b}\right)
}{
\Phi_b\left(\frac{{\widehat p}}{2\pi b}-\frac{iM_2}{2b}\right)
}
\frac{1}{2\cosh\frac{{\widehat x}}{2}}
e^{\left(\frac{1}{2}-\frac{M_2}{2k}\right){\widehat p}}
\frac{
\Phi_b\left(\frac{{\widehat p}}{2\pi b}-\frac{iM_2}{2b}+\frac{ib}{2}\right)
}{
\Phi_b\left(\frac{{\widehat p}}{2\pi b}+\frac{iM_2}{2b}-\frac{ib}{2}\right)
}\nonumber \\
&\quad \times e^{-\frac{M_2}{2k}{\widehat x}}
e^{-\frac{i\zeta_2}{k}{\widehat p}}
e^{\left(\frac{i\zeta_1}{k}+\frac{M_1+M_2}{2k}\right){\widehat x}}
e^{-\frac{M_1}{2k}{\widehat x}}
e^{\left(\frac{1}{2}-\frac{M_1}{2k}\right){\widehat p}}
\frac{
\Phi_b\left(\frac{{\widehat p}}{2\pi b}-\frac{iM_1}{2b}+\frac{ib}{2}\right)
}{
\Phi_b\left(\frac{{\widehat p}}{2\pi b}+\frac{iM_1}{2b}-\frac{ib}{2}\right)
}
\frac{1}{2\cosh\frac{{\widehat x}}{2}}
e^{\frac{M_1}{2k}{\widehat p}}
\frac{
\Phi_b\left(\frac{{\widehat p}}{2\pi b}+\frac{iM_1}{2b}\right)
}{
\Phi_b\left(\frac{{\widehat p}}{2\pi b}-\frac{iM_1}{2b}\right)
}\nonumber \\
&\quad \times e^{\left(-\frac{1}{2}+\frac{M_1}{2k}\right){\widehat x}}\nonumber \\
&\sim
e^{\pi\zeta_1}
e^{\left(-\frac{i\zeta_2}{k}+1-\frac{M_1+M_2}{2k}\right){\widehat p}}
\frac{
\Phi_b\left(\frac{{\widehat p}}{2\pi b}-\frac{iM_1}{2b}+\frac{ib}{2}\right)
}{
\Phi_b\left(\frac{{\widehat p}}{2\pi b}+\frac{iM_1}{2b}-\frac{ib}{2}\right)
}
\frac{
\Phi_b\left(\frac{{\widehat p}}{2\pi b}-\frac{iM_2}{2b}+\frac{ib}{2}+\frac{\zeta_1}{b}\right)
}{
\Phi_b\left(\frac{{\widehat p}}{2\pi b}+\frac{iM_2}{2b}-\frac{ib}{2}+\frac{\zeta_1}{b}\right)
}
\frac{1}{2\cosh\frac{{\widehat x}}{2}}\nonumber \\
&\quad \times e^{\left(\frac{i\zeta_2}{k}+\frac{M_1+M_2}{2k}\right){\widehat p}}
\frac{
\Phi_b\left(\frac{{\widehat p}}{2\pi b}+\frac{iM_1}{2b}\right)
}{
\Phi_b\left(\frac{{\widehat p}}{2\pi b}-\frac{iM_1}{2b}\right)
}
\frac{
\Phi_b\left(\frac{{\widehat p}}{2\pi b}+\frac{iM_2}{2b}+\frac{\zeta_1}{b}\right)
}{
\Phi_b\left(\frac{{\widehat p}}{2\pi b}-\frac{iM_2}{2b}+\frac{\zeta_1}{b}\right)
}
\frac{1}{2\cosh\frac{{\widehat x}}{2}}.
\end{align}
The last expression coincides with ${\widehat\rho}$ \eqref{rhowritteninPhib2} with the replacement $\left(\zeta_1,\zeta_2,{\widehat x},{\widehat p}\right)\rightarrow \left(\zeta_2,\zeta_1,{\widehat p},-{\widehat x}\right)$.
Since the change of variables in ${\widehat x},{\widehat p}$ is a canonical transformation, which can be realized by a similarity transformation, we conclude that ${\widehat\rho}_k\left(M_1,M_2,0,\zeta_2,\zeta_1\right)\sim {\widehat\rho}_k\left(M_1,M_2,0,\zeta_1,\zeta_2\right)$.

\section{Weyl transformations}
\label{220208app_weyltransformations}

We give the list of $2\cdot 4!$ Weyl transformations discussed in section in terms of the generators, realized also as matrices, arranged by length. A positive integer $i_1...i_k$ represents the product $s_{i_1}\cdots s_{i_k}$:
\resizebox{1\linewidth}{!}{
  \begin{minipage}{\linewidth}
\begin{align*}
&
21343=\left(
\begin{smallmatrix}
 1 & 0 & -1 & 0 & 0 \\
 0 & 1 & -1 & 0 & 0 \\
 1 & 1 & -1 & 0 & 0 \\
 0 & 0 & 0 & 1 & 0 \\
 0 & 0 & 0 & 0 & -1 \\
\end{smallmatrix}
\right),\,
213413=\left(
\begin{smallmatrix}
 \nicefrac{1}{2} & \nicefrac{1}{2} & -1 & 0 & -1 \\
 \nicefrac{1}{2} & \nicefrac{1}{2} & -1 & 0 & 1 \\
 1 & 1 & -1 & 0 & 0 \\
 0 & 0 & 0 & 1 & 0 \\
 \nicefrac{1}{2} & -\nicefrac{1}{2} & 0 & 0 & 0 \\
\end{smallmatrix}
\right),\,
213423=\left(
\begin{smallmatrix}
 \nicefrac{1}{2} & \nicefrac{1}{2} & -1 & 0 & 1 \\
 \nicefrac{1}{2} & \nicefrac{1}{2} & -1 & 0 & -1 \\
 1 & 1 & -1 & 0 & 0 \\
 0 & 0 & 0 & 1 & 0 \\
 -\nicefrac{1}{2} & \nicefrac{1}{2} & 0 & 0 & 0 \\
\end{smallmatrix}
\right),\,
\\
&
432134=\left(
\begin{smallmatrix}
 1 & 0 & 0 & 0 & 0 \\
 0 & 1 & 0 & 0 & 0 \\
 1 & 1 & -1 & 0 & 0 \\
 0 & 0 & 0 & 1 & 0 \\
 0 & 0 & 0 & 0 & -1 \\
\end{smallmatrix}
\right),\,
452134=\left(
\begin{smallmatrix}
 \nicefrac{1}{2} & \nicefrac{1}{2} & -1 & 1 & 0 \\
 \nicefrac{1}{2} & \nicefrac{1}{2} & -1 & -1 & 0 \\
 1 & 1 & -1 & 0 & 0 \\
 \nicefrac{1}{2} & -\nicefrac{1}{2} & 0 & 0 & 0 \\
 0 & 0 & 0 & 0 & -1 \\
\end{smallmatrix}
\right),\,
2134213=\left(
\begin{smallmatrix}
 0 & 1 & -1 & 0 & 0 \\
 1 & 0 & -1 & 0 & 0 \\
 1 & 1 & -1 & 0 & 0 \\
 0 & 0 & 0 & 1 & 0 \\
 0 & 0 & 0 & 0 & 1 \\
\end{smallmatrix}
\right),\,
\\
&
4532134=\left(
\begin{smallmatrix}
 \nicefrac{1}{2} & \nicefrac{1}{2} & 0 & 1 & 0 \\
 \nicefrac{1}{2} & \nicefrac{1}{2} & 0 & -1 & 0 \\
 1 & 1 & -1 & 0 & 0 \\
 \nicefrac{1}{2} & -\nicefrac{1}{2} & 0 & 0 & 0 \\
 0 & 0 & 0 & 0 & -1 \\
\end{smallmatrix}
\right),\,
2345134=\left(
\begin{smallmatrix}
 \nicefrac{1}{2} & \nicefrac{1}{2} & -1 & 0 & -1 \\
 \nicefrac{1}{2} & \nicefrac{1}{2} & -1 & 0 & 1 \\
 1 & 1 & -1 & 0 & 0 \\
 \nicefrac{1}{2} & -\nicefrac{1}{2} & 0 & 0 & 0 \\
 0 & 0 & 0 & 1 & 0 \\
\end{smallmatrix}
\right),\,
1345234=\left(
\begin{smallmatrix}
 \nicefrac{1}{2} & \nicefrac{1}{2} & -1 & 0 & 1 \\
 \nicefrac{1}{2} & \nicefrac{1}{2} & -1 & 0 & -1 \\
 1 & 1 & -1 & 0 & 0 \\
 \nicefrac{1}{2} & -\nicefrac{1}{2} & 0 & 0 & 0 \\
 0 & 0 & 0 & -1 & 0 \\
\end{smallmatrix}
\right),\,
\\
&
432134131=\left(
\begin{smallmatrix}
 \nicefrac{1}{2} & \nicefrac{1}{2} & 0 & 0 & -1 \\
 \nicefrac{1}{2} & \nicefrac{1}{2} & 0 & 0 & 1 \\
 1 & 1 & -1 & 0 & 0 \\
 0 & 0 & 0 & 1 & 0 \\
 \nicefrac{1}{2} & -\nicefrac{1}{2} & 0 & 0 & 0 \\
\end{smallmatrix}
\right),\,
452134131=\left(
\begin{smallmatrix}
 \nicefrac{1}{2} & \nicefrac{1}{2} & -1 & 1 & 0 \\
 \nicefrac{1}{2} & \nicefrac{1}{2} & -1 & -1 & 0 \\
 1 & 1 & -1 & 0 & 0 \\
 0 & 0 & 0 & 0 & -1 \\
 \nicefrac{1}{2} & -\nicefrac{1}{2} & 0 & 0 & 0 \\
\end{smallmatrix}
\right),\,
432134232=\left(
\begin{smallmatrix}
 \nicefrac{1}{2} & \nicefrac{1}{2} & 0 & 0 & 1 \\
 \nicefrac{1}{2} & \nicefrac{1}{2} & 0 & 0 & -1 \\
 1 & 1 & -1 & 0 & 0 \\
 0 & 0 & 0 & 1 & 0 \\
 -\nicefrac{1}{2} & \nicefrac{1}{2} & 0 & 0 & 0 \\
\end{smallmatrix}
\right),\,
\\
&
452134232=\left(
\begin{smallmatrix}
 \nicefrac{1}{2} & \nicefrac{1}{2} & -1 & 1 & 0 \\
 \nicefrac{1}{2} & \nicefrac{1}{2} & -1 & -1 & 0 \\
 1 & 1 & -1 & 0 & 0 \\
 0 & 0 & 0 & 0 & 1 \\
 -\nicefrac{1}{2} & \nicefrac{1}{2} & 0 & 0 & 0 \\
\end{smallmatrix}
\right),\,
4532134131=\left(
\begin{smallmatrix}
 \nicefrac{1}{2} & \nicefrac{1}{2} & 0 & 1 & 0 \\
 \nicefrac{1}{2} & \nicefrac{1}{2} & 0 & -1 & 0 \\
 1 & 1 & -1 & 0 & 0 \\
 0 & 0 & 0 & 0 & -1 \\
 \nicefrac{1}{2} & -\nicefrac{1}{2} & 0 & 0 & 0 \\
\end{smallmatrix}
\right),\,
2345134131=\left(
\begin{smallmatrix}
 1 & 0 & -1 & 0 & 0 \\
 0 & 1 & -1 & 0 & 0 \\
 1 & 1 & -1 & 0 & 0 \\
 0 & 0 & 0 & 0 & -1 \\
 0 & 0 & 0 & 1 & 0 \\
\end{smallmatrix}
\right),\,
\\
&
1345234131=\left(
\begin{smallmatrix}
 0 & 1 & -1 & 0 & 0 \\
 1 & 0 & -1 & 0 & 0 \\
 1 & 1 & -1 & 0 & 0 \\
 0 & 0 & 0 & 0 & -1 \\
 0 & 0 & 0 & -1 & 0 \\
\end{smallmatrix}
\right),\,
2345321341=\left(
\begin{smallmatrix}
 \nicefrac{1}{2} & \nicefrac{1}{2} & 0 & 0 & -1 \\
 \nicefrac{1}{2} & \nicefrac{1}{2} & 0 & 0 & 1 \\
 1 & 1 & -1 & 0 & 0 \\
 \nicefrac{1}{2} & -\nicefrac{1}{2} & 0 & 0 & 0 \\
 0 & 0 & 0 & 1 & 0 \\
\end{smallmatrix}
\right),\,
4532134232=\left(
\begin{smallmatrix}
 \nicefrac{1}{2} & \nicefrac{1}{2} & 0 & 1 & 0 \\
 \nicefrac{1}{2} & \nicefrac{1}{2} & 0 & -1 & 0 \\
 1 & 1 & -1 & 0 & 0 \\
 0 & 0 & 0 & 0 & 1 \\
 -\nicefrac{1}{2} & \nicefrac{1}{2} & 0 & 0 & 0 \\
\end{smallmatrix}
\right),\,
\\
&
2345134232=\left(
\begin{smallmatrix}
 0 & 1 & -1 & 0 & 0 \\
 1 & 0 & -1 & 0 & 0 \\
 1 & 1 & -1 & 0 & 0 \\
 0 & 0 & 0 & 0 & 1 \\
 0 & 0 & 0 & 1 & 0 \\
\end{smallmatrix}
\right),\,
1345234232=\left(
\begin{smallmatrix}
 1 & 0 & -1 & 0 & 0 \\
 0 & 1 & -1 & 0 & 0 \\
 1 & 1 & -1 & 0 & 0 \\
 0 & 0 & 0 & 0 & 1 \\
 0 & 0 & 0 & -1 & 0 \\
\end{smallmatrix}
\right),\,
1345321342=\left(
\begin{smallmatrix}
 \nicefrac{1}{2} & \nicefrac{1}{2} & 0 & 0 & 1 \\
 \nicefrac{1}{2} & \nicefrac{1}{2} & 0 & 0 & -1 \\
 1 & 1 & -1 & 0 & 0 \\
 \nicefrac{1}{2} & -\nicefrac{1}{2} & 0 & 0 & 0 \\
 0 & 0 & 0 & -1 & 0 \\
\end{smallmatrix}
\right),\,
\\
&
4321343213=\left(
\begin{smallmatrix}
 0 & 1 & 0 & 0 & 0 \\
 1 & 0 & 0 & 0 & 0 \\
 1 & 1 & -1 & 0 & 0 \\
 0 & 0 & 0 & 1 & 0 \\
 0 & 0 & 0 & 0 & 1 \\
\end{smallmatrix}
\right),\,
4521343213=\left(
\begin{smallmatrix}
 \nicefrac{1}{2} & \nicefrac{1}{2} & -1 & 1 & 0 \\
 \nicefrac{1}{2} & \nicefrac{1}{2} & -1 & -1 & 0 \\
 1 & 1 & -1 & 0 & 0 \\
 -\nicefrac{1}{2} & \nicefrac{1}{2} & 0 & 0 & 0 \\
 0 & 0 & 0 & 0 & 1 \\
\end{smallmatrix}
\right),\,
3213452134=\left(
\begin{smallmatrix}
 \nicefrac{1}{2} & \nicefrac{1}{2} & -1 & -1 & 0 \\
 \nicefrac{1}{2} & \nicefrac{1}{2} & -1 & 1 & 0 \\
 1 & 1 & -1 & 0 & 0 \\
 \nicefrac{1}{2} & -\nicefrac{1}{2} & 0 & 0 & 0 \\
 0 & 0 & 0 & 0 & 1 \\
\end{smallmatrix}
\right),\,
\\
&
23453213431=\left(
\begin{smallmatrix}
 1 & 0 & 0 & 0 & 0 \\
 0 & 1 & 0 & 0 & 0 \\
 1 & 1 & -1 & 0 & 0 \\
 0 & 0 & 0 & 0 & -1 \\
 0 & 0 & 0 & 1 & 0 \\
\end{smallmatrix}
\right),\,
13453213432=\left(
\begin{smallmatrix}
 1 & 0 & 0 & 0 & 0 \\
 0 & 1 & 0 & 0 & 0 \\
 1 & 1 & -1 & 0 & 0 \\
 0 & 0 & 0 & 0 & 1 \\
 0 & 0 & 0 & -1 & 0 \\
\end{smallmatrix}
\right),\,
45321343213=\left(
\begin{smallmatrix}
 \nicefrac{1}{2} & \nicefrac{1}{2} & 0 & 1 & 0 \\
 \nicefrac{1}{2} & \nicefrac{1}{2} & 0 & -1 & 0 \\
 1 & 1 & -1 & 0 & 0 \\
 -\nicefrac{1}{2} & \nicefrac{1}{2} & 0 & 0 & 0 \\
 0 & 0 & 0 & 0 & 1 \\
\end{smallmatrix}
\right),\,
\\
&
23451343213=\left(
\begin{smallmatrix}
 \nicefrac{1}{2} & \nicefrac{1}{2} & -1 & 0 & 1 \\
 \nicefrac{1}{2} & \nicefrac{1}{2} & -1 & 0 & -1 \\
 1 & 1 & -1 & 0 & 0 \\
 -\nicefrac{1}{2} & \nicefrac{1}{2} & 0 & 0 & 0 \\
 0 & 0 & 0 & 1 & 0 \\
\end{smallmatrix}
\right),\,
13452343213=\left(
\begin{smallmatrix}
 \nicefrac{1}{2} & \nicefrac{1}{2} & -1 & 0 & -1 \\
 \nicefrac{1}{2} & \nicefrac{1}{2} & -1 & 0 & 1 \\
 1 & 1 & -1 & 0 & 0 \\
 -\nicefrac{1}{2} & \nicefrac{1}{2} & 0 & 0 & 0 \\
 0 & 0 & 0 & -1 & 0 \\
\end{smallmatrix}
\right),\,
32134532134=\left(
\begin{smallmatrix}
 \nicefrac{1}{2} & \nicefrac{1}{2} & 0 & -1 & 0 \\
 \nicefrac{1}{2} & \nicefrac{1}{2} & 0 & 1 & 0 \\
 1 & 1 & -1 & 0 & 0 \\
 \nicefrac{1}{2} & -\nicefrac{1}{2} & 0 & 0 & 0 \\
 0 & 0 & 0 & 0 & 1 \\
\end{smallmatrix}
\right),\,
\\
&
1345321342131=\left(
\begin{smallmatrix}
 0 & 1 & 0 & 0 & 0 \\
 1 & 0 & 0 & 0 & 0 \\
 1 & 1 & -1 & 0 & 0 \\
 0 & 0 & 0 & 0 & -1 \\
 0 & 0 & 0 & -1 & 0 \\
\end{smallmatrix}
\right),\,
3213452134131=\left(
\begin{smallmatrix}
 \nicefrac{1}{2} & \nicefrac{1}{2} & -1 & -1 & 0 \\
 \nicefrac{1}{2} & \nicefrac{1}{2} & -1 & 1 & 0 \\
 1 & 1 & -1 & 0 & 0 \\
 0 & 0 & 0 & 0 & -1 \\
 -\nicefrac{1}{2} & \nicefrac{1}{2} & 0 & 0 & 0 \\
\end{smallmatrix}
\right),\,
2345321342132=\left(
\begin{smallmatrix}
 0 & 1 & 0 & 0 & 0 \\
 1 & 0 & 0 & 0 & 0 \\
 1 & 1 & -1 & 0 & 0 \\
 0 & 0 & 0 & 0 & 1 \\
 0 & 0 & 0 & 1 & 0 \\
\end{smallmatrix}
\right),\,
\\
&
3213452134232=\left(
\begin{smallmatrix}
 \nicefrac{1}{2} & \nicefrac{1}{2} & -1 & -1 & 0 \\
 \nicefrac{1}{2} & \nicefrac{1}{2} & -1 & 1 & 0 \\
 1 & 1 & -1 & 0 & 0 \\
 0 & 0 & 0 & 0 & 1 \\
 \nicefrac{1}{2} & -\nicefrac{1}{2} & 0 & 0 & 0 \\
\end{smallmatrix}
\right),\,
5432134521343=\left(
\begin{smallmatrix}
 1 & 0 & -1 & 0 & 0 \\
 0 & 1 & -1 & 0 & 0 \\
 1 & 1 & -1 & 0 & 0 \\
 0 & 0 & 0 & -1 & 0 \\
 0 & 0 & 0 & 0 & 1 \\
\end{smallmatrix}
\right),\,
13453213432131=\left(
\begin{smallmatrix}
 \nicefrac{1}{2} & \nicefrac{1}{2} & 0 & 0 & -1 \\
 \nicefrac{1}{2} & \nicefrac{1}{2} & 0 & 0 & 1 \\
 1 & 1 & -1 & 0 & 0 \\
 -\nicefrac{1}{2} & \nicefrac{1}{2} & 0 & 0 & 0 \\
 0 & 0 & 0 & -1 & 0 \\
\end{smallmatrix}
\right),\,
\\
&
32134532134131=\left(
\begin{smallmatrix}
 \nicefrac{1}{2} & \nicefrac{1}{2} & 0 & -1 & 0 \\
 \nicefrac{1}{2} & \nicefrac{1}{2} & 0 & 1 & 0 \\
 1 & 1 & -1 & 0 & 0 \\
 0 & 0 & 0 & 0 & -1 \\
 -\nicefrac{1}{2} & \nicefrac{1}{2} & 0 & 0 & 0 \\
\end{smallmatrix}
\right),\,
23453213432132=\left(
\begin{smallmatrix}
 \nicefrac{1}{2} & \nicefrac{1}{2} & 0 & 0 & 1 \\
 \nicefrac{1}{2} & \nicefrac{1}{2} & 0 & 0 & -1 \\
 1 & 1 & -1 & 0 & 0 \\
 -\nicefrac{1}{2} & \nicefrac{1}{2} & 0 & 0 & 0 \\
 0 & 0 & 0 & 1 & 0 \\
\end{smallmatrix}
\right),\,
32134532134232=\left(
\begin{smallmatrix}
 \nicefrac{1}{2} & \nicefrac{1}{2} & 0 & -1 & 0 \\
 \nicefrac{1}{2} & \nicefrac{1}{2} & 0 & 1 & 0 \\
 1 & 1 & -1 & 0 & 0 \\
 0 & 0 & 0 & 0 & 1 \\
 \nicefrac{1}{2} & -\nicefrac{1}{2} & 0 & 0 & 0 \\
\end{smallmatrix}
\right),\,
\\
&
32134521343213=\left(
\begin{smallmatrix}
 \nicefrac{1}{2} & \nicefrac{1}{2} & -1 & -1 & 0 \\
 \nicefrac{1}{2} & \nicefrac{1}{2} & -1 & 1 & 0 \\
 1 & 1 & -1 & 0 & 0 \\
 -\nicefrac{1}{2} & \nicefrac{1}{2} & 0 & 0 & 0 \\
 0 & 0 & 0 & 0 & -1 \\
\end{smallmatrix}
\right),\,
54321345213413=\left(
\begin{smallmatrix}
 \nicefrac{1}{2} & \nicefrac{1}{2} & -1 & 0 & -1 \\
 \nicefrac{1}{2} & \nicefrac{1}{2} & -1 & 0 & 1 \\
 1 & 1 & -1 & 0 & 0 \\
 0 & 0 & 0 & -1 & 0 \\
 -\nicefrac{1}{2} & \nicefrac{1}{2} & 0 & 0 & 0 \\
\end{smallmatrix}
\right),\,
54321345213423=\left(
\begin{smallmatrix}
 \nicefrac{1}{2} & \nicefrac{1}{2} & -1 & 0 & 1 \\
 \nicefrac{1}{2} & \nicefrac{1}{2} & -1 & 0 & -1 \\
 1 & 1 & -1 & 0 & 0 \\
 0 & 0 & 0 & -1 & 0 \\
 \nicefrac{1}{2} & -\nicefrac{1}{2} & 0 & 0 & 0 \\
\end{smallmatrix}
\right),\,
\\
&
54321345432134=\left(
\begin{smallmatrix}
 1 & 0 & 0 & 0 & 0 \\
 0 & 1 & 0 & 0 & 0 \\
 1 & 1 & -1 & 0 & 0 \\
 0 & 0 & 0 & -1 & 0 \\
 0 & 0 & 0 & 0 & 1 \\
\end{smallmatrix}
\right),\,
321345321343213=\left(
\begin{smallmatrix}
 \nicefrac{1}{2} & \nicefrac{1}{2} & 0 & -1 & 0 \\
 \nicefrac{1}{2} & \nicefrac{1}{2} & 0 & 1 & 0 \\
 1 & 1 & -1 & 0 & 0 \\
 -\nicefrac{1}{2} & \nicefrac{1}{2} & 0 & 0 & 0 \\
 0 & 0 & 0 & 0 & -1 \\
\end{smallmatrix}
\right),\,
543213452134213=\left(
\begin{smallmatrix}
 0 & 1 & -1 & 0 & 0 \\
 1 & 0 & -1 & 0 & 0 \\
 1 & 1 & -1 & 0 & 0 \\
 0 & 0 & 0 & -1 & 0 \\
 0 & 0 & 0 & 0 & -1 \\
\end{smallmatrix}
\right),\,
\\
&
54321345432134131=\left(
\begin{smallmatrix}
 \nicefrac{1}{2} & \nicefrac{1}{2} & 0 & 0 & -1 \\
 \nicefrac{1}{2} & \nicefrac{1}{2} & 0 & 0 & 1 \\
 1 & 1 & -1 & 0 & 0 \\
 0 & 0 & 0 & -1 & 0 \\
 -\nicefrac{1}{2} & \nicefrac{1}{2} & 0 & 0 & 0 \\
\end{smallmatrix}
\right),\,
54321345432134232=\left(
\begin{smallmatrix}
 \nicefrac{1}{2} & \nicefrac{1}{2} & 0 & 0 & 1 \\
 \nicefrac{1}{2} & \nicefrac{1}{2} & 0 & 0 & -1 \\
 1 & 1 & -1 & 0 & 0 \\
 0 & 0 & 0 & -1 & 0 \\
 \nicefrac{1}{2} & -\nicefrac{1}{2} & 0 & 0 & 0 \\
\end{smallmatrix}
\right),\,
543213454321343213=\left(
\begin{smallmatrix}
 0 & 1 & 0 & 0 & 0 \\
 1 & 0 & 0 & 0 & 0 \\
 1 & 1 & -1 & 0 & 0 \\
 0 & 0 & 0 & -1 & 0 \\
 0 & 0 & 0 & 0 & -1 \\
\end{smallmatrix}
\right).
\end{align*}
\end{minipage}
}

\bibliography{bunken_210824Nosaka.bib}
\end{document}